\documentclass[twocolumn, dvipsnames]{aastex631}
\usepackage{amsmath}
\usepackage{graphicx}
\usepackage{graphbox}
\usepackage{tikz}
\usepackage{float}
\usepackage{multirow}
\usepackage{color}
\usepackage{makecell}
\usepackage{wrapfig}
\usepackage{comment}

\usepackage[colorinlistoftodos]{todonotes}

\usepackage{xcolor}

\usepackage{enumitem}

\accepted{in ApJS - November 2023}

\shorttitle{Atmospheric variability of WASP-121b}
\shortauthors{Changeat, Skinner et al. 2023}

\begin{document}

	\title{Is the atmosphere of the ultra-hot Jupiter WASP-121\,b variable?}

	\correspondingauthor{Q. Changeat}
	\email{qchangeat@stsci.edu}
	\author[0000-0001-6516-4493]{Q. Changeat $^\dagger$}\altaffiliation{These authors      contributed equally to this work.}
        \affil{European Space Agency (ESA), ESA Office, 
        Space Telescope Science Institute (STScI), Baltimore MD 21218, USA.}
	    \affil{Department of Physics and Astronomy,
		University College London,
		Gower Street,WC1E 6BT London, United Kingdom}
    \author[0000-0002-5263-385X]{J. W. Skinner}\altaffiliation{These authors                contributed equally to this work.}
	    %\affil{School of Physics and Astronomy,Queen Mary University of London, Mile End Road, London, E1 4NS, United Kingdom}
        %\affil{Jet Propulsion Laboratory, California Institute of Technology, Pasadena, CA, USA}
        %\affil{Joint Institute for Regional Earth System Science and Engineering, University of California Los Angeles, Los Angeles, CA, USA}
        \affil{Division of Geological and Planetary Sciences, California Institute of Technology, Pasadena, California, USA}
        \affil{Martin A. Fisher School of Physics, Brandeis University, 
        Waltham MA 02453, USA}
    \author[0000-0002-4525-5651]{J. Y-K. Cho}
          \affil{Martin A. Fisher School of Physics, Brandeis University, Waltham MA 02453, USA}
	   \affil{Center for Computational Astrophysics,
		Flatiron Institute,
		162 Fifth Avenue, New York, NY 10010, USA}
    \author[0000-0002-3226-4575]{J. N\"attil\"a}
            \affil{Center for Computational Astrophysics,
		Flatiron Institute,
		162 Fifth Avenue, New York, NY 10010, USA}
        \affil{Physics Department and Columbia Astrophysics Laboratory, Columbia University, New York NY 10027, USA}
    \author[0000-0002-4205-5267]{I. P. Waldmann}
	\affil{Department of Physics and Astronomy,
		University College London,
		Gower Street,WC1E 6BT London, United Kingdom}
    \author[0000-0003-2241-5330]{A. F. Al-Refaie}
	   \affil{Department of Physics and Astronomy,
		University College London,
		Gower Street,WC1E 6BT London, United Kingdom}
    \author[0000-0001-7189-6463]{A. Dyrek}
	   \affil{Université Paris Cité, Université Paris-Saclay, CEA, CNRS, AIM, F-91191 Gif-sur-Yvette, France}
	\author[0000-0002-5494-3237]{B. Edwards}
        \affil{SRON, Netherlands Institute for Space Research, Niels Bohrweg 4, NL-2333 CA, Leiden, The Netherlands}
	   \affil{Department of Physics and Astronomy,
		University College London,
		Gower Street,WC1E 6BT London, United Kingdom}
  \author[0000-0001-5442-1300]{T. Mikal-Evans}
        \affil{Max Planck Institute for Astronomy, Heidelberg 69117, Germany}
        \author[0000-0002-1386-6378]{M. Joshua}
        \affil{Blue Skies Space Ltd., London EC2A 2BB, UK}
        \author[0000-0001-6193-0576]{G. Morello}
        \affil{Department of Space, Earth and Environment, Chalmers University of Technology, SE-412 96 Gothenburg, Sweden}
        \affil{Instituto de Astrof\'isica de Canarias (IAC), 38205 La Laguna, Tenerife, Spain}
        \author[0000-0002-9372-5056]{N. Skaf}
        \affil{Astronomy \& Astrophysics Department, University of California, Santa Cruz, CA 95064, USA}
        \author[0000-0003-3840-1793]{A. Tsiaras}
        \affil{Department of Physics and Astronomy,
		University College London,
		Gower Street,WC1E 6BT London, United Kingdom}
        \author[0000-0003-2854-765X]{O. Venot}
        \affil{Univ de Paris Cite and Univ Paris Est Creteil, CNRS, LISA, 
        F-75013 Paris, France}
        \author[0000-0002-9616-1524]{K. H. Yip}
	\affil{Department of Physics and Astronomy,
		University College London,
		Gower Street,WC1E 6BT London, United Kingdom}
%%%%%%%%%%%%%%%%%%%%%%%%%%%%%%%%%%%%%%%%%%%%%%%%%%%%%%%%%%%%%%%%%%%%%%%%%%%%%%%%

\renewcommand{\thefootnote}{\fnsymbol{footnote}}
\footnotetext[2]{ESA Research Fellow}
\renewcommand{\thefootnote}{\arabic{footnote}}

\begin{abstract}
We present a comprehensive analysis of the Hubble Space Telescope
observations of the atmosphere of WASP-121\,b, a ultra-hot Jupiter. 
After reducing the transit, eclipse, and phase-curve observations 
with a uniform methodology and addressing the biases from 
instrument systematics, sophisticated atmospheric retrievals are 
used to extract robust constraints on the thermal structure, 
chemistry, and cloud properties of the atmosphere. 
Our analysis shows that the observations are consistent with a 
strong thermal inversion beginning at $\sim$$10^4$\,Pa on the 
dayside, solar to subsolar metallicity $Z$ 
(i.e., $-0.77 < \log(Z) < 0.05$), and super-solar C/O ratio 
(i.e., $0.59 < \textrm{C/O} < 0.87$). 
More importantly, utilizing the high signal-to-noise ratio and 
repeated observations of the planet, we identify the following 
unambiguous time-varying signals in the data: {\it i}) a shift of 
the putative {\it hotspot} offset between the two phase-curves and 
{\it ii})~varying spectral signatures in the transits and eclipses.
By simulating the global dynamics of WASP-121\,b 
atmosphere at high-resolution, we show that the identified signals 
are consistent with quasi-periodic weather patterns, hence 
atmospheric variability, with signatures at the level probed by 
the observations ($\sim$5\% to $\sim$10\%) that change on a
timescale of $\sim$5\,planet days; in the simulations, the 
weather patterns arise from the formation and movement of storms 
and fronts, causing hot (as well as cold) patches of atmosphere 
to deform, separate, and mix in time. 
\end{abstract}


\vspace{0.3cm}

\section{Introduction}\label{intro}

Spectroscopic measurements of transiting exoplanets have provided a 
wealth of ``snapshot'' information about the thermal structure, 
chemistry, and cloud properties of exoplanet atmospheres \citep[e.g.,][]{charbonneau_2002, tinetti_water, kreidberg_w43, Kreidberg_GJ1214b_clouds, swain_2008_hd189_nicmos, swain_hd189_nicmos_tr, Stevenson_2014, Madhu_2014, Line_HD209_spectrum_em, sing_pop, tsiaras_30planets, tsiaras_h2o, Benneke_2019, mansfield_2021_pop, Edwards_2020_ares, skaf_aresII, pluriel_aresIII, Mugnai_2021, Line_2021_nature, Changeat_2022_five, 
edwards_pop, JWST_2023}.
However, temporally varying information has yet to be unambiguously 
obtained by observations. 
This is partly because, prior to the recently launched James Webb 
Space Telescope (JWST), exoplanet atmospheres have generally been 
studied with a single observation whose spectral feature 
signal-to-noise ratio (S/N) is too low. 
In an attempt to reduce the noise, the current standard practice is to 
average the signal from different observations; however,
the averaging removes any temporal variability that may be captured. 
On the other hand, when a single observation can achieve a high 
enough S/N, the observation of the planet is generally not 
repeated---due to observing time constraints. 

With the Spitzer telescope, a number of studies have analyzed 
repeated measurements of individual transiting exoplanets via 
photometric multi-epoch measurements of eclipses. 
Many of these studies did not detect atmospheric variability below 
a certain level, due to the quality of the data \cite[e.g.,][]{Agol_2010,  Crossfield_2012_HD209, Ingalls_2016, Morello_2016, Kilpatrick_2020, Murphy_2022}. 
Others, however, have suggested time-dependent shifts in phase-curve 
offsets for at least three exoplanets---HAT-P-7\,b, WASP-12\,b, 
and Kepler-76\,b---using either the Kepler or Spitzer telescopes \citep{armstrong_hatp7, Bell_2018, Jackson_2019, Wilson_2021, 
Ouyang_2023}. 
The latter studies have speculated that such changes might be due 
to varying cloud structures, but conclusive interpretations of the 
datasets have remained elusive \citep{Bell_2019, Lally_2022, 
Wong_2022}. 
Hence, presently there exists no unambiguous detection of 
atmospheric variability in the atmospheres of transiting exoplanets. 

In contrast, atmospheric variability is commonly reported for 
non-transiting exoplanets which are characterized by high-contrast 
imaging \citep[e.g.,][]{Artigau_2009, Biller_2015, Metchev_2015, 
Biller_2017, Manjavacas_2019, Vos_2022}. 
Among them, the $\sim$11--19\,Jupiter-mass planet VHS\,1256-1257\,b, 
which has recently been observed by the JWST-NirSpec and JWST-MIRI 
instruments as  part of the Early Release Science program 
\citep{Miles_2022}, exhibits one of the largest amplitudes of observed atmospheric variability; for example, a periodic brightness change of up to 38\% 
with a period of $\sim$15\,hours  is reported by \citet{Zhou_2022}. 
Many other planetary mass companions exhibit similar levels of 
variability.

Recently, intriguing possibility of time-variability for the 
ultra-hot Jupiter WASP-121\,b has been reported in two studies, 
\cite{Wilson_2021} and \cite{Ouyang_2023}. 
The former study compares spectra from a ground-based observation 
using Gemini-GMOS and a Hubble Space Telescope (HST) observation
using HST-STIS and finds differences in the two spectra, which 
could be associated with a temporal variability. 
The latter study uses ground-based data from SOAR-GHTS and finds 
a spectra that also does not match that of the previous HST 
observations. 
These studies associate the observed differences with the presence 
of enhanced scattering slope in the case of GMOS, which could be 
explained by clouds or hazes, 
and varying abundances of molecular TiO/VO in the case of GHTS. 
Additionally, phase-curve observations with the HST 
\citep{Evans_2022_diunarl}, Spitzer \citep{Morello_2023}, and 
JWST-NirSpec G395H \citep{evans_2023_jwst} also report different 
phase-curve characteristics (i.e., ``hotspot'' offset and shape), 
which could be indicative of moving hot regions in the atmosphere. 
However, the variability inferred in these works again relies on 
combining the constraints from different instruments and/or 
observing conditions. 

On the atmospheric dynamics modeling side, many hot Jupiter 
simulations in the past have suggested the presence of a single, 
stationary hot region eastward of the substellar 
point---particularly between the $\sim$$10^4$\,Pa to 
$\sim$$10^3$\,Pa pressure levels \citep[e.g.,][]{Cooper_2006, 
Dobbs-Dixon_2010, Parmentier_2018_w121photodiss, Komacek_2020}. 
However, at high-resolution, highly-dynamic, variably-shaped 
(and often multiple) hot regions emerge instead \citep{Skinner_2021, 
Cho_2021, Skinner_2021_modons,Skinner_2022_cyclogenesis}. 
In these simulations, a long-lived giant storm-pair forms near the substellar point, drifts initially toward one of the terminators, 
then rapidly translate westward thereafter---traversing the 
nightside and ultimately breaking up or dissipating near the 
eastern terminator; this cycle is quasi-periodic.
Throughout each cycle, hot (as well as cold) patches of air are 
chaotically mixed over large areas and distances by the storms 
and sharp fronts around them.  
Similar mixing due to storms and fronts has initially been 
predicted in high-resolution simulations (with a different 
initial condition than in the above studies) by \cite{Cho_2003}, 
who suggested that weather patterns would lead to potentially 
observable variability on hot exoplanets. 

In this backdrop, we present here results from an in-depth study 
of WASP-121\,b atmosphere---focusing on its variability. 
WASP-121\,b is one of the best targets for atmospheric 
characterization because it is characterized by a high S/N 
{\it and} has been observed multiple times.
It has been observed four times with the HST Wide Field Camera 3 
Grism~141 (WFC3-G141): one transit in June 2016, one eclipse in 
November 2016, and two phase-curves in March 2018 and February 2019. 
Significantly, we utilize the entirety of these observations in 
this work.
Previous analyses from a combination of facilities---HST, TESS, 
Spitzer, JWST and ground-based facilities \citep{ tsiaras_30planets, Evans_wasp121_t2, Evans_wasp121_e3, Sing_2019, Cabot_2020, 
Ben-Yami_2020, hoeij_w121, Borsa_2021, daylan_tess_wasp121, 
Evans_2022_diunarl, Changeat_2022_five, Gibson_2022, Silva_2022, evans_2023_jwst}---have detected the presence of water vapour, 
absorbers of visible radiation (VO and TiO), hydrogen ions (H$^-$), 
and atomic species (Ba, Ca, Cr, Fe, H, K, Li, Mg, Na, V, and Sr); 
but, atmospheric variability has not yet been detected.

The outline of this paper is as follows.
Our basic methodology and codes are presented in Section~2. 
The reduction procedure for the construction of a consistent 
set of spectra from the raw observations is described in Section~3.
Then, the results from our state-of-the-art atmospheric retrievals, 
which use the newly-developed ``1.5D phase-curve retrieval'' 
models \citep{Changeat_2020_phasecurve1, changeat_2021_phasecurve2} 
and one-dimensional (1D) models, are presented in Section~4 and 
Section 5, respectively; the 1.5D models enable mapping of the 
atmospheric properties (i.e. chemistry, cloud, and thermal 
structure) as a function of longitude using the entire phase-curve 
data, while the 1D models are used to extract information from each 
of the recovered spectrum individually.
In Section 6, we present the findings from the highest resolution, 
three-dimensional (3D) dynamics simulations of WASP-121\,b 
atmosphere to date.
Finally, discussion and conclusion are presented in Section 7.
Additional material is included in an Appendix as well.

\begin{figure*}
    \vspace*{0.3cm}
    \includegraphics[width = 0.99\textwidth]{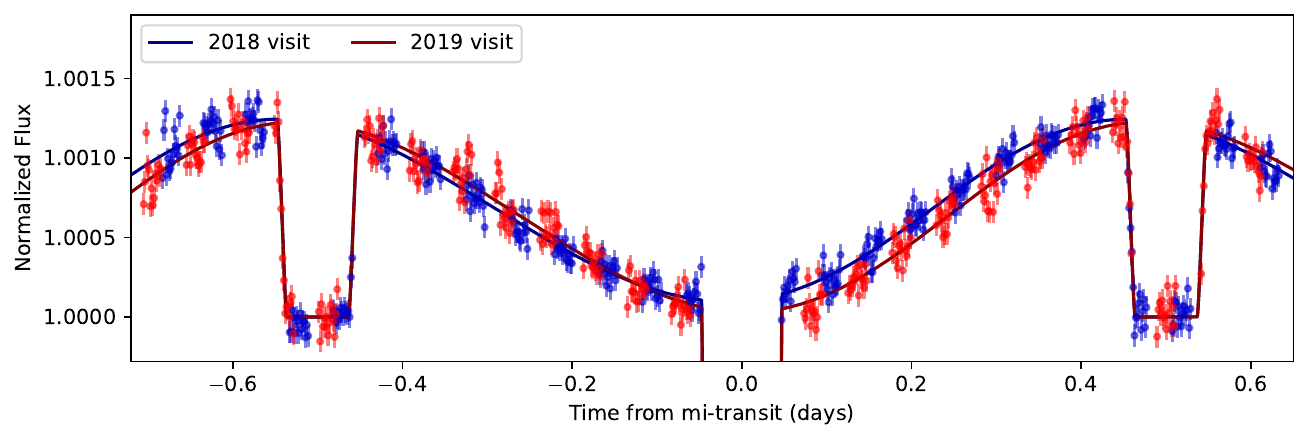}
    \caption{Corrected white light curves when the two WASP-121\,b 
    phase-curve visits are fitted individually. 
    The two observations, while close, present differences that 
    could come from either atmospheric variability or  
    instrument systematics. 
    Note that the second-order coefficients from the sine 
    phase-curve models are fixed to the best-fit values from 
    the combined fit (see Figure~\ref{fig:white_corrected}). }
    \label{fig:two_runs_lc}
\end{figure*}

\section{Methodology}
 
The WASP-121\,b data we have analyzed in this work is obtained 
using HST WFC3-G141. 
This data is publicly available at the Mikulski Archive for Space 
Telescopes (MAST)\footnote{\url{https://archive.stsci.edu/}\ \ ; \\
data from HST proposals P14468 (1 transit, 02/06/2016, \\ 
PI: Evans), P14767 (1 eclipse, 10/11/2016, PI: Sing), and \\ 
P15134 (2 phase-curves, 12/03/2018 and 03/02/2019, \\
PI: Evans) }.
Importantly, we have chosen to not include the observations from 
other telescopes or instruments, since the combination of 
multiple instruments is known to produce incompatible results \citep{Yip_2020_LC,Yip_2021_W96, edwards_pop}.
The analyzed data is from one eclipse, one transit, and two 
phase-curves---comprising a total observing time of about 90~hours; 
each phase-curve observation contains two eclipses and one transit. 
For consistency, we have used the same data reduction pipeline, 
\textsc{Iraclis}, and have adopted identical assumptions for all 
of the observations. 
The individual eclipse and transit events which are not from
the phase-curve observations have been previously extracted using \textsc{Iraclis} by \citet{Changeat_2022_five} [hereafter C22],
and the phase-curve data has been recently analyzed 
by \citet{Evans_2022_diunarl} [hereafter ME22].
However, the phase-curves have not been extracted using 
\textsc{Iraclis}. 
Therefore, we have re-analyzed the data with \textsc{Iraclis} 
in order to ensure consistency of treatment with that by C22.

Equipped with a consistently treated set of transit, eclipse, 
and phase-curve spectra, we use a suite of atmospheric retrieval 
codes and a high-resolution atmospheric dynamics model code 
which is extensively tested and validated specifically for hot 
exoplanet simulations.
Our aim here is to perform a robust extraction of the thermal 
structures and chemical abundance profiles in WASP-121\,b's 
atmosphere, which allows us to estimate planetary formation 
markers and investigate potential time variability. 
Broadly, our methodology can be grouped into four main 
activities, or parts:
\begin{description}\itemsep0em 
    
\item[{\small Part 1}] extracting a consistent set of WFC3-G141 
light curves for the transit, eclipse, and phase-curve data 
using \textsc{Iraclis}; fitting the light curves for the 
phase-curve data, using the \textsc{PoP} (Pipeline of Pipes) 
code (see description in Materials and Methods); and, testing 
various assumptions to model the instrument systematics in the 
literature.

\item [{\small Part 2}] analyzing the recovered phase-curve 
dataset with the 1.5D retrievals developed by \citet{Changeat_2020_phasecurve1} and 
\citet {changeat_2021_phasecurve2} in the \textsc{TauREx3.1} 
framework \citep{al-refaie_2021_taurex3.1, 
2019_al-refaie_taurex3}, permitting time-independent, global 
properties (e.g., mean metallicity Z and C/O ratio as well as 
mean thermal profiles at different longitudes) to be extracted.

\item [{\small Part 3}] analyzing the transit and eclipse 
data using 1D retrievals; and, incorporating the 
constraints (e.g., chemical parameters) obtained in 
Part 2 to reduce the degeneracies between temperature 
and chemistry so that observations can be analyzed 
individually rather than just in sum.

\item [{\small Part 4}] performing high-resolution, global
atmospheric dynamics simulations with the psudospectral code 
\textsc{BoB} (Built On Beowolf) \citep[e.g.,][]{Skinner_2021, 
Polietal14}, suitably optimized and forced with $T$--$p$ 
profiles obtained in Part 2; and, interpreting the observed 
variability informed by these simulations. 

\end{description}
A more detailed description of each part is provided in the 
Materials and Methods section of the Appendix. 

%%%%%%%%%%%%%%%%%%%%%%%%

\section{Spectrum extraction for the phase-curve data}

\subsection{Combined white light curve correction}

Performing the extraction from the raw full-frame images of the 
two observed phase-curves with \textsc{Iraclis}, we obtain very 
similar results to ME22 (a comparison is provided in 
Figure~\ref{fig:white}). 
We then correct and fit the reduced light curves with 
\textsc{PoP} (see description in Materials and Methods). 
Here, the two observations are fitted together, sharing orbital 
(mid-transit time $t_\mathrm{mid}$, inclination $i$, and semi-major axis $a$) and model (planet-to-star radius ratios $R_p$/$R_s$, and phase-coefficients: $C_0$, $C_1$, 
$C_2$, $C_3$, $C_4$) parameters, as well as the parameters for 
the short-term HST systematics. 
The parameters for the long-term HST systematics are not 
shared to accommodate for the six different observation segments. 
Fitting the white light curves, we explore the effects of different 
short and long-term HST systematics on our results. 

For the short-term ramps, we have attempted two models: a simple 
exponential (e.g., as in \citet{tsiaras_hd209} [hereafter 
2-param IS$_\mathrm{short}$]) and a double exponential (e.g., as in 
\citet{DeWit_2018} and \citet{Evans_2022_diunarl} [hereafter 
4-param IS$_\mathrm{short}$]). 
For the long-term systematics, three options are possible: 
linear, quadratic, and hybrid (e.g., quadratic for the first 
segments of each visit and linear for the others, as in ME22). 
The comparison of these runs can be found in 
Figures~\ref{fig:spec_white} and \ref{fig:corner_white}. 
Overall, we conclude that assuming simple or double 
exponential short-term ramp does not change our results 
for this dataset. 
For consistency with ME22, we therefore adopt the double 
exponential ramp model for the remainder of the study; for 
the long-term ramp, however, the corrected phase-curve 
observations depend on the assumption of linear, hybrid, 
or quadratic option. 

Comparing the Bayesian evidence ($\mathrm{E}$), a linear, hybrid, and 
quadratic correction give log-values, $\ln({\rm E})$, of 5733, 
5797, and 5803, respectively. 
While the quadratic case gives a slightly higher log-evidence, 
such an assumption is too flexible for the second segment 
(i.e., the data around transit), generating artificial 
nightside flux and causing large degeneracies (see posterior 
distribution in Figure~\ref{fig:corner_white}). 
We therefore adopt the more conservative approach and adopt a 
hybrid long-term ramp, as was done in ME22. 
The final recovered white light curve is corrected from the 
instrument systematics and compared with the results of ME22 in Figure~\ref{fig:white_corrected}. 
The residuals, in particular, demonstrate good agreement 
with the previous literature results.

\subsection{Individual white light curve corrections}
\label{subsec:individual}

To investigate the variability of the atmosphere and the 
instrument systematics in the available phase-curves 
observations, we have reproduced our white light curve analysis 
for each visit individually. 
Due to the lower amount of information, we choose to fix the 
orbital and second-order phase-curve parameters ($C_3$ and 
$C_4$) to the ones of the combined fit. 
For those runs, we have also employed the hybrid trend for the 
long-term ramps and the double exponential model for the 
detector short-term systematics. 
Figure~\ref{fig:two_runs_lc} shows the corrected white 
light curves for the individual fits; 
Figure~\ref{fig:corner_white_two_obs} shows the corresponding 
posterior distributions. 

The recovered phase-curves in Figure~\ref{fig:two_runs_lc}
exhibit clear differences. 
For instance, the first-order phase of the sinusoidal model is 
larger in the 2018 visit (0.19 rad) than in the 2019 visit 
(0.03 rad), or a variation in white transit depth is found. 
These differences could come from a combination of atmospheric 
variability (e.g., movement of hot/cold regions or changing cloud 
coverage)---a possibility we explore further in 
Section~\ref{sec:dynamics}---and instrument systematic effects. 
In any case, these results suggest that combining HST 
phase-curve observations may not be straightforward.

\subsection{Spectral light curve fitting}

The spectral extraction is done using two different binning 
strategies, at ``low'' (i.e., as in ME22) and ``medium'' (i.e., as 
in C22) resolutions; see Materials and Methods for more details. Figures~\ref{fig:spec_corrected} and~\ref{fig:spec_corrected_HR} 
show the corrected spectral light curves for the two cases,
respectively. 
Inspecting the light curves, both strategies lead to similar 
residuals. 
We therefore moved on to the extraction of the spectra from 
the corrected light curves. 
Using 16 temporal bins (about 1.5h of observation), the final extracted spectra compared to 
that of ME22 are shown in Figure~\ref{fig:spec_comparison}. 
Overall, the spectra agree very well---except for phase 0.07, 
where the reductions of our paper are consistent with each other 
but show larger flux than for phase 0.05 of ME22; this is 
despite the similar flux for phase 0.93, when compared with 
phase 0.95 of ME22. This slight difference could arise from the fact that ME22 includes the in-transit planetary flux (after removal of the transit signal) for bins 0.07 and 0.93 (those bins labeled 0.05 and 0.95 in ME22 span 3h), while we chose to ignore the in-transit planetary flux for consistency and simplicity.

\begin{figure*}
    %\vspace*{0.3cm}
    \begin{interactive}{animation}{phase_curve_retrieval.mp4}
    \includegraphics[width = 0.85\textwidth]{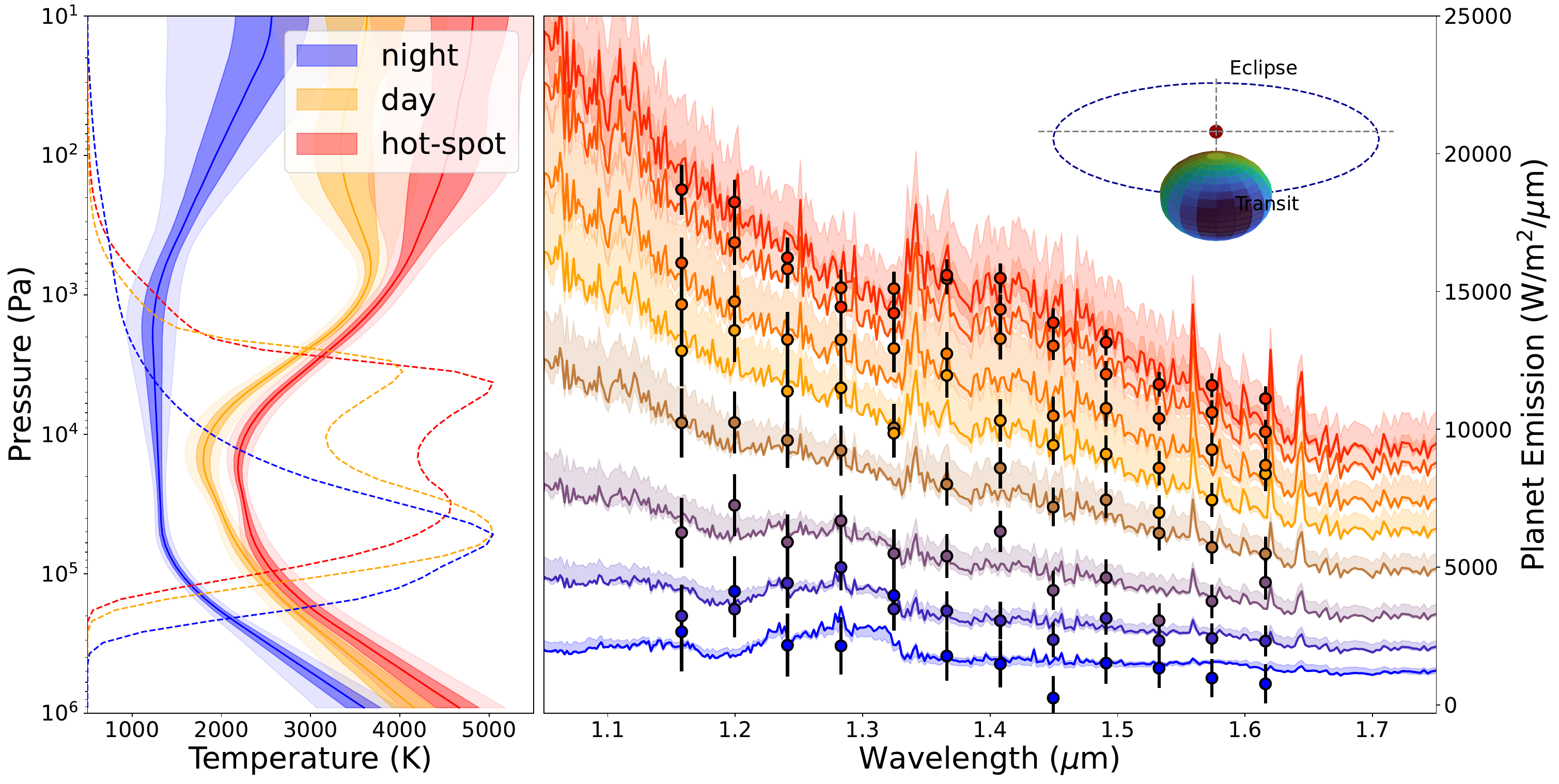}
    \end{interactive}
    \caption{Recovered temperature--pressure ($T$--$p$) profiles 
    (left) and best-fit spectra (right) for the phases from 0.05 
    (blue) to 0.5 (red), obtained from the phase-curve 
    atmospheric retrieval. 
    In the $T$--$p$ plot, the shaded regions correspond to one 
    and three sigma confidence regions (dark to light, 
    respectively). 
    The  radiative contribution function is also shown in 
    dashed line, colored for each region: hotspot (red), 
    dayside (orange), and nightside (blue). 
    These retrievals show good agreement with the observed data 
    and demonstrate a strong dayside thermal inversion, with 
    the presence of a hotter region (e.g. hotspot). 
    The best-fit $T-p$ profiles (solid lines, left) are used 
    to thermally force the atmospheric dynamics simulations. 
    This figure is accompanied by a 15\,s video, available online at the journal, showing the evolution of WASP-121\,b emission (from blue to red) and the corresponding thermal structure as a function of phase. As the planet moves from transit to eclipse, absorption features in the data are replaced by emission features. These spectral variations enable the characterization of the thermal structure and chemistry across WASP-121\,b's atmosphere.}
    \label{fig:spectra_tp_retrieval}
\end{figure*}
\begin{figure*}
    %\vspace*{0.5cm}
    \centering
    \includegraphics[width = 0.32\textwidth]{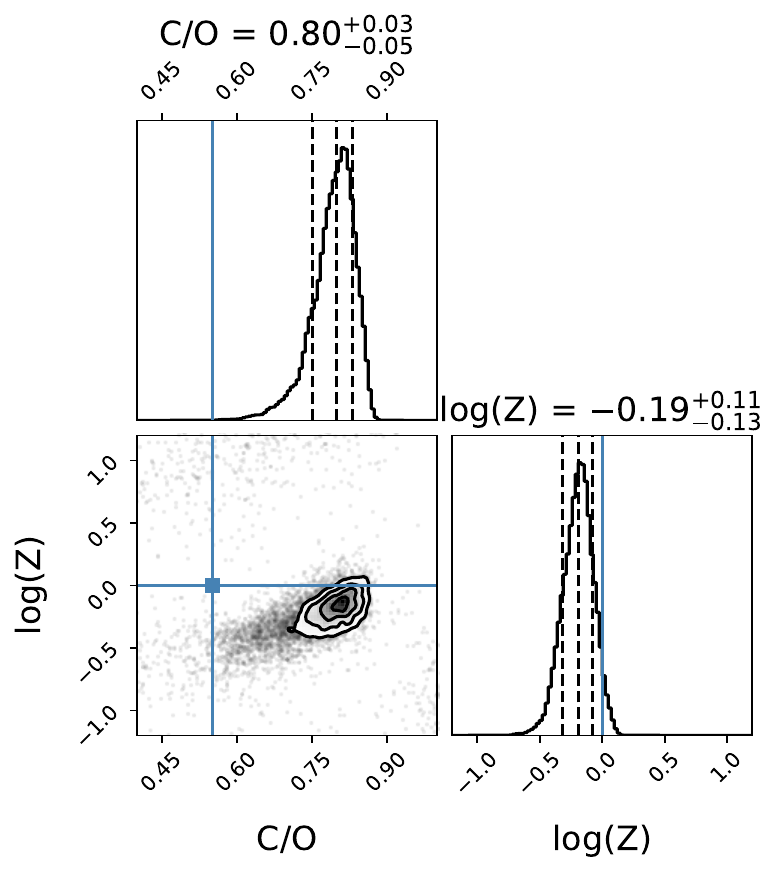}
    \includegraphics[width = 0.32\textwidth]{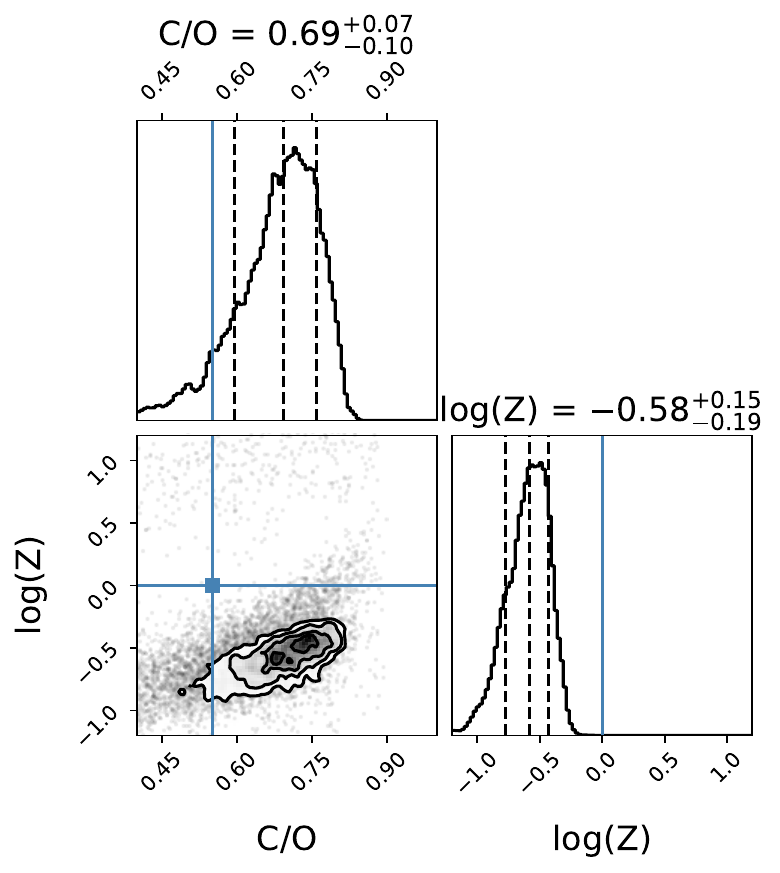}
    \includegraphics[width = 0.32\textwidth]{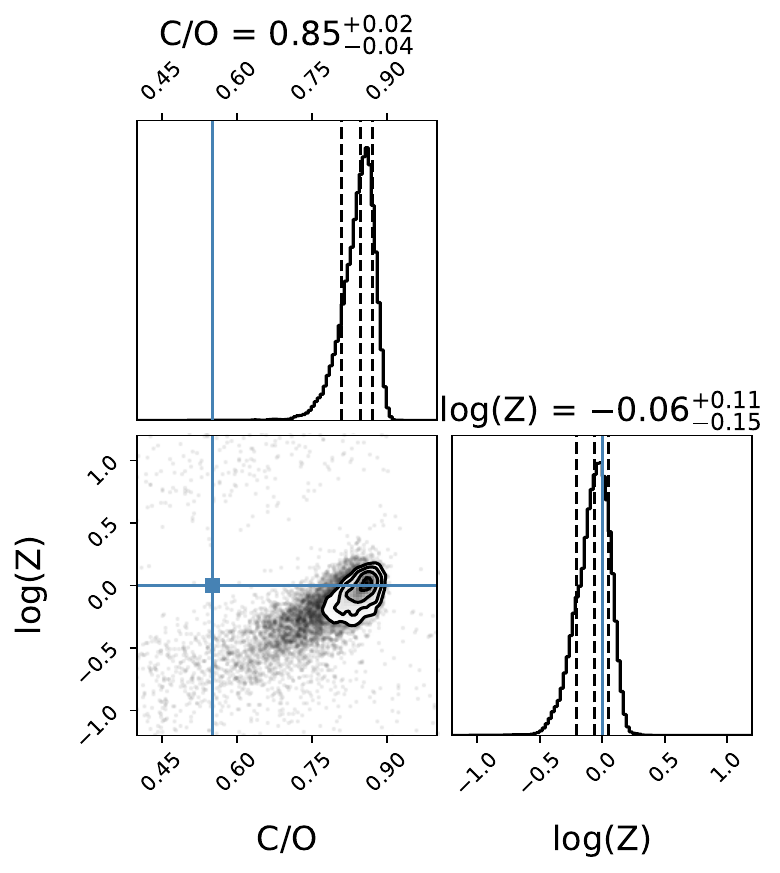}
    \caption{Posterior distribution of the chemistry retrieved from the 
    phase-curve data using the ``Low'' resolution spectra (Left), the 
    ``Medium'' resolution spectra (Middle) and the spectra from 
    ME22 (Right). 
    The constraints obtained on metallicity Z and C/O ratio are 
    consistent with a solar to slightly sub-solar metallicity and a 
    super-solar C/O ratio, indicating the formation of the planet 
    likely occurred beyond the snow-line. 
    The blue line indicates the solar values for Z and C/O ratio. 
    The full posterior distributions for the {\it low}-resolution 
    retrieval are available in Figure~\ref{fig:posteriors_full_compa}}
    \label{fig:corner_chemistry}
\end{figure*}

\section{Phase-curve atmospheric retrievals}

One of the goals of this study is to robustly characterize
the bulk properties of WASP-121\,b atmosphere using the combined 
phase-curve data. 
As described in Materials and Methods, we use the 1.5D atmospheric 
retrievals to interpret the observed data.\footnote{See 
\citet{Changeat_2020_phasecurve1}, 
\citet{changeat_2021_phasecurve2}, \\
and \citet{changeat_2021_w103} for additional examples of the 1.5D \\ method.} 
This retrieval technique analyzes the two phase-curve observations 
using a single unified atmospheric model (e.g., a single likelihood);
hence, it efficiently exploits all the information content---as 
opposed to the more traditional 1D retrieval performed individually 
for each phase \cite[see, e.g.,][]{Stevenson_2017}. 
Since the ``hotspot'' offset $D_{\mathrm{HS}}$ and (angular) size 
$A_\mathrm{HS}$ are difficult to constrain from HST data alone \citep{changeat_2021_phasecurve2}, 
we have tested different combinations and found that
$(D_\mathrm{HS},\,A_\mathrm{HS})  = (30^{\circ},\,50^{\circ})$ 
leads to the highest Bayesian evidence.
Therefore, we here focus on this case.

Figure~\ref{fig:spectra_tp_retrieval} (and the associated animation) 
shows the best-fit spectra and recovered thermal structure from 1.5D 
retrievals of the {\it low}-resolution data. 
Here, we obtain consistent $T$--$p$ profiles by reproducing the retrievals 
on the ``Medium'' resolution spectra as well as those from ME22 (see Figures~\ref{fig:tp_reductions} and~\ref{fig:BV_profiles}), 
demonstrating that this information is independent of the data 
reduction.
The chemical parameters (i.e., metallicity and C/O ratio) 
are slightly dependent on the spectral resolution (especially in terms of precision, see Figure~\ref{fig:corner_chemistry}), which could be due to reduced correlation between the spectral channels at lower resolution, or information dilution occurring during the fit due to the lower S/N of the light curves at higher resolution. 
Since our ``Low'' resolution reduction is consistent with ME22, 
we focus the rest of our discussion on this case; nevertheless, full posterior 
distributions for all three retrievals are provided in 
Figure~\ref{fig:posteriors_full_compa}.

Importantly, due to the resolving power of phase-curve data, 
our retrievals allow precise thermal and chemical estimates at 
different locations in the planet's atmospheres to be obtained. 
Importantly, for ultra-hot Jupiters, the presence of absorption features for water, refractory species (TiO, VO, and FeH), and hydrogen ions in WFC3 allows to break the degeneracies between metallicity and C/O ratio \cite[see also][]{changeat_2021_w103}. 
Given our retrieval assumptions, we find a strong thermal 
inversion on the dayside with the hottest region (here
labeled {\it hotspot} in Figure~\ref{fig:spectra_tp_retrieval}) 
being $\sim$300\,K hotter than the rest of the dayside between 
10$^5$\,Pa and 10$^3$\,Pa (see red and orange profiles). 

The thermal inversion could be caused by two different mechanisms.
One mechanism is the production of energy at high altitudes by the 
presence of refractory molecules and H$^-$ (the latter from H$_2$ 
thermal dissociation): the temperature in the inversion region of the 
dayside is indeed hot enough to dissociate most molecules---including 
water and even more stable volatiles (CO and CO$_2$) and refractory 
molecules (FeH, TiO, and VO), along with H$_2$ (see the retrieved 
chemical profiles of Figure~\ref{fig:chemical_profiles}). 
At lower pressures, the atmosphere could partially be ionized, with 
an increased abundance of free electrons creating a continuum H$^-$ 
opacity, as suggested for other similar ultra-hot Jupiters 
\citep{Edwards_2020_ares, pluriel_aresIII, Changeat_2021_k9, 
changeat_2021_w103}. 
Another possible mechanism is heat deposition of breaking or 
saturating planetary and gravity waves launched from the atmospheric 
region below \citep[e.g.,][]{Watkins_2010,cho_2015}.
Both mechanisms likely contribute to the observed thermal inversion layer.
The retrievals we performed include a ``gray'' cloud model (i.e., 
constant opacity cloud deck); 
however, large cloud patches are not favored by the data 
(see Figure~\ref{fig:posteriors_full_compa}) despite the temperatures at 
the nightside being potentially suitable for silicate cloud formation \citep{Powell_2018, Gao_2021}.

Comparing the results from all the reductions (see also recovered 
$T$--$p$ profiles and posteriors for other {\it hotspot} 
characteristics in Figures~\ref{fig:tp_reductions_ADCompa} and 
\ref{fig:posteriors_full_compa_ADCompa}), we conclude that the data 
is consistent with a solar to slightly sub-solar metallicity Z 
and a super-solar C/O ratio. 
For the {\it low}-resolution case, we estimate the mean chemical 
characteristics of the planet to be 
$\log({\rm Z}) = -0.19^{+0.11}_{-0.13}$ and  
${\rm C/O}\, =\, 0.80^{+0.03}_{-0.05}$. 
A more conservative estimate, encompassing the uncertainties from 
all the reductions and retrievals tested in this work is
$-0.77 < \log({\rm Z}) < 0.05$ and $0.59 < {\rm C/O} < 0.87$. 
As a byproduct, this enables us to also speculate on the 
formation history of this planet; specifically, the obtained Z 
and the super-solar C/O ratio are suggestive of an early formation 
via significant gas accretion (i.e., without significant planetesimal 
pollution) and beyond the snowline of the proto-planetary 
disk \citep[e.g.,][]{Oberg_2011, Mordasini_2016, madhu_formation, 
Brewer_2017, Cridland_2019, Shibata_2020, Turrini_2021, Pacetti_2022}.

To further support the above conclusions, we have performed additional 
sensitivity tests, which are shown in Figure~\ref{fig:sensitivity}. 
We apply $\pm3\sigma$ departures, where $\sigma$ is the retrieved 
uncertainty on the modified parameter, to the best-fit Z and C/O 
from the {\it low}-resolution fit and compare the simulated spectra 
with the observations. 
Introducing these departures leads to spectra that do not properly 
explain the observations, confirming the magnitude of the recovered 
uncertainties for the atmospheric chemistry of WASP-121\,b.

Due to the high constraints on the mean atmospheric properties 
of this planet obtained via the phase-curve data, the parameters 
extracted at this stage (e.g., thermal profiles and chemistry) 
can serve as priors for the subsequent parts of our analysis. 
In particular, assuming that Z and C/O ratio remains spatially 
homogeneous and constant in time allows us to re-use the retrieved 
values to analyze the transits and eclipse data individually;
the thermal profile, chemistry, and cloud properties extracted from 
individual transit and eclipse observations are known to be much 
more degenerate on their own. 
Additionally, the recovered thermal profiles provide important 
information for dynamics calculations. 
For example, they can be re-introduced as an observation-driven 
forcing to enable physically realistic and case-specific 
simulations.

\begin{figure*}
    \vspace*{0.5cm}
    \includegraphics[width = 0.50\textwidth]{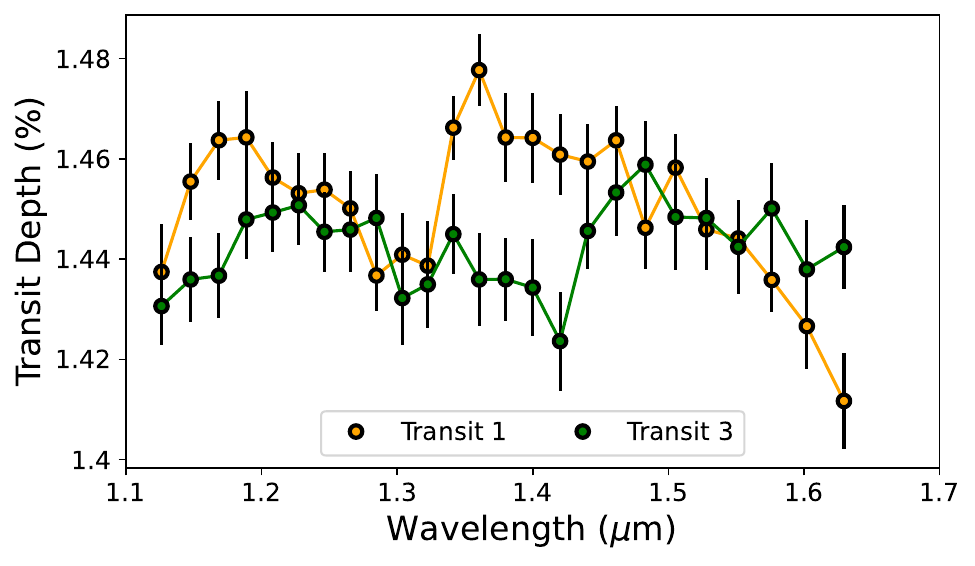}
    \includegraphics[width = 0.49\textwidth]{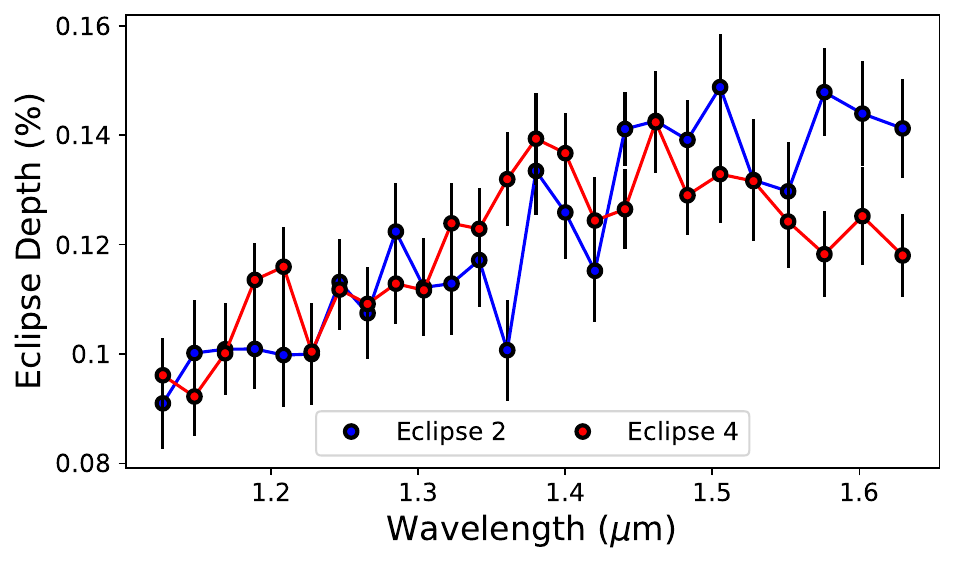}
    \caption{Examples of observed transit (left) and eclipse (right) 
    spectra of WASP-121\,b analyzed in this work. 
    The spectra shown are corrected for vertical offsets to highlight 
    the differences in spectral shapes. 
    Variability in the planet's weather patterns could create 
    such variations in the spectroscopic data; this is 
    strongly suggested in high-resolution simulations carried out 
    for this planet in this work (see, e.g., 
    Figures~\ref{fig:dynamics_time} and \ref{fig:time_variab}). 
    For example, during the transits, the observed temporal variations could be interpreted as a formation of intermittent clouds and/or hazes; during the eclipse, the observed variations may be due to subtle changes in the thermal structure of the atmosphere---induced by motions of hot/cold regions from the planet's atmospheric dynamics.}
    \label{fig:spectra_ec_tr_comp}
\end{figure*}

%%%%%%%%%%%%%%%%%%%%%%%%

\section{Transit and eclipse atmospheric retrievals}

As mentioned, C22 has previously reduced the individual transit 
and eclipse datasets with the \textsc{Iraclis} pipeline; therefore, 
we make use of the spectra from that work directly. 
Already interesting differences appear in the transit spectra, 
although the eclipse spectra look more alike (Figure~\ref{fig:spectra_ec_tr_comp}). 
For both transits and eclipses, we perform 1D  retrievals using 
the standard \textsc{TauREx3} models. 
We have first attempted to retrieve all the free parameters of 
the models without particular priors; but, as expected for HST 
data, the degeneracies between thermal structure, chemistry, 
and cloud properties were difficult to break from individual 
HST transit/eclipse spectra: we could not extract a consistent 
picture. 
However, since Z and C/O ratio are expected to remain 
time-independent\footnote{ On considered timescales, the atmosphere is 
essentially a closed system. Note however that variable cloud formation via condensation can remove oxygen from the gas phase and locally change the C/O ratio.} and have better constraints from the 
phase-curve data, we have decided to re-inject this information 
from the 1.5D retrievals.
Therefore, the chemistry is fixed to the median value from the 
retrieval on the {\it low}-resolution spectra; this has allowed 
to obtain a consistent fit of the spectra from all the visits (Figure~\ref{fig:spectra_ec_tr}). 
For completeness, the posterior distributions are presented in Figures~\ref{fig:posteriors_transits} and
\ref{fig:posteriors_eclipses} for the transits and eclipses, 
respectively.

For the transits, the three individual observations show the 
presence of covering hazes or clouds. 
Specifically, the transit spectra captured during the two 
phase-curves (blue and green) are fully cloudy (i.e., 
consistent with featureless), while the observation obtained 
in 2016 (orange) shows clear spectral modulations from water 
vapor. 
Note that the red end of this spectrum, however, cannot be 
fit properly using the chemical equilibrium assumption; 
however, free chemistry retrievals, which use H$_2$O, VO, 
and H$^-$ opacities beyond equilibrium, could achieve a 
better match. 

Nevertheless, Transit 1 (orange) is not consistent with a 
featureless spectrum, as shown by $\Delta\ln({\rm E}) = 18.3$, 
and possesses strong water vapor absorption features. 
Transit 2 (blue) displays an interesting multi-modal solution. 
The spectrum is best explained by either very high 
temperatures ($T \approx 3000$\,K) at the terminator, 
forcing dissociation of the main molecules and leaving a flat 
contribution from the H$^-$ continuum, or a more cloudy/hazy 
atmosphere with a slight slope towards the blue end of the 
spectrum. 
Given the un-physical nature of the high-temperature solution, 
we suggest that this second observation is consistent with the 
presence of hazes, especially as a lower dimension featureless 
fit achieves a similar Bayesian evidence, 
$\Delta\ln({\rm E}) = 0.5$. 
For Transit 3 (green), we find the observation consistent with 
a flat spectrum, which would be well explained by clouds and/or 
hazes, given $\Delta\ln({\rm E}) = -0.6$. 
In transit, stellar activity (i.e., unocculted stellar spots and faculae) can can cause important spectral variations between repeated observations \citep{Rackham_2018, Thompson_2023}. However, long-term monitoring campaign from the ground \citep{Delrez_Wasp121b_em, Evans_wasp121_t2} suggests that WASP-121 is a very quiet and stable star.
If confirmed, these results would indicate a transient formation 
of cloud/haze structures at the terminator of WASP-121\,b.

We note that gray clouds, which were only considered for the night-side, were not recovered by the retrievals on the phase-curve 
data.
Many reasons could explain this result: {\it i}) the emitted 
signal in phase-curve does not probe the same altitudes as the 
transit data, {\it ii}) the observed clouds are poorly described 
by the gray cloud model, or 
{\it iii}) the clouds are located around the terminator region 
and not covering large patches of WASP-121\,b.
Additionally, unambiguously determining the presence of gray 
clouds from emission data is more challenging due to the 
degeneracies with the vertical temperature profile and the 
shorter geometrical path length through the atmosphere.

\begin{figure*}
    \vspace*{0.5cm}
    \centering
    \includegraphics[height = 0.37\textheight]{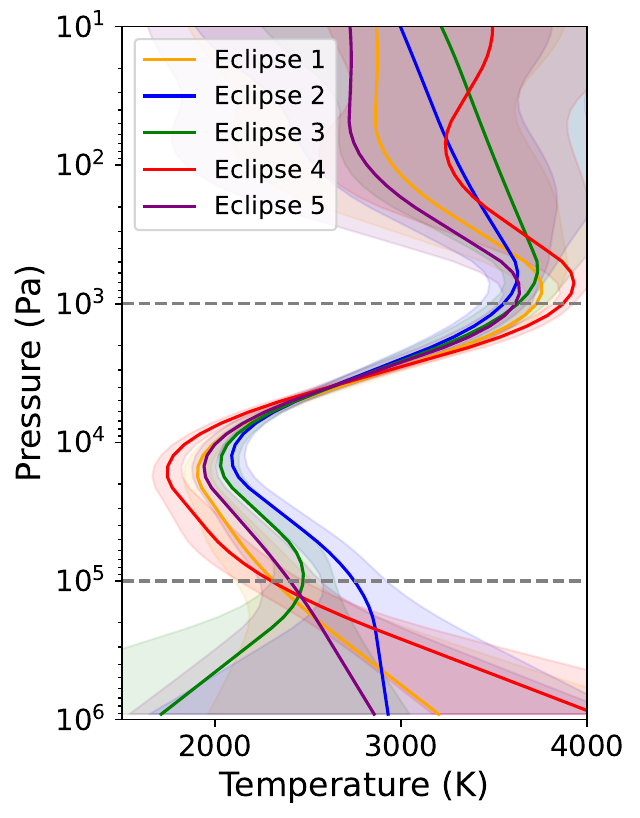}
    \includegraphics[height = 0.37\textheight]{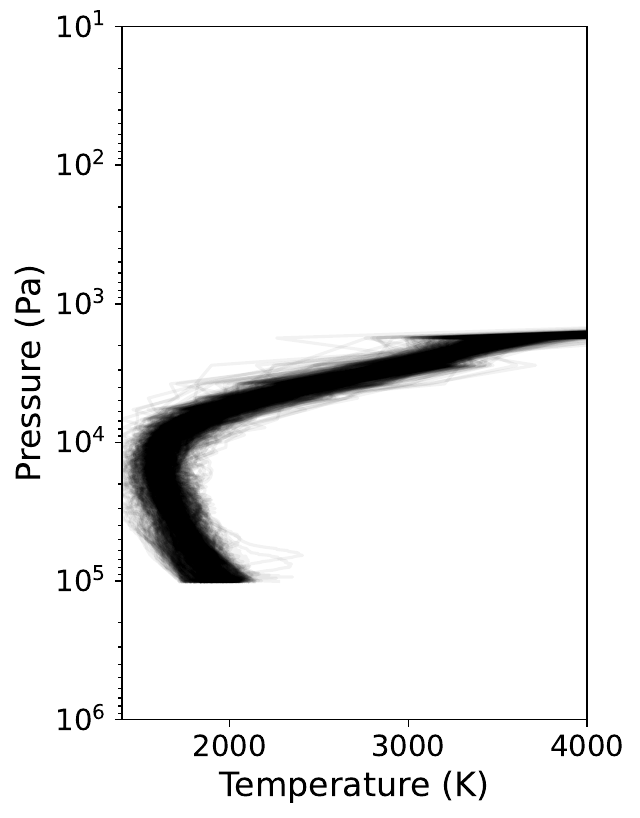}
    \caption{One-dimensional (1D) thermal structure recovered by 
    our retrieval analysis of the five eclipse observations with 
    one sigma confidence region (left), and $T$--$p$ profiles 
    from multiple times ($t \in [40, 185]$\,days) at the substellar 
    point from a three-dimensional (3D) atmospheric dynamics 
    simulation (right). 
    The magnitude of variability in $p \in [10^5, 10^3]$\,Pa is 
    $\sim$300\,K, which is consistent with the variation predicted 
    by the 3D simulation.
    Dashed gray lines show the vertical extent of the atmosphere 
    modeled by the simulations in this study. 
    Note, while these profiles are not like-for-like comparable 
    because the retrieved thermal structure is global and 
    substellar temperature predictions are local. }
    \label{fig:tp1d}
\end{figure*}

For the eclipse data, the spectral differences are much smaller 
and difficult to infer by visual inspection of the spectra. 
We have conducted atmospheric retrievals and extracted the 
thermal structure for the five different eclipses individually (see left panel of Figure~\ref{fig:tp1d}). 
The thermal profiles are overall consistent across the five 
observations (i.e., similar thermally inverted structure), 
but we find variations in the mean temperature of 310\,K when 
averaging the profiles over the 10$^5$\,Pa to 10$^3$\,Pa 
region. 
Importantly, this range is much larger than the average 
one-sigma uncertainty of the profiles in the same region (the 
averaged standard deviation is $108 \,{\rm K}$).
For instance, in Figure~\ref{fig:tp1d}, the thermal profiles 
extracted from Eclipse 2 (blue) and Eclipse 4 (red) are not 
consistent within the retrieved uncertainties.
Similar to the phase-curve data, the observed differences in 
eclipse could be attributed to temporal variations of hot/cold 
regions in the planet's dayside and/or a changing thermal 
structure of the substellar point. 

While instrument systematics could remain (see Section above), 
the observed differences in hot region offset from the phase-curves, 
cloud coverage from the transits, as well as dayside thermal structure 
from the eclipses are all plausibly explained by the presence of 
atmospheric temporal variations. 
The observed differences are in fact expected from a theoretical 
atmospheric dynamics viewpoint due to the intense stellar heating 
contrast from the planet's parent star WASP-121.
To investigate more precisely the possible origins of atmospheric 
temporal variations on the planet WASP-121\,b and verify if they can 
affect our data to observable levels, we model its atmosphere 
with high-resolution dynamics simulations, to which we now turn.

%%%%%%%%%%%%%%%%%%%%%%%%

\section{Dynamics modeling}\label{sec:dynamics}

We simulate the dynamics of WASP-121\,b atmosphere with the 
\textsc{BoB} code at ``T682L50'' resolution, where $T = 682$ is 
the triangular truncation wavenumber (i.e., number of total and 
zonal modes each in the spherical harmonics) and $L = 50$ is the 
number of vertical layers (uniformly space in $p$); see Sec~\ref{sec:meth:dyn}, as well as \citet{Skinner_2021} and \citet{Polietal14}, for detailed 
descriptions of the numerical model and simulation parameters. 
The use of \textsc{BoB} at this resolution---to directly guide 
the retrieval interpretation with numerical robustness and 
verisimilitude---is a significant feature of this study.
The simulations are performed to obtain a broad idea and insights 
into the variability plausible on planets like WASP-121\,b, when 
the flow is adequately resolved: it has recently been shown 
explicitly that hot-exoplanet simulations are not converged if 
the resolution employed is much below that in this work 
\citep{Skinner_2021}.
As in most past studies, the atmosphere initially at rest 
is set in motion via a thermal relaxation to $T$--$p$ profiles. 
Here the forcing profiles are obtained from the retrievals 
described above (e.g., Fig.~\ref{fig:spectra_tp_retrieval}) 
and prescribed. 
Profiles at many different times, which have deviated from the 
prescribed one due to the nonlinear atmospheric motion, are 
compiled in Fig.~\ref{fig:tp1d} (right panel): they should be 
compared with observations (left panel).

\begin{figure*}
    \vspace*{1cm}
    \centering
    \includegraphics[width = 0.325\textwidth]{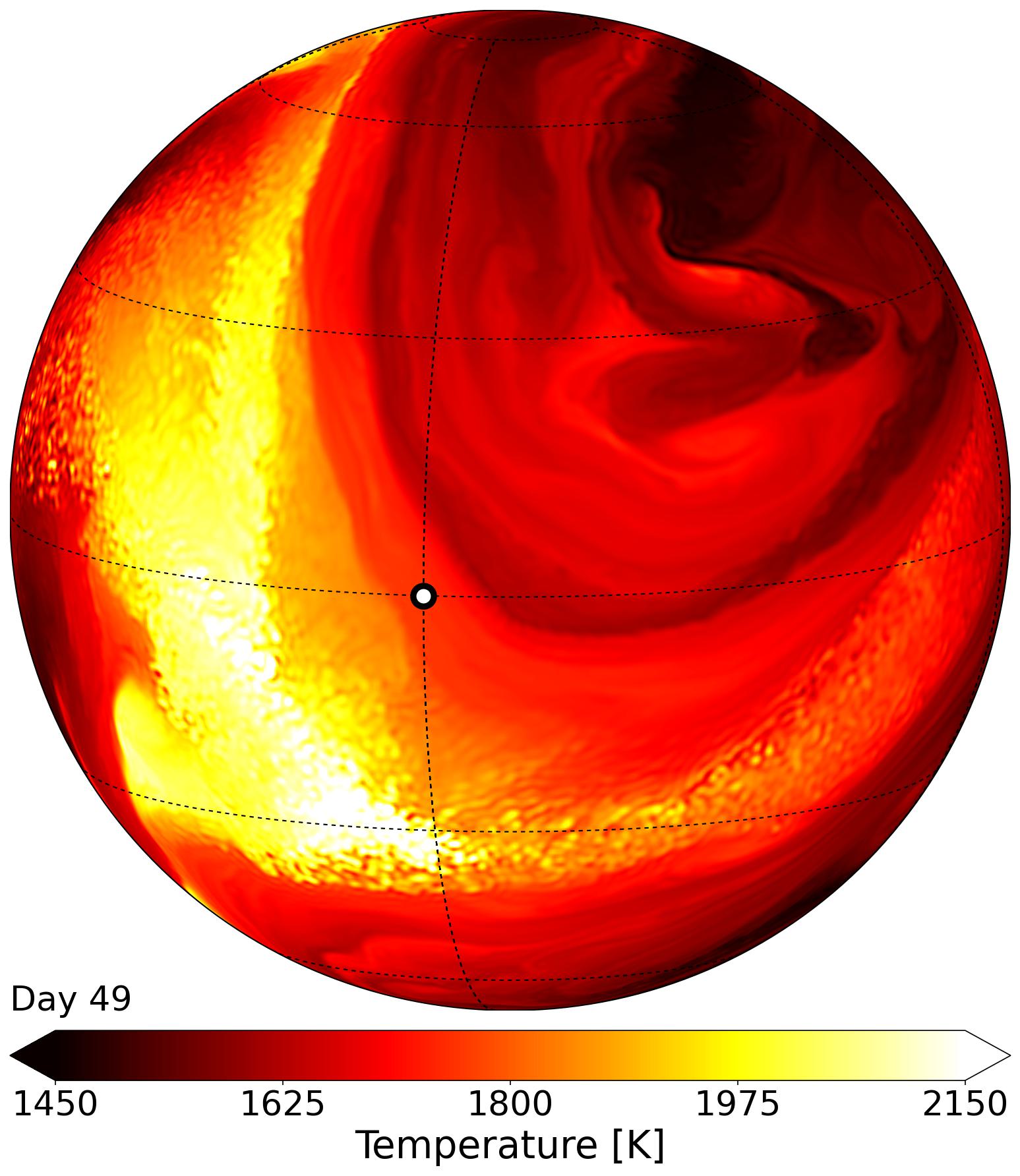}
    \includegraphics[width = 0.325\textwidth]{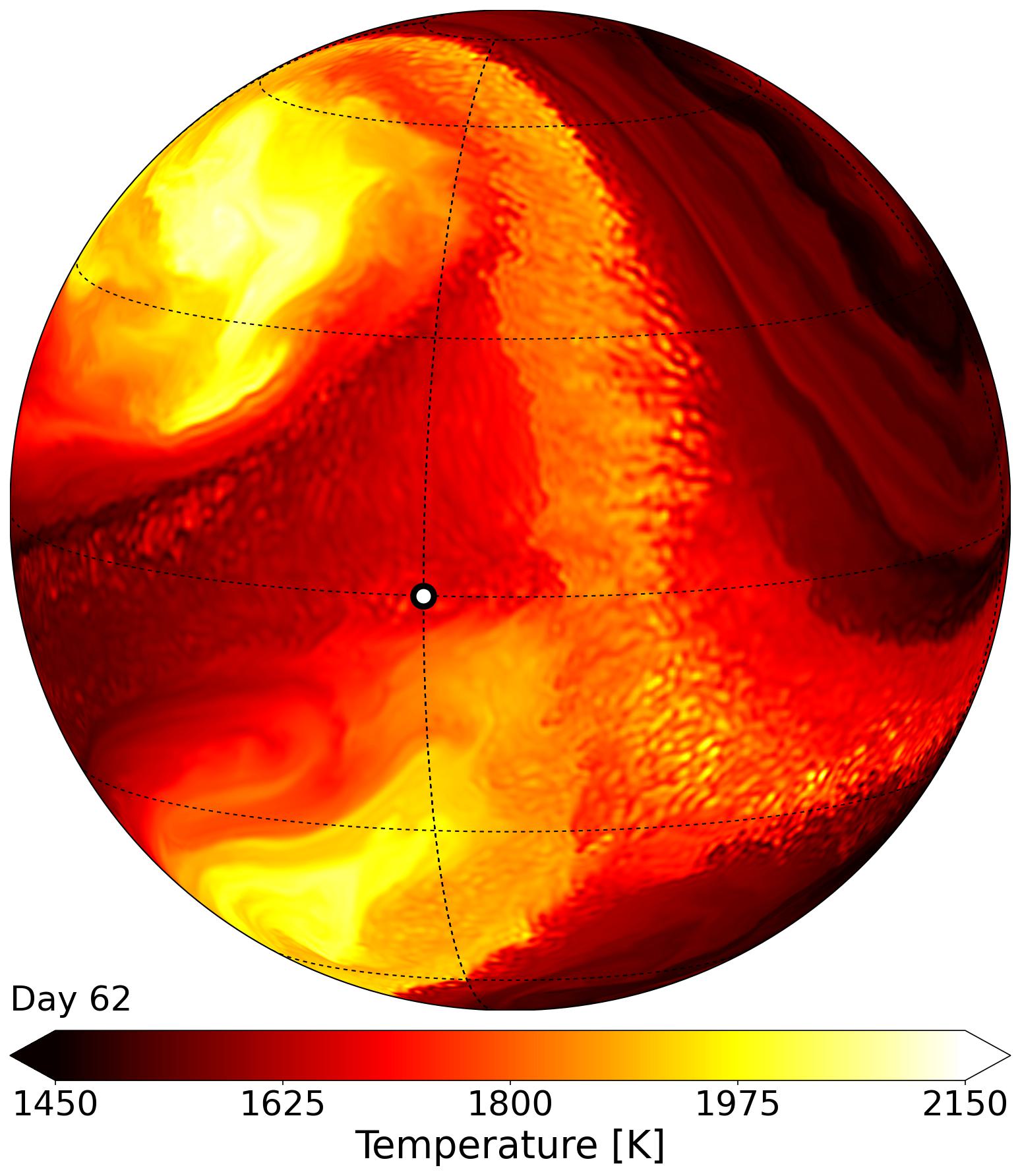}
    \includegraphics[width = 0.325\textwidth]{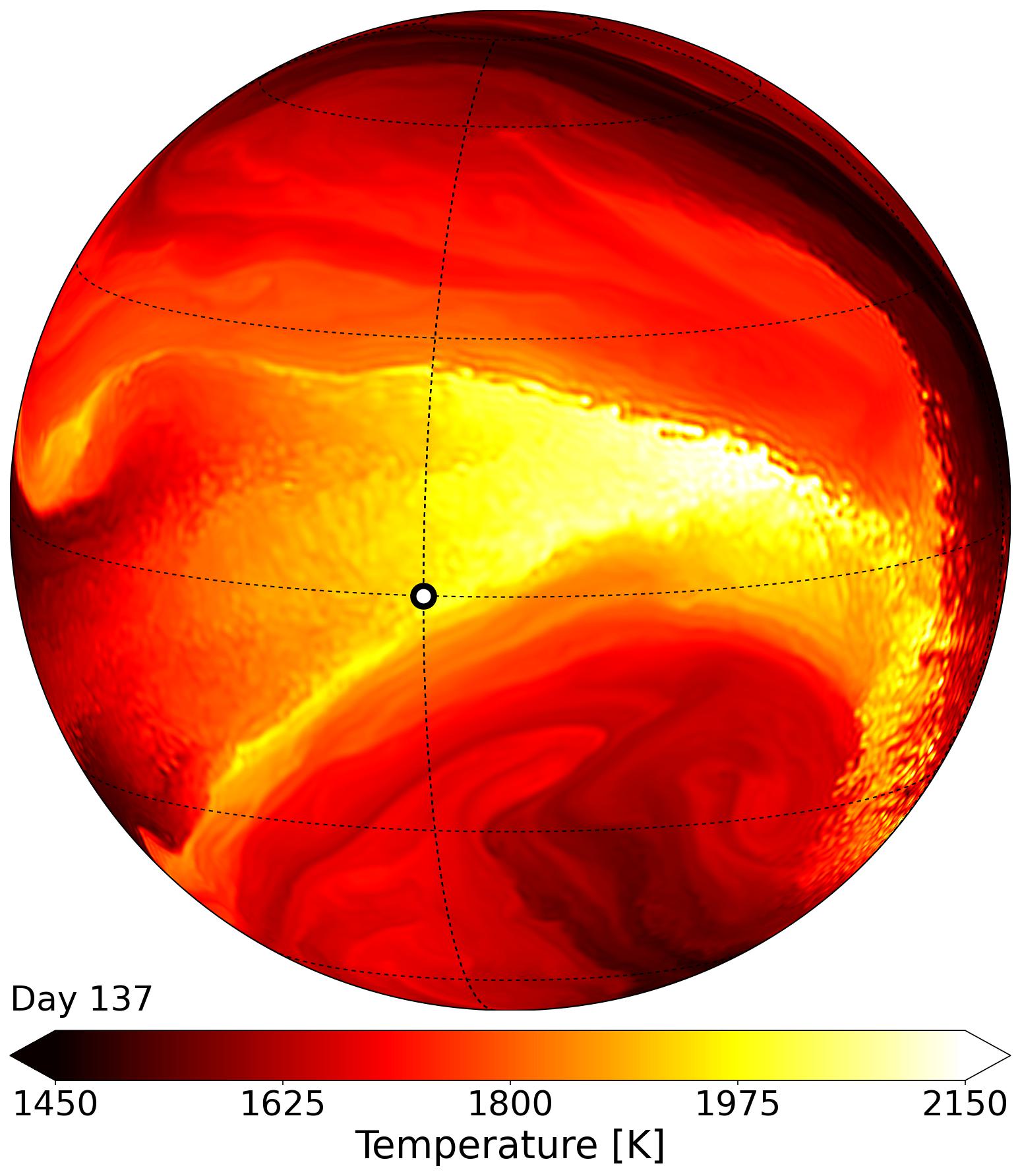}
    \caption{Instantaneous spatial temperature maps 
    $T(\lambda,\phi,p,t)$, where $\lambda$ is the longitude, 
    $\phi$ is the latitude, $p$ is the pressure, and $t$ is the 
    time); $p = 10^5$\,Pa level at $t = \{49, 62, 137\}$\,planetary 
    days after the start of the simulation are shown.
    The maps demonstrate that very different temperature distributions 
    arise at different times. 
    Maps are in spherical orthographic projection, where the white 
    circle marks the substellar point. 
    The fields result from simulations which are thermally forced by 
    the $T$--$p$ profiles in Fig~\ref{fig:spectra_tp_retrieval}.
    The planet's ``hotspot'' is highly variable not only in location 
    and time but also in shape. 
    These storms are planetary-scale, coherent, and exhibit repetitive 
    behaviors, leading to high amplitude and identifiable periodic 
    signature in the planet's flux.}
    \label{fig:dynamics_time}
\end{figure*}

Figure~\ref{fig:dynamics_time} shows temperature maps at 
$p = 10^5$\,Pa from three widely separated times in the 
simulation.
The maps demonstrate the highly variable nature of the 
planet's temperature field at a pressure level from which 
observable flux would originate. 
Snapshots of the temperature at two different pressure levels, 
$p = 5\!\times\! 10^3$\,Pa and $p = 10^5$\,Pa, show the vertical 
distribution, which is strongly barotropic---i.e, vertically 
aligned (Figure~\ref{fig:dynamics_press}); a full movie of 
this simulation is provided in the Supplementary Materials.
We also show chemical species maps, which are simply 
post-processed using the instantaneous density and temperature 
distributions from the simulation 
(Figures 
\ref{fig:dynamics_chemistry_maps1}--\ref{fig:dynamics_chemistry_maps3}) 
at $p = 1\,{\rm bar} = 10^5\,{\rm Pa}$ and  
$p = 50\,{\rm mbar} = 5\times 10^3\,{\rm Pa}$ (left two columns 
and right two columns, respectively), at $t = 49$\,days and 
$t = 62$\,days (left and right columns, respectively, for 
each $p$-level); distributions of the main relevant molecules 
(H$_2$, H, H$^-$, e$^-$, H$_2$O, CO, CO$_2$, CH$_4$, TiO, VO, and 
FeH) are shown.
Not surprisingly, variations (vertical, horizontal, and temporal) 
induced by WASP-121\,b's atmospheric dynamics impacts the chemistry 
as well as temperature, the latter of which is discussed more in 
detail below. 
Note that here a simple chemical equilibrium assumption is 
made, which may not be valid everywhere in the modeled 
domain---particularly if the reaction time is comparable to the 
advection time. 

As can be seen in the figures, the complex motion of the 
atmosphere---and the organized structures therein---cause hot 
{\it and} cold regions to chaotically mix in time. 
Generally, the hottest region periodically forms slightly eastward 
of the substellar point, but it always moves away from the point 
of emergence.
Depending on when the atmosphere is observed, the hottest region 
can even be located on the west side of the substellar 
point---either sequestered in a long-lived storm or generated near hyperbolic flow points between the storms (middle and left panels 
in Figure~\ref{fig:dynamics_press}, respectively).

Interestingly, the dynamical behavior here is reminiscent of that 
of WASP-96\,b in the $p \gtrsim 10^4$\,Pa vertical region.
\citet{Skinner_2022_cyclogenesis} have recently reported a 
quasi-periodic generation of giant cyclonic storms moving away 
westward from the point of emergence.
This is due to ``deep heating'' (i.e. strong heating at 
$\sim$$10^5$\,Pa), which may be experienced by some hot Jupiters.
Here the similarity in behavior is likely due to the 
morphologically similar $T$--$p$ profiles in the aforementioned 
region on WASP-121\,b and WASP-96\,b. 
The strength of the dayside--nightside difference is greater on 
WASP-121\,b, but the profiles drive---and steer---the dynamics in 
a qualitatively similar way to WASP-96\,b.

\begin{figure*}
    \vspace*{0.7cm}
    \centering
    \includegraphics[width = 0.99\textwidth]{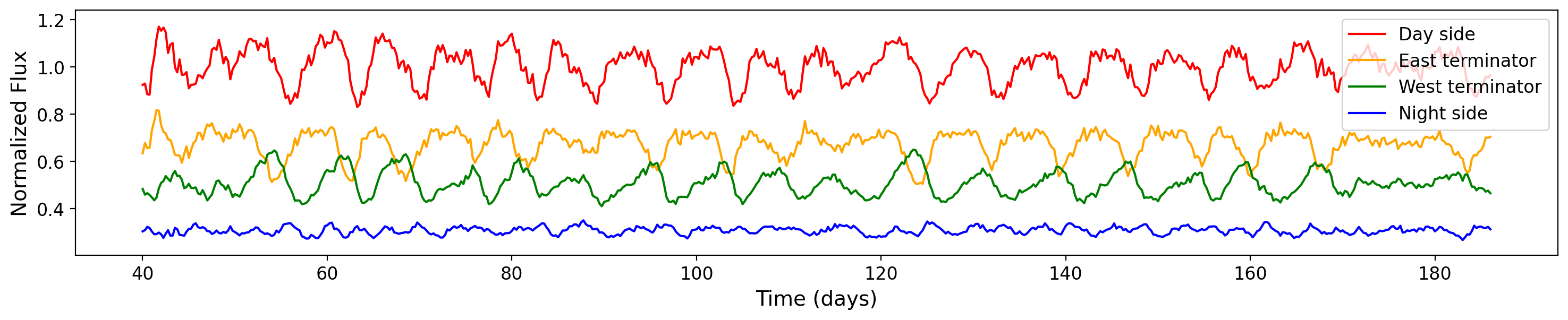}
    \caption{Disk-averaged black-body flux variations at 1.3$\mu$m 
    using Planck's law for a single layer in the middle 
    of our modeled domain ($10^4$\,Pa), and centered on four 
    regions: dayside (substellar point), eastern terminator, 
    western terminator, and nightside (antistellar point). 
    The flux is normalized by the mean value of the dayside.
    The fluxes are quasi-periodic on periods of $\sim$5\,days, with variations of $\sim$5 to $\sim$10\%. 
    The normalized flux for the west terminator is generally lower 
    than that of the east terminator, indicating that a phase-curve 
    observation of this planet is more likely to have an 
    eastward-shifted phase offsets.
    The level of variability is compatible with the uncertainties 
    of our HST observations; however, note that the HST instrument 
    systematics makes absolute flux measurements at such $p$-level 
    highly uncertain \citep[see, e.g.,][]{Yip_2021_W96, edwards_pop}.}
    \label{fig:time_variab}
\end{figure*}
 
The similar dynamical behavior also leads to qualitatively 
similar disk-averaged flux signatures at $p = 10^5$\,Pa, for 
example; cf.\ Figure~\ref{fig:time_variab} with Figure~4 in \citet{Skinner_2022_cyclogenesis}\footnote{NB., the former 
presents a black-body flux at 1.3$\mu$m and the \\
latter presents a much simpler flux 
($\,\equiv\!\sigma_\mathrm{SB} T^4$, where $\sigma_\mathrm{SB}$ 
is \\ the Stefan–Boltzmann constant).}. 
In both figures, the fluxes exhibit quasi-periodic variations, 
with excess flux at the substellar (``Day side'' in Figure~\ref{fig:time_variab}) and east terminator regions.
Hence, the flux is persistently ``shifted to the east of the 
substellar point''---{\it when averaged over the disk}.
It is important to note that, when not averaged, the hottest 
{\it regions} are rarely located near the equator and
often situated westward and at higher latitude of the substellar 
point (middle and left panels in Figure~\ref{fig:dynamics_time},
respectively); the regions are not always vertically aligned 
either (cf.\ temperature maps from two pressure levels at 
Day~49 in Figure~\ref{fig:dynamics_chemistry_maps1}).
Movies of the simulation show both clearly.
The movies also show that the timescale of the variability is 
$\sim$5\,planet days with sharp flux changes occurring in 
much shorter ($\lesssim 0.5$\,day) windows---as was reported 
for WASP-96\,b--like atmospheres by 
\citet{Skinner_2022_cyclogenesis}.
The above behavior overall may also explain why most infrared 
observations to date---except by \citet{Dang_2018}, 
\citet{Bell_2019}, and \citet{Morello_2023}---have reported 
only ``eastward-shifted hotspots''.

As expected, the magnitude of the variability in the model domain 
varies with the $p$-level: it is greatest at $p \sim 10^5$\,Pa.
Above $p \approx 2\times 10^4$\,Pa, the dayside maintains a strong 
thermal inversion associated with an atmosphere that nevertheless 
remains highly variable. 
Given our model assumptions, this altitude-dependent behavior 
originates from the greater intensity of stellar irradiation 
on WASP-121\,b compared to that of a typical hot Jupiter, leading 
to a much shorter thermal relaxation time, as well as the presence 
of the visible light absorbers H$^-$, arising from the 
dissociation of H$_2$ (Figure~\ref{fig:dynamics_chemistry_maps1}). 
On WASP-121\,b, while the stellar irradiation can still penetrate 
as deep as 10$^5$\,Pa (as in a typical hot Jupiter), the upper layers 
of the atmosphere likely maintain much stronger dayside--nightside 
temperature gradient overall because of the shorter relaxation time 
\cite[which is $\propto T^{-4}$:][]{Andrews_1987, Salby_1996_Chapt8, Cho_2008}.
The thermal profile at the substellar point shows a short period 
($\sim$5\,days) temporal variation of the order of 
$\pm 200$\,K, which could be captured by the HST 
observations.

To round out our investigation, we compare the temperature 
profiles and fluxes obtained in our dynamics simulations with the 
information obtained from the data. 
In Figure~\ref{fig:tp1d}, we have already shown the agreement 
between the extracted temperature profiles from the observed 
eclipses (left panel) and the typical profiles resulting from 
simulations at the substellar point (right panel).
Significantly, the predicted and observed spread of the 
temperature profiles at the substellar point in time is very 
similar, suggesting a qualitative agreement between observation 
and theory. 
From the simulations, we find that the three-$\sigma$ 
altitude-averaged temperature range is $680$\,K between p $\in [10^5,10^3]$\,Pa (e.g., assuming a normally distributed temporal spread of the $T$--$p$ profiles, this value means that altitude averaged $T$--$p$ profiles 
$340$\,K hotter or cooler than the mean should be 
considered outliers).
In comparison, we find from the retrievals that the five eclipses 
have an average temperature range of $311$\,K, when averaged over 
the same pressure domain range. 
Despite the possibility of our data being affected by small 
HST systematics still remaining and the difficulty of directly comparing inherently different models (i.e, 3D vs 1D), the scale of the temperature 
variations from the eclipse observations and from the theory is 
compatible. 

\begin{figure}
  %\vspace*{0.5cm}
    \centering
    \includegraphics[width = 0.45\textwidth]{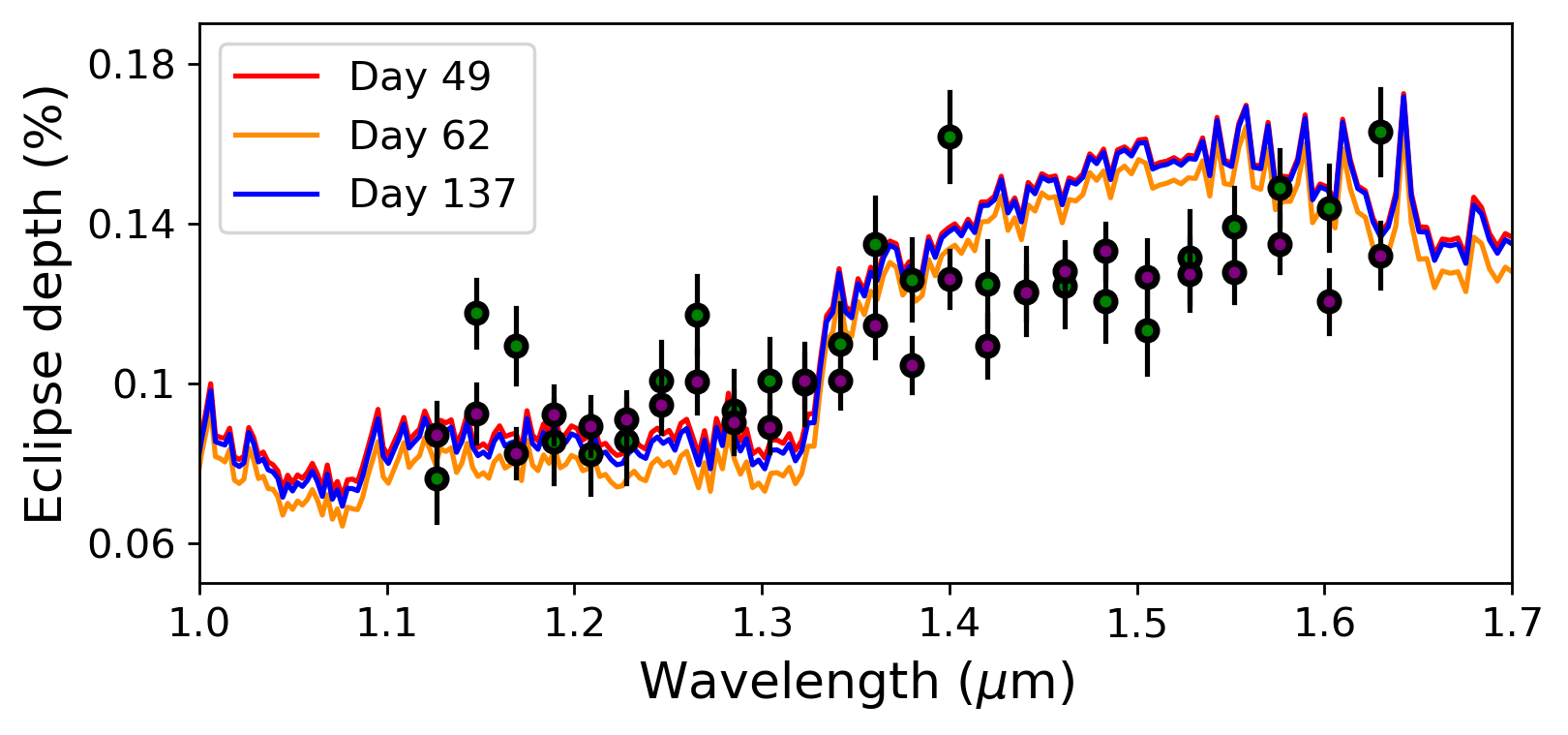}
    \includegraphics[width = 0.45\textwidth]{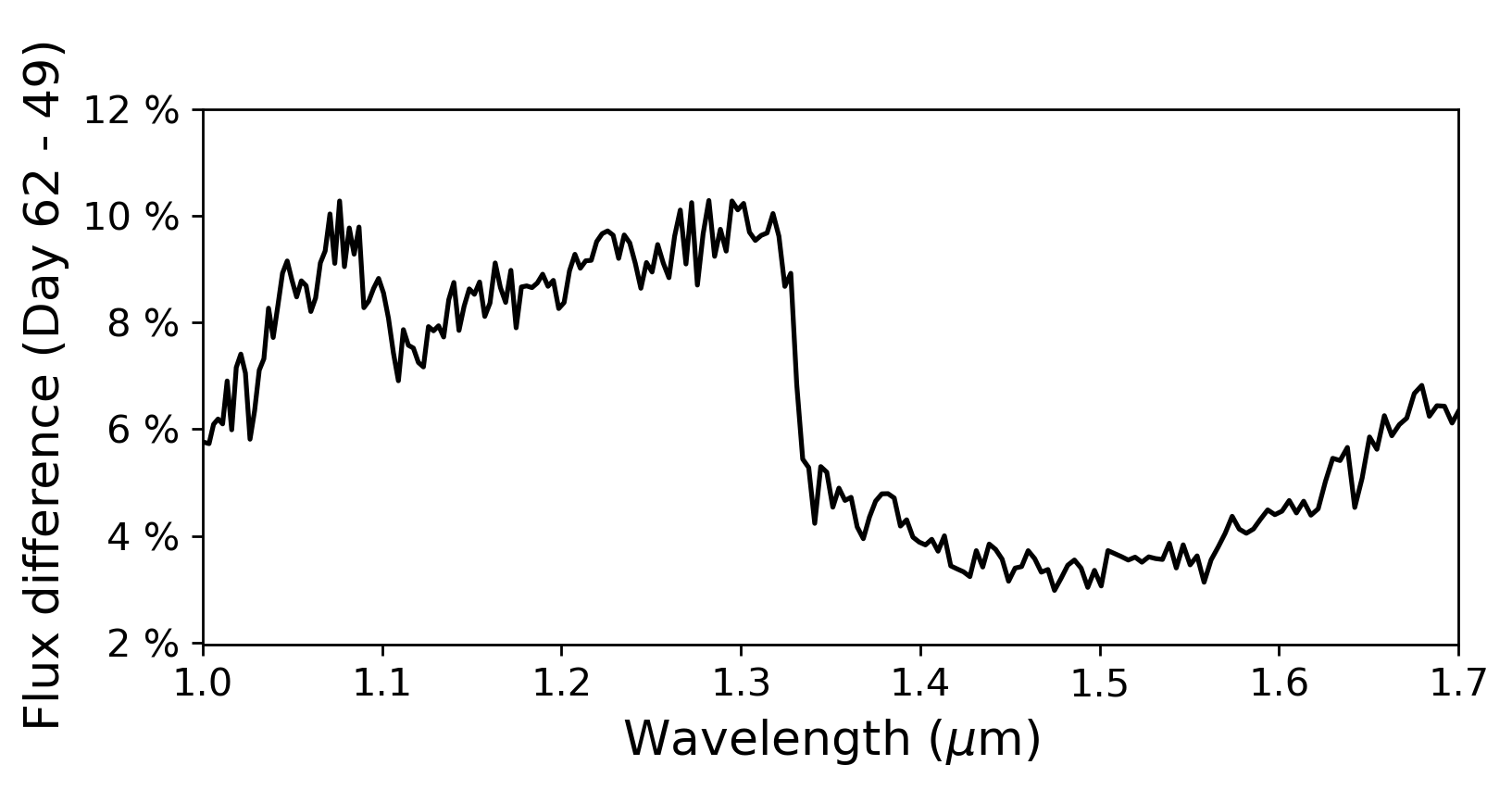}
    \caption{Spectroscopic eclipse flux obtained at three different 
    times, $t = \{49,62,137\}$\,days, when the dynamics calculations 
    are post-processed (top), and spectroscopic differences between 
    two times, $t = 62$\,days and $t = 49$\,days. 
    In the top panel, the eclipse spectra obtained for Observation 3 
    (green) and Observation 4 (purple) are also shown for reference. 
    The simulation shows that the atmospheric variability should 
    be wavelength dependent since, e.g., H$^-$ is highly sensitive 
    to thermal dissociation (impacting the short wavelengths) 
    while H$_2$O remains more chemically stable.}
    \label{fig:flux_variation_postprocess}
\end{figure}

As explained in the Materials and Method section of the Appendix, 
for three time frames, $t = \{49,62,137\}$\,days, we have 
post-processed the simulation outputs to obtain the emitted 
spectral flux. 
Figures 
\ref{fig:dynamics_chemistry_maps1}---\ref{fig:dynamics_chemistry_maps3} 
demonstrate the wide range of physical and chemical conditions that 
could appear on WASP-121\,b, which could strengthen the changes in 
cloud coverage and refractory chemistry claimed by \citet{Wilson_2021} 
and \citet{Ouyang_2023} respectively. 
In addition, we present a comparison of the modeled spectra in Figure~\ref{fig:flux_variation_postprocess}, 
which shows that variability of up to 10\% in the observed flux 
is compatible with the simulations. 
Such a level of flux variability agrees with theoretical works 
on other objects \citep{Skinner_2022_cyclogenesis}, and is within 
the capabilities of HST---provided that its instrument systematics 
are kept under control. 
Additionally, this level of variability should easily be captured 
by repeated observations of similar planets from JWST and Ariel.

Note that we obtain a lower level of variability when performing 
full radiative transfer calculation for the presented frames in Figure~\ref{fig:flux_variation_postprocess}. 
This could be because of a selection bias (with the chosen frames) 
and/or because of the baroclinicity (vertical non-alignment) of the atmosphere, which smooth the variability when the flux is integrated 
over a large pressure range. 
Moreover, we find that the variability is wavelength dependent, 
which we believe is caused by the changes in the chemistry. 
In particular, in the case of WASP-121\,b, H$_2$ thermally 
dissociates faster than H$_2$O; hence, we expect larger variability 
signals for wavelengths between 1.0\,$\mu$m and 1.3\,$\mu$m. \\

%%%%%%%%%%%%%%

\section{Discussion and Conclusion} \label{sec:dnc}

The extreme atmospheric conditions of WASP-121\,b make it an 
ideal laboratory to test our understanding of physical and 
chemical processes in the atmospheres of exoplanets.
Until now, detecting and studying weather patterns on exoplanets 
have remained elusive because of the lack of either adequate S/N 
or repeated, cross-verifiable observations that can be directly 
compared.
However, multiple, comparable observations with the HST for 
WASP-121\,b exist---and now permit progress to be made.
Moreover, truly high-resolution, {\it numerically converged} 
simulations also permit those observations to be interpreted 
with some fidelity, since the governing equations are now much 
more accurately solved than have been in the past.
The uniform treatment in the analysis of observation, together 
with informed guidance from numerically accurate and validated 
simulations, are the salient features of this study: without at 
least both, variability cannot be confidently assessed at 
present.
Here, bolstered by state-of-the-art simulations, we identify 
spectroscopic variability in the HST observations of 
WASP-121\,b. 

While some caution must still be exercised in interpreting HST 
data, given the well-known high level of systematics and the 
large number of assumptions required in reducing spectroscopic 
observations, we demonstrate here a strong potential evidence 
for variability associated with weather on WASP-121\,b.
Here the weather is inferred  from the following: {\it i}) the 
movement of the peak emission in two phase-curves, {\it ii}) 
the changing depth of the water feature in three transits, and 
{\it iii}) the variable retrieved thermal profiles in five 
eclipses. 
On WASP-121\,b, the large dayside--nightside temperature 
gradient---which is not necessarily fixed in space and 
time\footnote{due to the feedback from atmospheric dynamics 
or to the as yet incompletely understood thermal and/or orbital 
coupling with the host star}---is expected to power (as well as 
steer) dynamical, thermal, and chemical processes.
These include vortex instability, gravity wave and front 
generation, thermal dissociation, chemistry changes, and 
potential cloud/haze formation (e.g., silicate clouds), whose 
consequences are observable. 
Other studies using ground-based data have also suggested 
variable atmospheric conditions on WASP-121\,b \citep{Wilson_2021, 
Ouyang_2023}. 

A new finding in this study is that high-resolution dynamics 
simulations forced by $T(p)$ information retrieved from 
observations show that ultra-hot Jupiters, such as WASP-121\,b, 
likely have hot regions that are generally situated slightly 
eastward of the substellar point {\it when disk-averaged}---but 
whose actual shape and location are markedly variable in time. 
The variability is particularly visible in the modeled 
region, $p \in [10^3, 10^5] \, {\rm Pa}$, for which we have 
given the lion’s share of focus in this study.
This is in stark contrast with nearly all past hot Jupiter 
simulations, which show a fixed location and shape for a 
singular hot region; well-resolved simulations consistently 
indicate otherwise \citep[e.g.][]{Cho_2003, Skinner_2021}.
More specifically, our simulations indicate that variability 
for WASP-121\,b should be $\sim$5\% to $\sim$10\% in the 
disk-averaged flux with a frequency of $\sim$5\,days.
Hence, the signatures generated by these quasi-periodic 
weather patterns should be detectable with current as well 
as future instruments---e.g., by the JWST \citep{Greene_2016} 
and Ariel \citep{Tinetti_2021_redbook} telescopes---if 
repeated, high-quality observations are obtained.

\section*{Acknowledgements}

This work resulted from an initial discussion at the Exoplanet 
Symposium 2022, organized at the Flatiron Institute (FI).

Q.C. is funded by the European Space Agency (ESA) under the 
2022 ESA Research Fellowship Program.
J.W.S. is supported by the Jet Propulsion Laboratory, California 
Institute of Technology, under a contract with the National 
Aeronautics and Space Administration (80NM0018D0004).
O.V. is funded by the Centre National d’\'Etudes Spatiales (CNES), 
the CNRS/INSU Programme National de Plan\'etologie (PNP), and the 
Agence Nationale de la Recherche (ANR), EXACT ANR-21-CE49-0008-01. 
J.N. is funded by the Columbia University/Flatiron Institute (FI) 
joint Research Fellowship program.
G.M. is funded by the European Union Horizon 2020 program, Marie 
Sk\l{}odowska-Curie grant agreement.
This research received support from the UK Science and Technology 
Funding Council (STFC, ST/S002634/1 and ST/T001836/1), from the 
UK Space Agency (UKSA, ST/W00254X/1), and from the European 
Research Council (ERC, Horizon 2020 758892 ExoAI project).
B.E. and A.T. received travel support for this work under the ESA 
Science Faculty Funds, funding reference ESA-SCI-SC-LE117 and 
ESA-SCI-SC-LE118.

The FI, DIRAC, and OzSTAR facilities provided the computing resources. 
This work utilized the Cambridge Service for Data-Driven 
Discovery (CSD3), part of which is operated by the University 
of Cambridge Research Computing on behalf of the STFC DiRAC HPC 
Facility (www.dirac.ac.uk). 
The DiRAC component of CSD3 was funded by BEIS capital funding 
via STFC capital grants ST/P002307/1 and ST/R002452/1 and STFC 
operations grant ST/R00689X/1. 
DiRAC is part of the National e-Infrastructure.  
The OzSTAR program receives funding in part from the Astronomy 
National Collaborative Research Infrastructure Strategy (NCRIS) 
allocation provided by the Australian Government.

\section*{Data and materials availability}

This work is based upon observations with the NASA/ESA Hubble 
Space Telescope, obtained at the Space Telescope Science 
Institute (STScI) operated by AURA, Inc. The raw data used in 
this work are available from the Hubble Archive which is part 
of the Mikulski Archive for Space Telescopes. The specific observations analyzed can be accessed via \dataset[DOI: 10.17909/7an2-2m33]{https://doi.org/10.17909/7an2-2m33}.
We are thankful to those who operate these telescopes and their 
corresponding archives, the public nature of which increases 
scientific productivity and accessibility \citep{Peek_2019}.
Our HST reduction pipeline, 
\textsc{Iraclis}\footnote{\url{https://github.com/ucl-exoplanets/Iraclis}}, our light curve fitting code, \textsc{PoP}\footnote{\url{https://github.com/QuentChangeat/PoP_public}},
and our atmospheric retrieval code 
\textsc{TauREx3}\footnote{\url{https://github.com/ucl-exoplanets/TauREx3\_public}}, 
are publicly available on Github and can be installed from \textsc{pypi}. 
The chemical code \textsc{GGChem}\footnote{\url{https://github.com/ucl-exoplanets/GGchem}} in its \textsc{TauREx3} plugin version is 
also publicly available.
The atmospheric dynamics code \textsc{BoB} is also publicly 
available from NCAR \citep{Scotetal04}. 

\clearpage
\bibliographystyle{aasjournal}
\bibliography{main}
  
\clearpage

%\bigskip
%\bigskip

\renewcommand\thesection{\Alph{section}}
\renewcommand\thesubsection{\thesection.\arabic{subsection}}
\value{section} = 0

%\renewcommand\thesection{\Alph{section}}
%\renewcommand\thesubsection{\thesection.\arabic{subsection}}
%\value{section} = 0

\section{Materials and Methods}

\setcounter{figure}{0}
\renewcommand{\thefigure}{A\arabic{figure}}
\renewcommand{\theHfigure}{A\arabic{figure}}

\twocolumngrid

\subsection{Data reduction with \textsc{Iraclis}} 

\begin{figure}[H]
    \includegraphics[width = 0.49\textwidth]{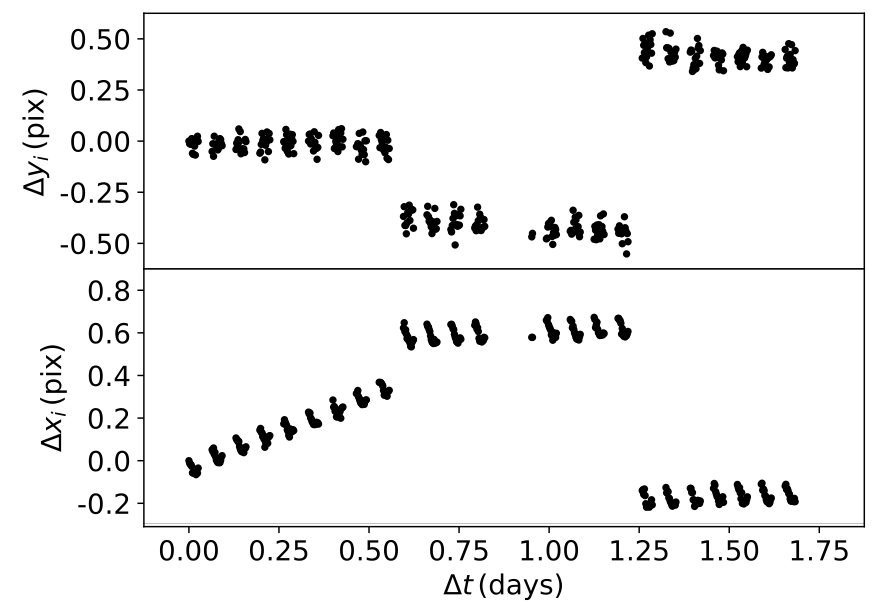}
    \caption{Displacement of the raw images onto the detector for the 
    2019 observation. The three segments, separated by the re-acquisition events, are clearly visible on those diagnostics.}
    \label{fig:offset}
\end{figure}

\begin{figure*}
    \includegraphics[width = 0.99\textwidth]{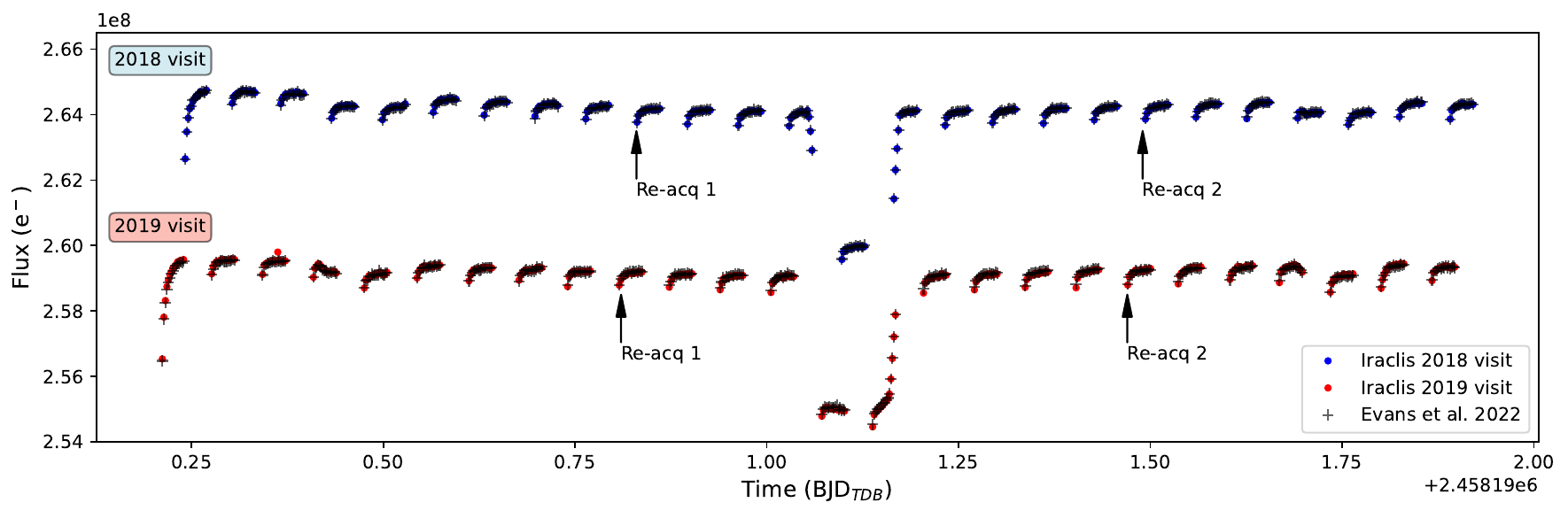}
    \caption{White light curves for the two WASP-121\,b phase-curve visits 
    obtained with our \textsc{Iraclis} extraction and compared to the ones in 
    ME22. The two visits were offset vertically (5$\times 10^6$ e$^-$) and 
    in time (by -257 orbital periods) for clarity. The observation 
    re-acquisitions (Re-acq) are also annotated.}
    \label{fig:white}
\end{figure*}

For the reduction and extraction of the spatially scanned spectroscopic 
images, we use the dedicated and publicly available 
pipeline \textsc{Iraclis} \citep{tsiaras_hd209, 
tsiaras_55cnce, tsiaras_30planets}. 
The individual transit and eclipse events have already been extracted in the population study of C22, so we obtain those outputs from this study. For the phase-curves, however, the data has not previously been extracted with  \textsc{Iraclis}, so we perform all the extraction steps described in \cite{tsiaras_h2o} as implemented in C22. Those steps consisted in zero-read subtraction, reference-pixels correction, non-linearity correction, dark current subtraction, gain conversion, sky background subtraction, calibration, flat-field correction, and bad-pixels/cosmic-rays correction. We then use \textsc{Iraclis} to extract the white (1.088-1.68 $\mu$m) and spectral light curves from the reduced images, taking into account the geometric distortions caused by the design of the WFC3 tilted detector. The two observed phase-curves had re-acquisition events, separating the observations in three distinct segments (see Figure~\ref{fig:offset}). Re-acquisitions cause displacements of the images onto the detector  and induce larger systematics at the beginning of each event. Pixels with higher flux rates are affected more, causing wavelength-dependent long-term ramps that require an additional treatment.
 
The extracted white light curves for both visits are compared to the ones obtained by ME22 in Figure~\ref{fig:white}. While completely independent, both reductions lead to very similar results. To extract the spectral light curves and study the impact of spectral binning on our conclusions, we explore two strategies: \\

1) {\it A medium spectral binning}. To ensure consistency, we test the same binning as other \textsc{Iraclis} reductions \citep{tsiaras_30planets, Changeat_2022_five, edwards_pop}, adopting an HST template composed of 18 wavelength bins \cite[see: ][]{tsiaras_h2o}. \\

2) {\it The low spectral binning}. For better comparisons with ME22, we test their proposed 12 bins strategy.  

\subsection{Light-curve analysis: the \textsc{PoP} pipeline}

To fit the light curves, we have developed a new pipeline \textsc{PoP}\footnote{Link to \textsc{PoP}: \url{https://github.com/QuentChangeat/PoP_public}} (Pipeline of Pipes), leveraging the flexibility of the \textsc{TauREx 3.1} framework \citep{al-refaie_2021_taurex3.1}. While \textsc{TauREx} was originally designed for atmospheric retrievals \citep{Waldmann_taurex1, Waldmann_taurex2}, its last version provides generic classes that can easily be used to solve optimization problems outside of its original scope. 

In terms of code structure, \textsc{PoP} combines an observation pipeline with a scientific model pipeline. Here, pipelines refer to a series of units representing individual transformation steps. In the context of this study, the ``observation'' pipeline has a unit to load the \textsc{Iraclis} reduced light curves and a sequence of other units to correct for the HST instrument systematics. The ``model'' pipeline contains an idealized light curve model. Note that the ``model'' pipeline can also possess additional transformation steps to convert the model's outputs to the ``observation'' pipeline format. During optimization, the free parameters are marginalized over likelihood for both the observation and the model pipelines. This structure makes rapid modifications of the pipeline easy and creates a clear separation between astrophysical models and instrument models. \\

\underline{Model Pipeline:} In the case of WASP-121\,b, we model 
the transit and eclipse events using the \textsc{Pylightcurve} package \citep{tsiaras_plc} as done in past works employing \textsc{Iraclis}, 
such as C22. 
The transit light curve drop (LC$_T$) and the normalized eclipse 
light curve drop (LC$_E$) from \textsc{Pylightcurve} are combined 
with a description of phase-curve variations (LC$_S$). 
Following the standard practice in the literature for 
HST \cite[see e.g.,][]{Evans_2022_diunarl}, we model the 
phase-curve variations using a first or second-order sinusoidal 
function: 
\begin{equation}
    LC_S\ =\ C_0 + 
    \frac{C_1}{2}\Big[1 - \mathrm{cos} (\Phi - C_2)\Big] + 
    \frac{C_3}{2}\Big[1 - \mathrm{cos}(2 \Phi - C_4)\Big],
\end{equation}
where $\Phi$ is the orbital phase, $C_0$ is the minimum of the 
nightside flux, $C_1$ is the maximum of the dayside flux, and 
$C_2$ is the phase-curve offset. $C_3$ and $C_4$ correspond to 
the second-order terms of the sine phase variations. 
As in \cite{Dang_2018}, the full phase-curve model M$_{PC}$ is 
constructed by:
\begin{equation}
    M_{PC}\ =\ LC_T + LC_S \times LC_E
\end{equation}
In transit, we computed the limb-darkening coefficients with the \cite{Claret_2000} law using the stellar \textsc{ATLAS} models \citep{Kurucz_1970, Howarth_2011, Espinoza_2015} included as part of the \textsc{ExoTETHyS} package \citep{Morello_2020,Morello2020_joss}. Those coefficients were fixed during the light curve fits. \\

\underline{Observation Pipeline:} With HST, the instrument systematics typically consists of a long-term trend affecting each visit and short-term ramps affecting each orbit. For the long-term trend ($IS_\mathrm{long}$), the standard practice is to assume a linear or a quadratic behavior \citep{tsiaras_hd209, Kreidberg_w103, arcangeli_w18_phase}. In the case of WASP-121\,b, since re-acquisition events had occurred, we were required to fit the long-term trends separately for each segment of the observations (here segments are noted with the index $i$ with $i \in \{1, 2, 3\}$). The corresponding long-term trend is: 
\begin{equation}
    IS_\mathrm{long}\ =\ (1 - A_0^i \times (t - t_s^i) - 
    A_1^i \times (t - t_s^i)^2) \times N^i 
\end{equation}
where $A_0^i$ is the linear coefficient of the segment $i$, $A_1^i$ the quadratic coefficient of the segment $i$, and $N^i$ is a normalization factor for the segment $i$. The time $t_s$ refers to the time at the beginning of each observation segment. For the detector short-term orbital ramps ($IS_\mathrm{short}$), previous studies have modeled its behavior using a combination of exponentials. Most studies discard the first frame of each orbit due to the larger systematics. Additionally, most methods also discard the first orbit, which is usually affected by a much larger and distinct exponential ramp. When discarding the first orbit and the first frame of each orbit, a simple exponential common to all the visits can be used \citep{kreidberg_w43, Stevenson_2014, tsiaras_hd209}:
\begin{equation}
    IS_\mathrm{short}\ =\ 1 + 
    B_0 \times \mathrm{exp} \left( \frac{t - t_o^j}{B_1}\right)  
\end{equation}
where $B_0$ and $B_1$ are the exponential factors, common to all visits in each observation and $t_o^i$ is the time at the beginning of each orbit $j$.
More recently, an evolution of this approach for the short-term ramps of HST has emerged \citep{Zhou_2017, DeWit_2018}. Their physically motivated model combines two exponential functions and is able to recover the first orbit. It is given by:
\begin{equation}
    IS_\mathrm{short}\ =\ 1 + 
    B_0 \times \mathrm{exp} \left( \frac{t - t_o^j}{B_1 R}\right)
\end{equation}
where $R$ is the correcting function mainly acting on the first orbit and given by:
\begin{equation}
    R\ =\ 1 + 
    B_2 \times \mathrm{exp} \left( \frac{t - t_v}{B_3}\right).
\end{equation}
$B_2$ and $B_3$ are the exponential factors of $R$ and $t_v$ is the time at the beginning of the visit. \\

In this paper, we explore the impact of various instrument systematics assumptions on the recovered data. The approach, however, is the same in all cases and followed C22 for consistency. We start by fitting the white light curve with the goal to infer the free parameters of the system that are wavelength independent -- i.e., the mid-transit time ($t_\mathrm{mid}$), the inclination ($i$), and the semi-major-axis ($a$). The other orbital parameters are fixed to literature values \citep{bourrier}, as they have better constraints from complementary observations. The fit is conducted with the nested sampling algorithm \textsc{MultiNest} \citep{Feroz_multinest, Buchner_2014}, using 1024 live points and an evidence tolerance of 0.5. An initial estimate of the observational noise is made by \textsc{Iraclis} \cite[see ][]{tsiaras_hd209}. However, we account for additional systematic noise or other unaccounted effects by re-scaling the uncertainties according to the RMS of the residuals, and perform a second fit using the updated uncertainties. In practice, this step has little effect for those WASP-121\,b observations as the RMS of the residuals was very close to 1 at the first stage. Since nested sampling computes accurate Bayesian log-evidence, ln(E), we note that the Bayes Factor can be used to perform model selection. 

To fit the spectral light curves, we employ a divide-white strategy similar to \cite{Kreidberg_GJ1214b_clouds, tsiaras_hd209, Evans_2022_diunarl} and we divide each spectral light curve by the corresponding white light curve. Since the detector ramps are correlated between wavelengths, this step essentially removes the short-term ramps and reduces the baseline drift in the data. As such, the observation pipeline for the spectral light curve only contains the long-term trend correction $IS_\mathrm{long}$ (i.e., $IS_\mathrm{short}$ is not needed). To match the divide-white observations, the model pipeline is also modified with an additional step. This step normalizes the modeled spectral light curves, dividing by the median white light curve obtained during the prior fit. Each spectral light curve is then fitted separately with this model, keeping $t_\mathrm{mid}$, $i$ and $a$ fixed to the best-fit value from the white. As with the white light curves, the errors are re-scaled to match the RMS of the residuals.

Finally, we extract the phase-curve spectra from the binned corrected light curves using 16 different phase bins ($\Phi$) of equal dimensions 0.05, giving us: $\Phi \in\{$0.07, 0.12, 0.17, 0.23, 0.28, 
0.32, 0.38, 0.43, 0.57, 0.62, 0.68, 0.72, 0.78, 0.82, 0.88, 0.93$\}$. Except for $\Phi \in \{0.07, 0.93 \}$, this matches the bins adopted in ME22. For those bins the planetary flux around $\Phi = 0.0$ is not included as it is blended during the transit, and a consistent bin size of 0.05 is used instead of 0.1 in ME22. Additionally, for this binning step, the finite integration time is not corrected for as the expected photometric distortions are below 5 ppm in the 0.05 intervals \citep{Morello_2022_binning}. For the transit spectra, self-blend (i.e, contamination by planetary emission) is not corrected for as the effect only affects the transit data by a few ppm in the case of WASP-121\,b \citep{2019_morello_w43, Evans_2022_diunarl, Morello_2021_blend}.

\subsection{Phase-curve retrievals}

We analyze the information contained in the phase-curve data using atmospheric retrievals. We employ the \textsc{TauREx 3.1} framework \citep{2019_al-refaie_taurex3, al-refaie_2021_taurex3.1}, following the previously established methodology detailed in \cite{Changeat_2020_phasecurve1,changeat_2021_phasecurve2,changeat_2021_w103}. We use the 1.5D phase-curve model, which is specifically designed to handle this type of observation, to simultaneously fit all the spectra in a Bayesian retrieval framework. The 1.5D model is composed of three different regions, referred to as hotspot, dayside, and nightside. Each region possesses independent properties allowing us to resolve large-scale atmospheric features from the data. The contribution of each region to the emitted flux at each phase is computed using a quadrature integration scheme \citep{Changeat_2020_phasecurve1}. For this study, the structure of the planet is defined by 90 layers equally spaced in log space from $p \in [10^6, 10^{-1}]$\,Pa. To first-order, such a model accounts for the main one-dimensional biases discussed in \cite{Feng_2016_inomogeneou,Taylor_2020,changeat_2021_phasecurve2}. 

As described in \cite{changeat_2021_phasecurve2}, the hotspot region is parameterized by two free parameters (hotspot size, $A_\mathrm{HS}$, and hotspot location, $D_\mathrm{HS}$). We have first attempted to retrieve those parameters but due to the large degeneracies between those parameters, the thermal structure, and the chemistry, this led to un-physical solutions \cite[a similar behaviour was found in][]{changeat_2021_phasecurve2}. Therefore, we have decided to fix those parameters and instead explored fixed values. For the hotspot size, we test $A_\mathrm{HS} \in \{30, 50\}$ degree cases. For the hotspot location, we test $D_\mathrm{HS} \in \{10, 20, 30, 40\}$ degree eastward-shifted cases (those choices are also motivated by the results in ME22). 

To parameterize the thermal structure, the temperature-pressure ($T$--$p$) profiles are created by linearly interpolating between $T$--$p$ nodes \cite[the pros and cons of such description are discussed in detail in the Appendix D of][]{changeat_2021_phasecurve2}. For the hotspot region and dayside, we use seven nodes at fixed pressures ($p \in \{10^6, 10^5, 10^4, 10^3, 100, 10, 0.1\}$\,Pa), while for the nightside, since the information content is reduced due to the lower planetary emission, we choose to only use five nodes ($p \in \{10^6, 10^5, 10^3, 10, 0.1\}$\,Pa). For the chemistry, we employ the \textsc{GGChem} \citep{Woitke_2018_GG} chemical equilibrium code in its \textsc{TauREx 3.1} plugin and we couple the two only free parameters (metallicity and C/O ratio) between the three different regions. Previous works have shown the importance of dis-equilibrium processes for hot Jupiters \citep{Moses_2011,Drummond_2020,venot_2020_w43, venot_chem_HJ, Al-Refaie_2022_FRECKLL}, however, given the temperatures of WASP-121\,b, chemical reactions should be fast, favoring chemical equilibrium for the investigated species \citep{Parmentier_2018_w121photodiss, kitzmann_k9}. The radiative transfer includes absorption from the main expected opacity sources via ExoMol line lists \citep{tennyson2012exomol, Chubb_2021_exomol}, namely: H$_2$O \citep{polyansky_h2o}, CO \citep{li_co_2015}, CO$_2$ \citep{Yurchenko_2020}, CH$_4$ \citep{ExoMol_CH4_new}, TiO \citep{McKemmish_TiO_new}, VO \citep{McKemmish_2016_vo}, FeH \citep{Bernath_2020_FeH}, and H$^-$ \citep{John_1988_hmin, Edwards_2020_ares}. We also consider Collision Induced Absorption (CIA) by H$_2$-H$_2$ and H$_2$-He pairs \citep{abel_h2-h2, abel_h2-he, fletcher_h2-h2} and Rayleigh scattering \citep{cox_allen_rayleigh}. A fully opaque cloud deck (referred here as gray clouds) was also included on the night side of the planet but we were not able to find evidence for clouds from this phase-curve data. The parameter space of the model is explored using the nested sampling optimizer \textsc{MultiNest} \citep{Feroz_multinest, Buchner_2014}, with 500 live points and an evidence tolerance of 0.5. To explore the parameter space, priors were chosen to be non-informative (i.e., uniform priors). Specifically, for the temperature points, we explored the space from $T \in [300, 6000]$\,K. For the chemistry, the metallicity was explored in log space from Z $\in [0.1, 100]$ times solar, while the C/O ratio was explored from C/O $\in [0.1, 2]$. From this retrieval analysis, we were able to extract averaged chemical properties of WASP-121\,b as well as the thermal structure of the three considered regions. 

\subsection{Transit and eclipse retrievals}

To evaluate the variability of WASP-121\,b's atmosphere, we also 
analyze each eclipse and transit spectra individually, using 1D 
atmospheric retrievals with \textsc{TauREx 3.1}. 
Previous studies have shown the difficulty of extracting reliable constraints from single HST visits \citep{Changeat_2020_K11} due 
to degeneracies between chemical abundances and thermal properties. 
To break those degeneracies, we use our most accurate estimate of 
the time-independent parameters from our phase-curve retrieval as 
priors for our individual retrievals. 
Using the {\it low}-resolution results, the metallicity Z of the atmosphere is therefore fixed at 
log(Z) = -0.19, while the carbon-to-oxygen ratio C/O is fixed at 
C/O = 0.80. Note that this simplification is not expected to always be correct, for instance cloud condensation can locally (and temporally) change the C/O ratio of the gas phase, which we do not model here (i.e., \textsc{GGChem} is used without condensation). However, without additional knowledge or constraints on condensates in WASP-121\,b, this remains a reasonable and necessary assumption. \\

{\it Transits:} Since HST transits probe a narrow pressure 
range, and because it is mainly affected by the atmospheric scale 
height \citep{Rocchetto_2016}, we consider a simple isothermal profile with a unique free parameter $T$ for those observations (spectra shown in Figure~\ref{fig:spectra_ec_tr}). As with the phase-curve data, the temperature is explored with uniform priors ($T \in [300, 6000]$\,K).
However, transit is much more sensitive to clouds than eclipse 
or phase-curve, so we have used a more complex representation 
of clouds from \cite{Lee_2013_clouds}, for which particle size 
and mixing ratio were fitted. This cloud model was favored by 
the Bayesian evidence compared to the gray-cloud case. \\

{\it Eclipses:} As planetary radius is known to be
degenerate with temperatures in HST eclipses 
\citep{Edwards_2020_ares, pluriel_aresIII}, we fix this 
parameter to the literature value. For each spectrum (see Figure~\ref{fig:spectra_ec_tr}), we retrieve a thermal profile 
with a similar parameterization to the phase-curve case. Namely, the $T$--$p$ profile is parameterized by linearly interpolating between seven freely moving $T$--$p$ nodes. The pressure of each node is fixed to log-spaced values of pressures (i.e., $p \in \{10^6, 10^5, 10^4, 10^3, 100, 10, 0.1\}$\,Pa) and we retrieve the temperature of each point individually using the same uniform non-informative priors. 
We refer to \cite{changeat_2021_phasecurve2, Rowland_2023} for a more 
complete discussion on thermal structure parameterizations 
and their trade-offs. 
A simpler 3-point thermal profile (where the pressure levels of each node are left free) was also tested, which did 
not change our overall conclusions. For the eclipse spectra, 
we also decided to run a simple blackbody planet fit, which 
served as our comparison baseline. Following this procedure, 
and because of the additional chemistry priors from our 
phase-curve analysis, we have obtained well-defined thermal 
structures for each of the five eclipses.

\subsection{Dynamics modeling}\label{sec:meth:dyn}

We model the atmospheric dynamics of WASP-121\,b using the 
pseudospectral dynamical core, \textsc{BoB} 
\citep[e.g.,][]{Rivietal02,Scotetal04, Polietal14, Skinner_2021, Skinner_2021_modons}. 
The core has been outfitted and set up in 
\citet{Skinner_2021_modons}, \citet{Skinner_2021}, and 
\citet{Cho_2021} especially for high-resolution hot Jupiter 
simulations; and, we refer the reader to those works for a more 
complete description of the code and governing equations solved. 
However, for the readers' convenience, we provide a brief 
summary of the key features of the model here -- especially as they 
pertain to modeling of WASP-121\,b atmosphere.  Other numerical 
models have been used to study hot Jupiter atmospheric dynamics in 
the past \citep[e.g.][and references therein]{Showman_Guillot_2002, 
Cho_2003, Cho_2008, cho_2008-mhd, Dobbs-Dixon_2010, Thrastarson_2010, Polietal14, Heng_2014, Mayne14, Mendonca_2018_w43, Parmentier_2018_w121photodiss}, and we 
direct the reader to consult those works for instructive context.

\textsc{BoB} calculates the large-scale dynamics of the atmosphere 
by numerically solving the traditional and hydrostatic primitive 
equations in (longitude, latitude, pressure) = ($\lambda, \phi, p$) coordinates an in vorticity-divergence--potential-temperature 
formulation. 
In the vertical ($p$) direction, \textsc{BoB} employs a second-order 
finite differencing scheme with free-slip boundary conditions at 
the top and bottom pressure surfaces. 
The present study builds upon extensive testing and validation 
of \textsc{BoB}, 
including simulations at high-resolution and under numerically 
challenging conditions resembling those found on WASP-121b \citep{PoliCho12, Polietal14, cho_2015, Skinner_2021}.

For the physical setup of WASP-121\,b we use the parameters in \cite{Delrez_Wasp121b_em}. 
To simulate irradiation from the planet's host star we implement 
an ``idealized'' thermal forcing using the Newtonian relaxation 
scheme which accelerates the initially resting atmosphere 
($\mathbf{u} = 0$) towards specified {\it hotspot} and nightside 
equilibrium $T$--$p$ profiles that are obtained from the phase-curve 
retrievals of the planet (see Fig.~\ref{fig:spectra_tp_retrieval}).
The initial temperature is the average of the {\it hot spot} and 
nightside equilibrium temperatures.
Due to the planet's close proximity to its star, the effect of 
its spherical geometry on stellar irradiation deposition is 
accounted for using a cosine profile to graduate the {\it hot spot} temperature from the substellar point to the terminators.
The radiative cooling time $\tau_r(p)$ is computed from the 
initial temperature profile following \cite{Cho_2008};  
$\tau_r(p)$ is approximately linear in $\log(p)$, ranging from
$\mathcal{O}(10^6)$s at $p = 10^5$ Pa to $\mathcal{O}(10^2)$s 
at $p = 10^3$ Pa. 

For the numerical setup, we use high horizontal and high vertical resolutions: T682 and L50 respectively. 
In the former, `T' denotes the highest wavenumber retained in the 
spherical harmonic expansion (the truncation wavenumber) is 
$n_t = 682$ and the latter `L' denotes the number of vertical 
levels which are distributed linearly in $p$-coordinate. By 
``high-resolution'' we mean our simulations are above the 
minimum resolution required for numerical convergence of flow 
solutions on hot, tidally synchronized exoplanets 
\citep{Skinner_2021}.
Note that high vertical resolution is also necessary to ensure 
the baroclinic region of the dayside and hotspot temperature 
profiles ($p \lesssim 10^4$ Pa) is well represented in $p$.
A small timestep size of $\Delta t = 6$ seconds is used 
concomitantly with the fine grid spacing to ensure flows at 
the maximum sound speed (i.e., $c_s \sim 4880$ m/s in the 
hottest region of the atmosphere) are well captured.
Hence, the simulations maintain a Courant--Friedrichs--Lewy 
\cite[CFL:][]{Courant_1928} condition of well below unity. 
Simulations are time-integrated for 200\,planet days, to 
ensure they reach a quasi-stable state long after their 
initial ramp-up period of $\sim$40 WASP-121\,b\,days. 

For the domain size, we model the expected $p$ range (from 
$p_t = 10^3$ Pa to $p_b=10^5$ Pa) from which radiation originates, 
as indicated by HST data (see Figure~\ref{fig:spectra_tp_retrieval} 
and Figure~\ref{fig:tp1d}). 
We have verified that the model equations are valid for this region 
by confirming that the retrieved thermal forcing profiles are stably stratified (i.e. that the Brunt–Väisälä frequency $\mathcal{N} = 
\sqrt{\left(g /\rho \right) \left({\rm d}\rho/{\rm d}z\right)}$ 
is real). This is shown in Fig.~\ref{fig:BV_profiles}. 
Below this ($p > 10^5$), the nightside temperature profile 
exhibits a jump in stratification. Although this could be due 
to increased uncertainty on the retrieved thermal structure in 
this region.

Finally, for numerical dissipation we use a high-order 
hyper-viscosity $\nabla^{16}$ with a small corresponding 
artificial viscosity coefficient of $\nu_{16} = 10^{48}$. 
This prevents excessive kinetic energy removal from small-scale 
flows and hence ensures the dynamics of large-scale flows are
well represented \citep[see e.g.][]{Cho_1996, Skinner_2021}. 
In addition, a very weak Robert--Asselin filter with coefficient
$\epsilon = 0.02$ is used to filter the additional computational 
mode arising from the models leapfrog time integration scheme \citep{Asselin_1972, Thrastarson_2010}. 
Besides this, no other drags are applied as these can coerce the 
flows to dynamically unphysical states \citep{Polietal14}. 
The simulations are allowed to evolve freely under thermal 
forcing from the retrieved temperature profiles.

For the post-processing of the three-dimensional atmospheric 
dynamics simulations, we employ two different approaches for 
evaluating the evolution of WASP-121,b from an observational 
perspective. 
The first is a one-dimensional time-series analysis of flux 
emitted by a single layer at the mid-region of our computational 
domain (0.1\,bar) in order to evaluate the qualitative behavior 
of the planet's weather over hundreds of planet days. Here the 
black-body emission is calculated from \textsc{BoB} temperature 
maps, centered on key regions of interest (i.e., the substellar 
point, eastern terminator, western terminator, and antistellar 
point).
While this approach does not account for the entire domains 
contribution towards atmospheric variability, it enables the 
key dynamical processes in the atmosphere to be isolated and 
studied in detail.

For second phase of our post processing, we link the outputs of 
\textsc{BoB} with the \textsc{TauREx3} library to simulate observables 
and produce detailed 3D chemical maps of the planet's atmosphere. 
Due to the large size of the \textsc{BoB} simulation grid cube we 
isolate frames from the \textsc{BoB} calculations with significant 
spatial temperature differences that are likely observable for 
this analysis (e.g., $t = 50$ and $t = 62$ days). 
First, we derive the chemical maps over the entire 
$(\lambda, \phi, p) = (2048, 1024, 50)$ (i.e., grid cube) by 
performing calculations with the \textsc{GGChem} \citep{Woitke_2018_GG} 
chemical equilibrium code using the $T$--$p$ values from each 
grid square of the \textsc{BoB} simulations. We maintain fixed 
values for the metallicity and C/O ratio based on the median 
values obtained during our atmospheric retrieval exploration 
(log(Z) = -0.19 and C/O = 0.80).
The flux emitted by the planet is then computed using the 
\textsc{TauREx3} plane-parallel radiative transfer model, 
modified for our three-dimensional grid and to account for 
the changing viewing angles. That is, for each column of the 
computational grid, the flux is propagated upwards from 
0.5$\mu$m to 50$\mu$m at resolution R = 15,000 and then 
summed by taking into account the viewing angle of each grid 
element. In this calculation, the same opacities as during the 
retrievals were used: molecular absorption via \textsc{ExoMol} 
cross-sections, H$^-$ opacity, Collision Induced Absorption 
(CIA) and Rayleigh scattering. We computed the planetary flux 
in those frames as if the planet were observed in eclipse 
(i.e., phase 0.5).

\onecolumngrid

\clearpage
\section{Complementary Figures to the Section 3}
\renewcommand{\floatpagefraction}{.99}%

This appendix contains the complementary figures to the main article Section 3, Figure~\ref{fig:spec_white} to Figure~\ref{fig:spec_comparison}.

\setcounter{figure}{0}
\renewcommand{\thefigure}{B\arabic{figure}}
\renewcommand{\theHfigure}{B\arabic{figure}}

\vfill
\begin{figure}[H]
\centering
    \includegraphics[width = 0.93\textwidth]{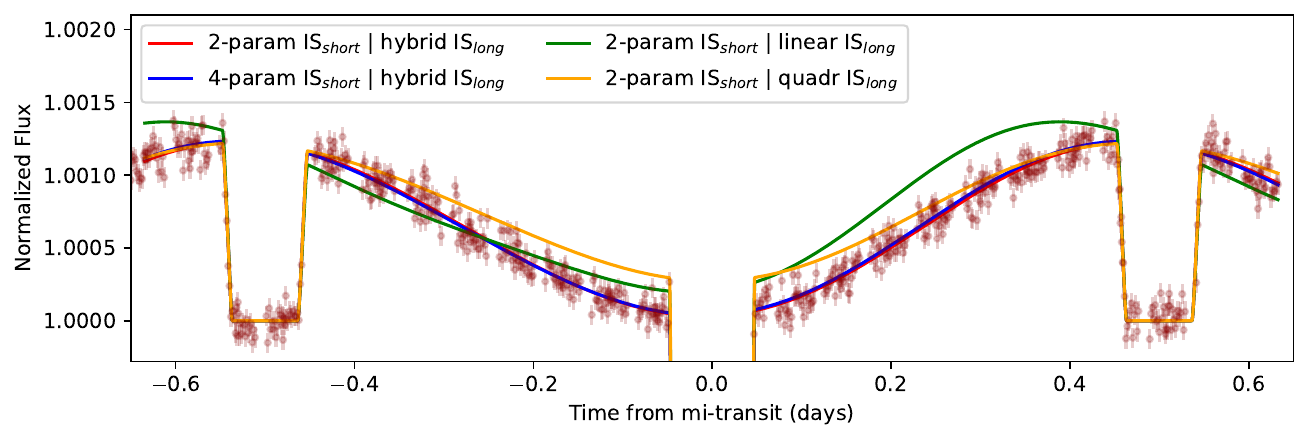}
    \caption{Comparison of recovered white light curves, fitted with the \textsc{PoP} when using different assumptions for the detector and long-term ramp models. We show in solid line the median of each tested model, and for our preferred set of assumption, we show the resulting observations. Red: Two parameters exponential ramp model from \cite{tsiaras_hd209} and hybrid long-term ramp; Blue: Four parameters exponential ramp model from \cite{DeWit_2018} and hybrid long-term ramp; Green: Two parameters exponential ramp model and linear long-term ramp for each segment; Orange: Two parameters exponential ramp model and quadratic long-term ramp for each segment. The hybrid long-term ramp consists in a quadratic trend for the first segment of each visit and a linear ramp for the remaining four segments, as done in ME22. This, as well as Figure~\ref{fig:corner_white}, demonstrate that the recovered light curve depends on the assumption for the long-term ramp.}
    \label{fig:spec_white}
\end{figure}
\vfill
\clearpage

\begin{figure}[H]
\centering
    \includegraphics[width = 0.97\textwidth]{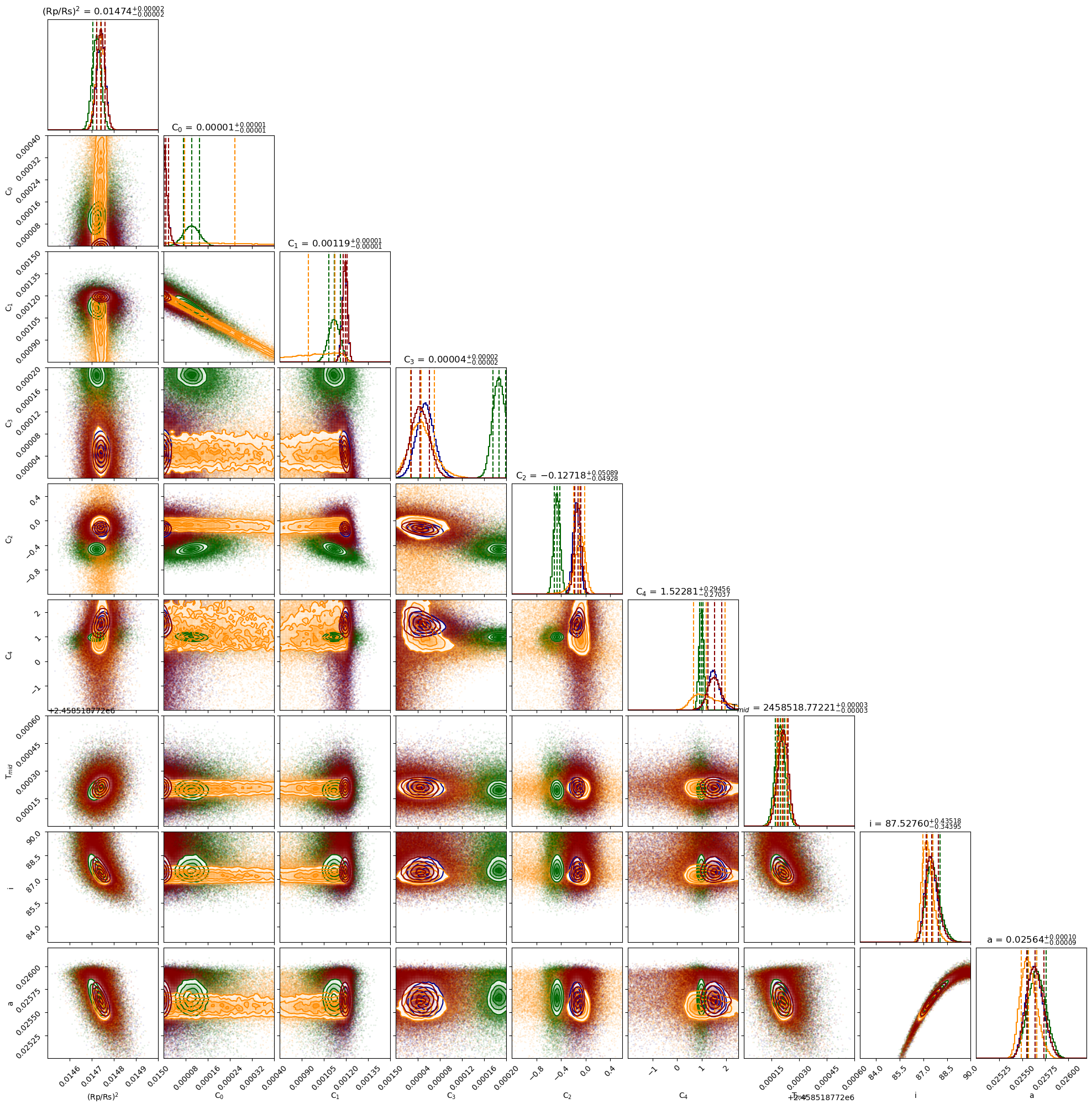}
    \caption{Posterior distributions of the white light curve fits 
    with \textsc{PoP} using four different detector ramp models (see Figure~\ref{fig:spec_white}). 
    Red: Two parameter exponential ramp model from \cite{tsiaras_hd209}; 
    Blue: Four parameters exponential ramp model from \cite{DeWit_2018}. 
    Green: Two parameters exponential ramp model and linear long-term 
    ramp for each segment; 
    Orange: Two parameters exponential ramp model and quadratic 
    long-term ramp for each segment. The recovered orbital 
    parameters are independent of the detector short-term ramp, 
    but they are impacted by the choice of the long-term trend.}
    \label{fig:corner_white}
\end{figure}
\clearpage

\vfill
\begin{figure}[H]
\centering
    \includegraphics[width = 0.93\textwidth]{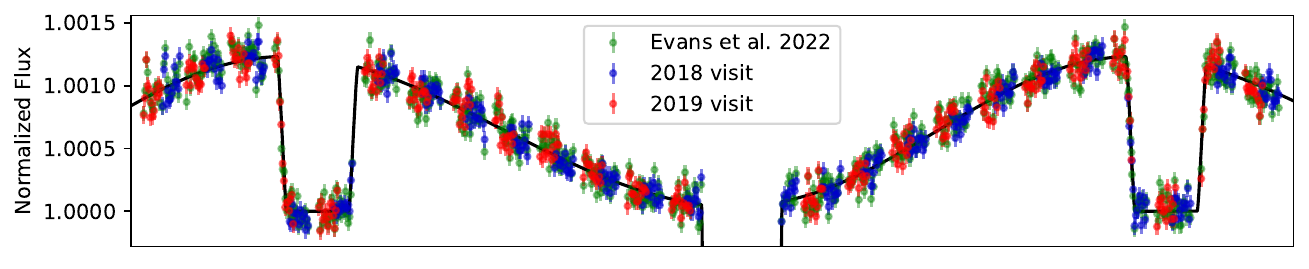}
    \includegraphics[width = 0.93\textwidth]{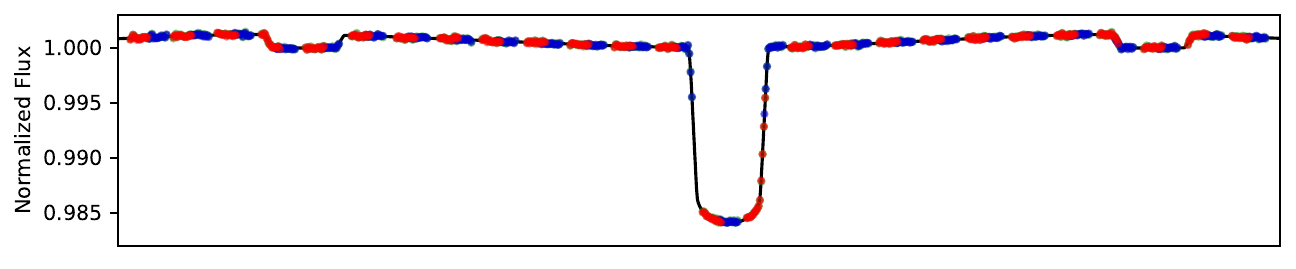}
    \includegraphics[width = 0.93\textwidth]{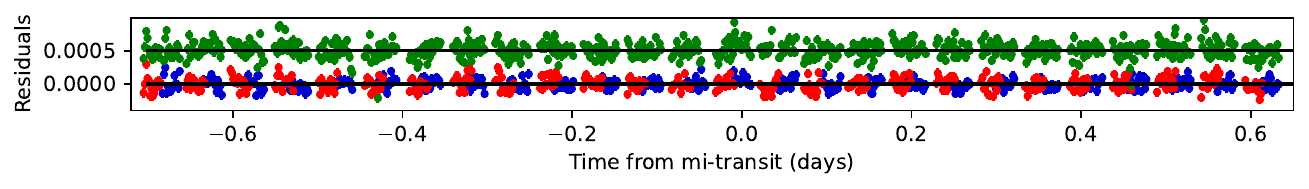}
    \caption{Corrected white light curves when the two WASP-121\,b 
    phase-curve visits are fitted together. We show our results with 
    the \textsc{Iraclis} extraction and compared to the ones in \cite{Evans_2022_diunarl}. 
    Top: Corrected light curves zoomed in the eclipses; Middle: Full 
    corrected light curves; Bottom: Residuals between our best-fit 
    light curve model and corrected data with red and blue for the 
    \textsc{Iraclis} reductions and green for the reduced data in \cite{Evans_2022_diunarl}.}
    \label{fig:white_corrected}
\end{figure}
\vfill
\clearpage

\begin{figure}[H]
\centering
    \includegraphics[width = 0.97\textwidth]{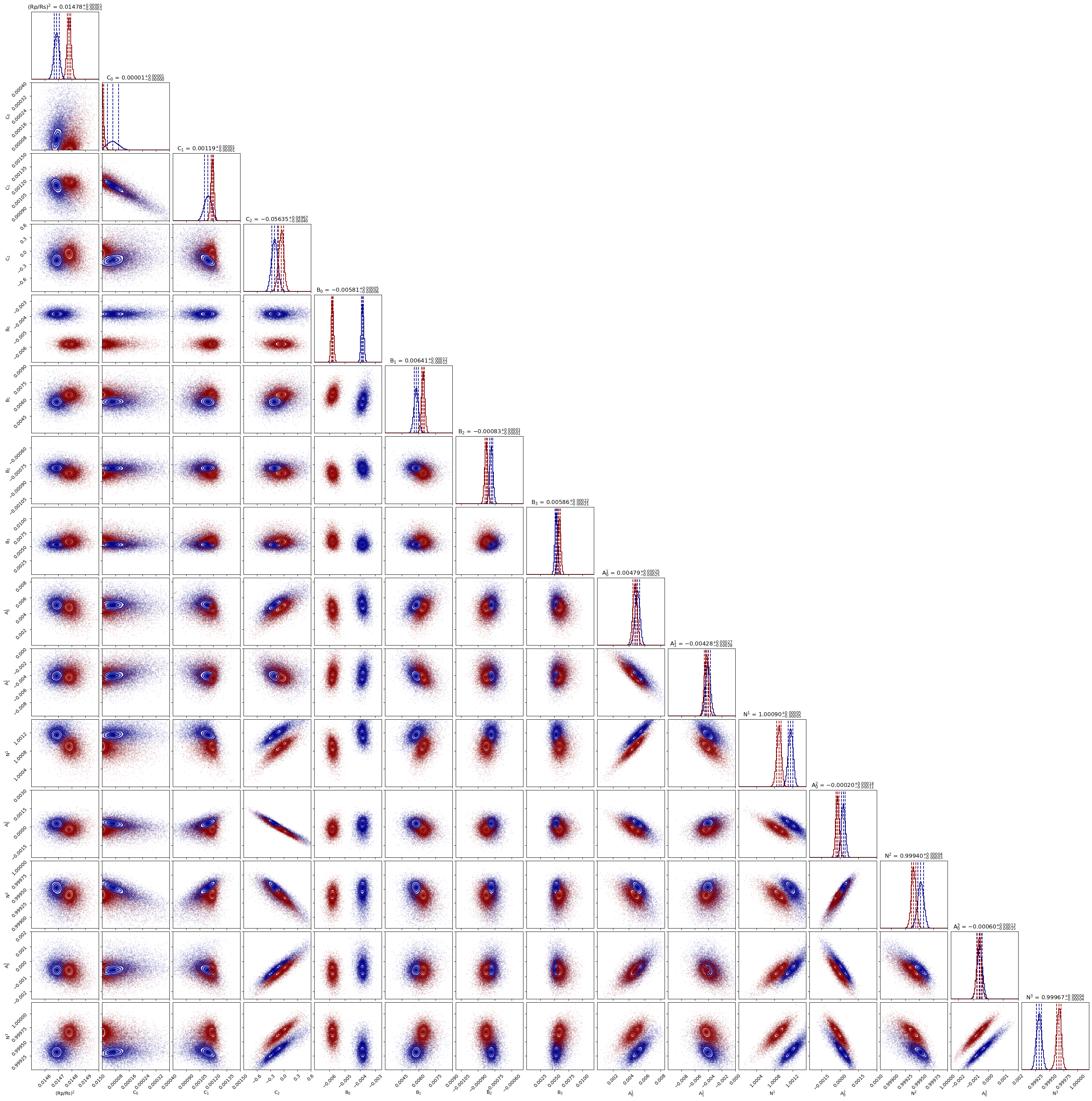}
    \caption{Posterior distributions of the white light curve fits with 
    \textsc{PoP} when fitting the two phase-curve visits independently. 
    The recovered parameters show differences in both the phase-curve 
    model and the instrument systematics.}
    \label{fig:corner_white_two_obs}
\end{figure}
\clearpage

\begin{figure}[H]
    \includegraphics[width = 0.679\textwidth]{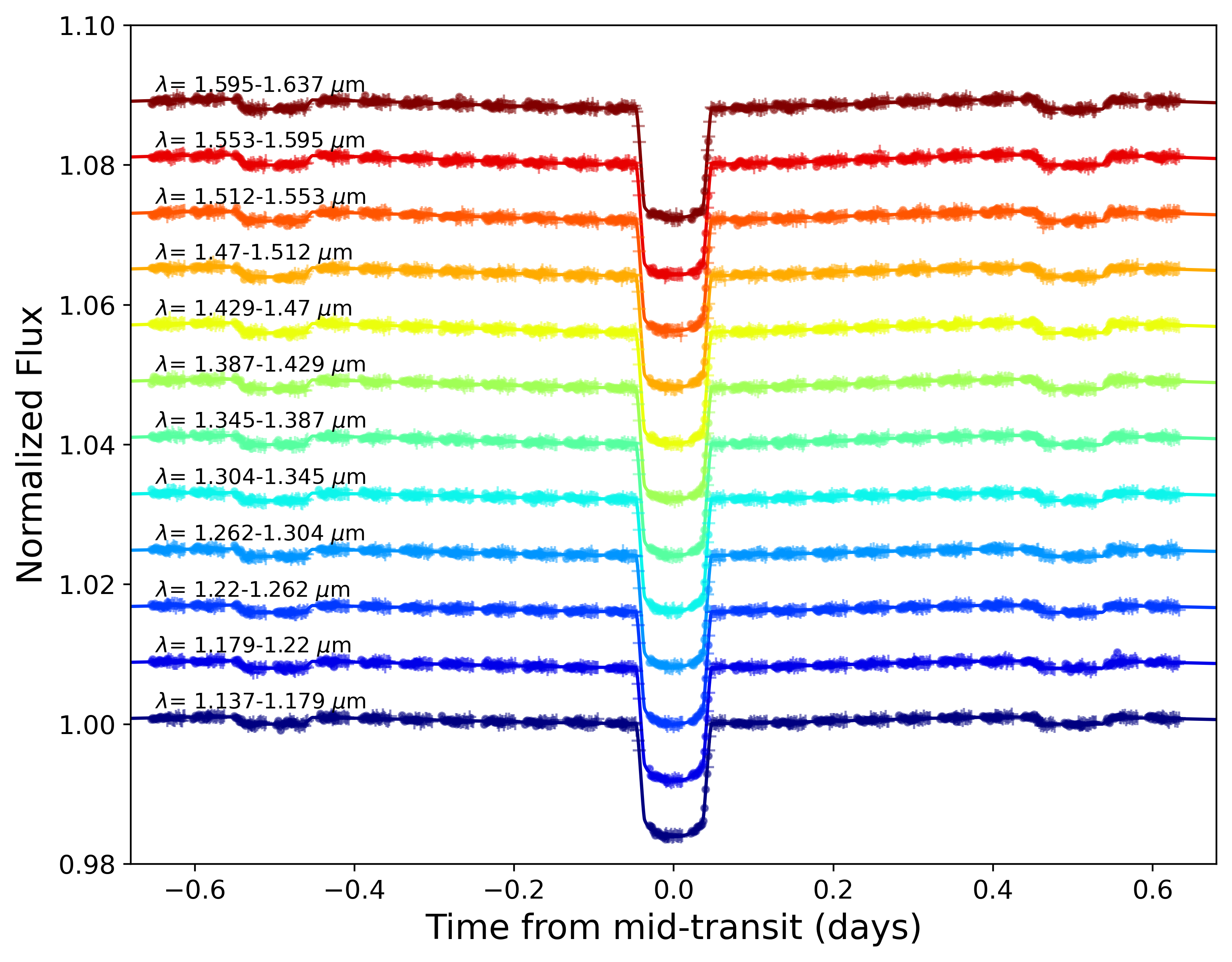}
    \includegraphics[width = 0.32\textwidth]{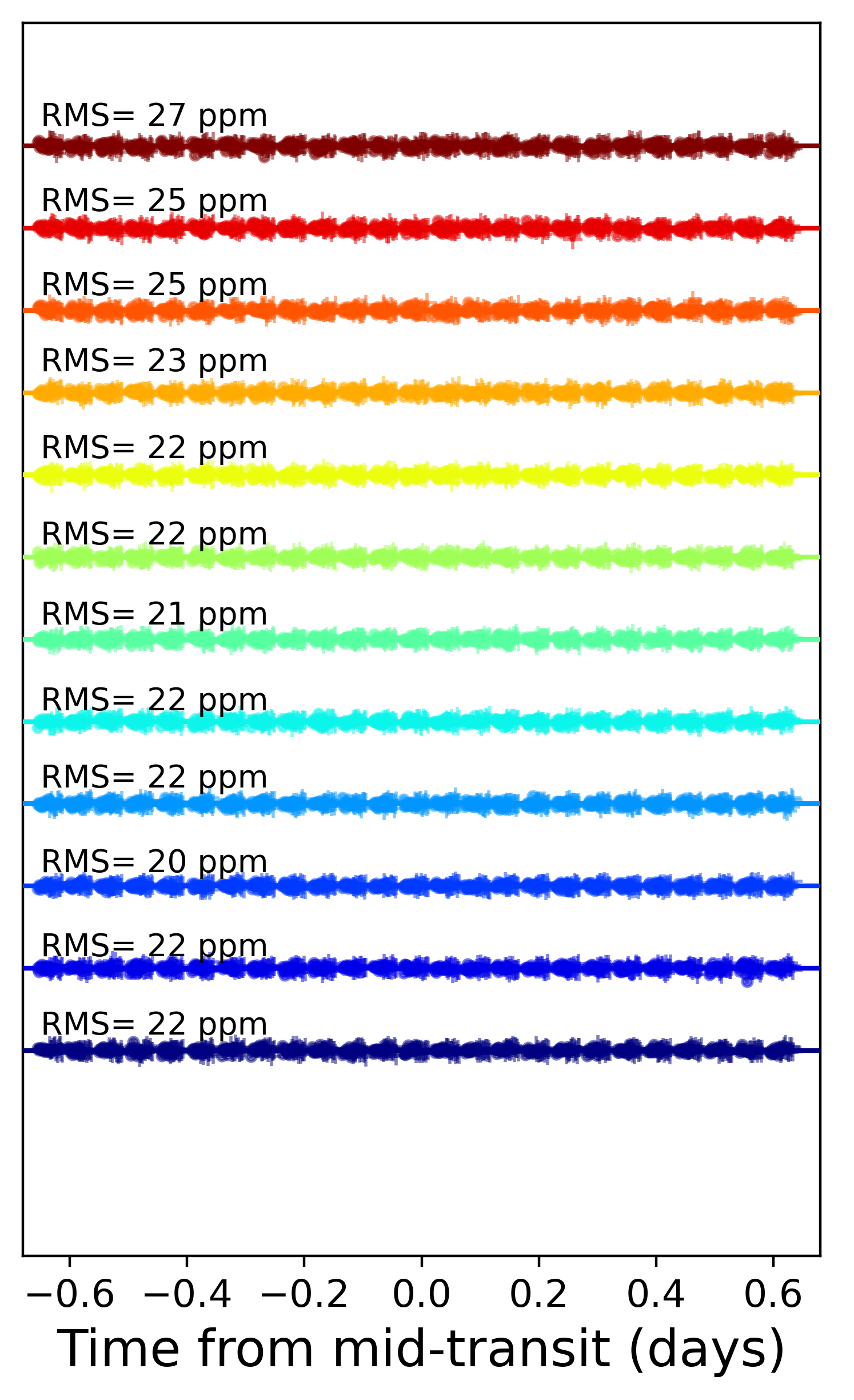}
    \caption{Corrected spectral light curves (left) and corresponding 
    residuals (right) for the {\it low} spectral binning. 
    This is the same binning  employed in \cite{Evans_2022_diunarl}. 
    Higher binning resolution fits (e.g. {\it medium} resolution) are 
    also performed and presented in Figure~\ref{fig:spec_corrected_HR}. }
    \label{fig:spec_corrected}
\end{figure}

\begin{figure}[H]
    \includegraphics[width = 0.679\textwidth]{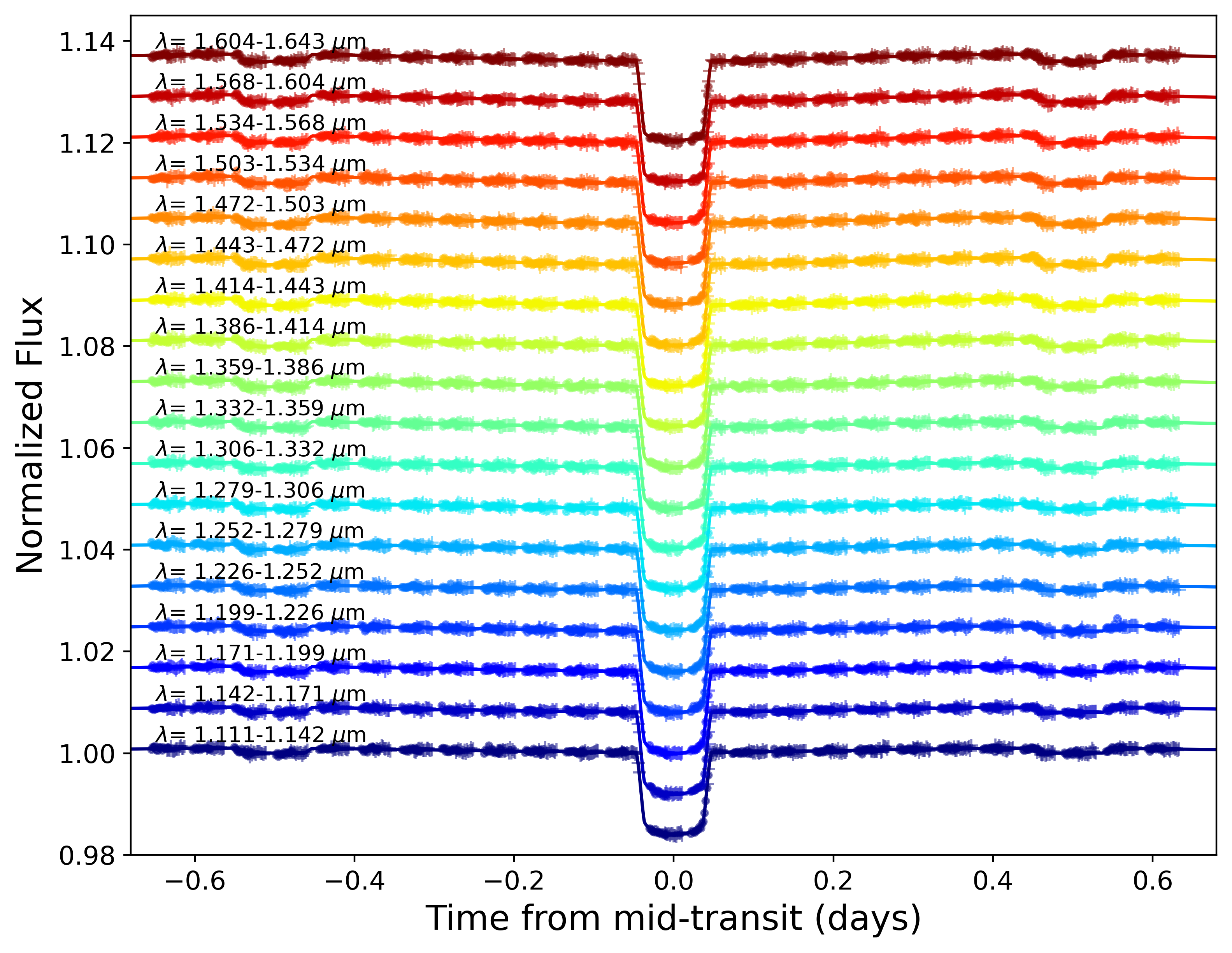}
    \includegraphics[width = 0.32\textwidth]{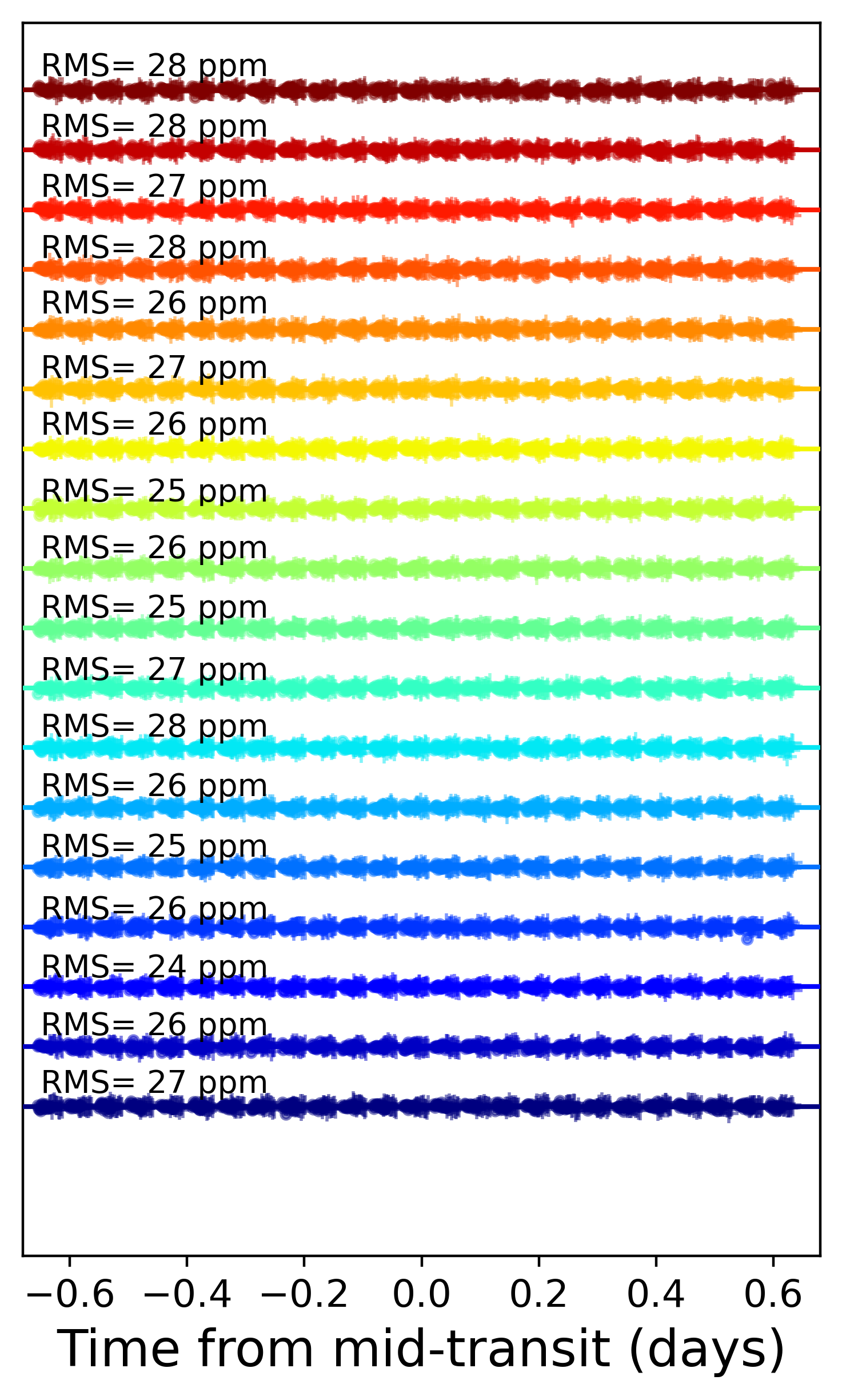}
    \caption{Corrected spectral light curves (left) and corresponding 
    residuals (right) for the {\it medium} spectral binning. 
    The binning is similar to previous works using reductions from 
    \textsc{Iraclis}, such as \cite{tsiaras_30planets, Changeat_2022_five, edwards_pop}.}
    \label{fig:spec_corrected_HR}
\end{figure}

\clearpage

\begin{figure}[H]
    \includegraphics[width = 0.99\textwidth]{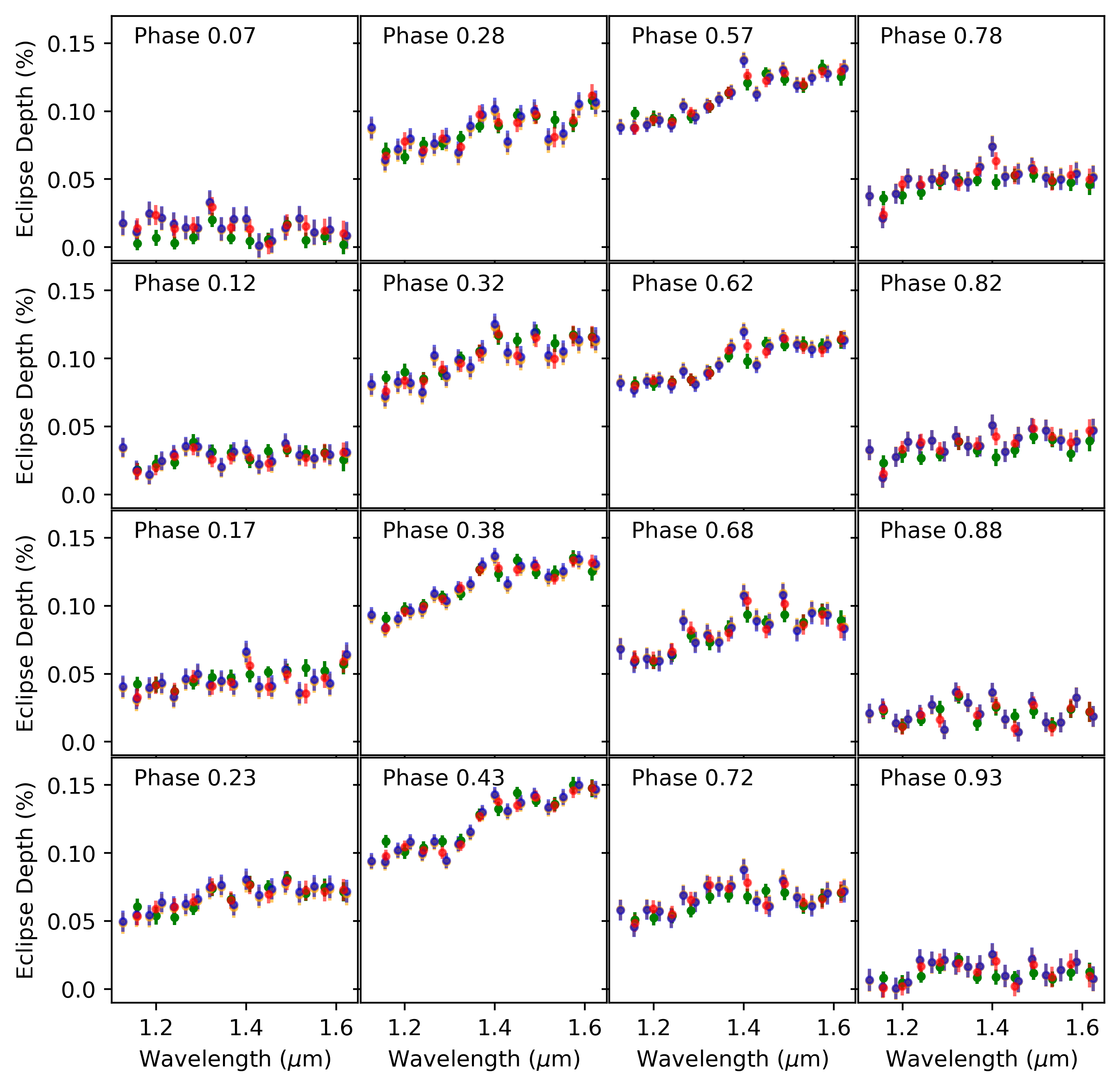}
    \caption{Emission spectra of WASP-121\,b, obtained at different 
    phases from various reduction methods. 
    Green: spectra at {\it low} resolution from ME22. 
    Red: spectra at {\it low} resolution obtained using the 4-short 
    instrument systematic model. 
    Blue: spectra at {\it medium} resolution obtained using the 4-short 
    instrument systematic model. 
    Orange: spectra at {\it medium} resolution obtained using the 
    2-short instrument systematic model. 
    All the reductions are consistent with each other. 
    Note that phases 0.07 and 0.93 do not exist in ME22; hence,  
    we instead plot their phase 0.05 and 0.095.}
    \label{fig:spec_comparison}
\end{figure}
\clearpage

\section{Complementary Figures to the Section 4}

This appendix contains the complementary figures to the main article 
Section 4, Figure~\ref{fig:tp_reductions} to Figure~\ref{fig:sensitivity}.

\setcounter{figure}{0}
\renewcommand{\thefigure}{C\arabic{figure}}
\renewcommand{\theHfigure}{C\arabic{figure}}

\begin{figure}[H]
\centering
    \includegraphics[width = 0.26\textwidth]{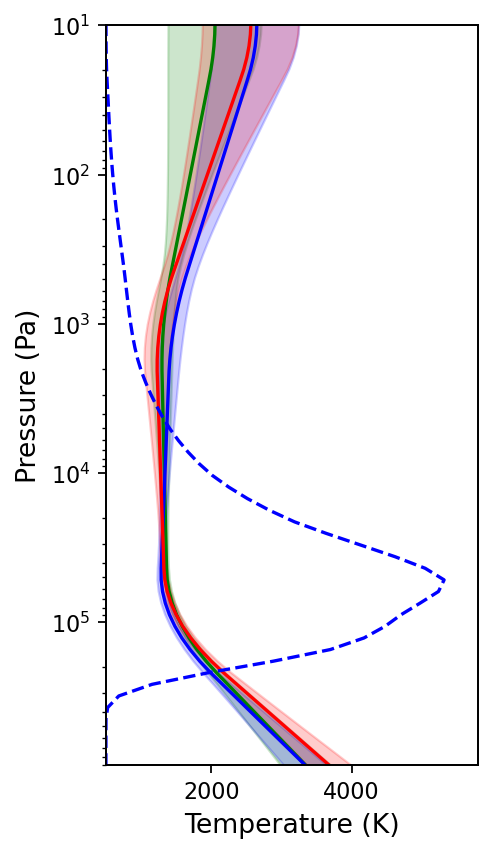}
    \includegraphics[width = 0.26\textwidth]{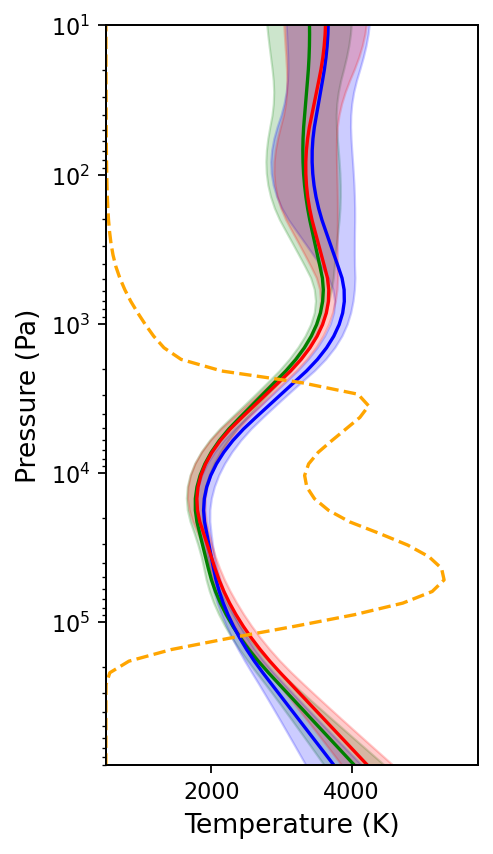}
    \includegraphics[width = 0.26\textwidth]{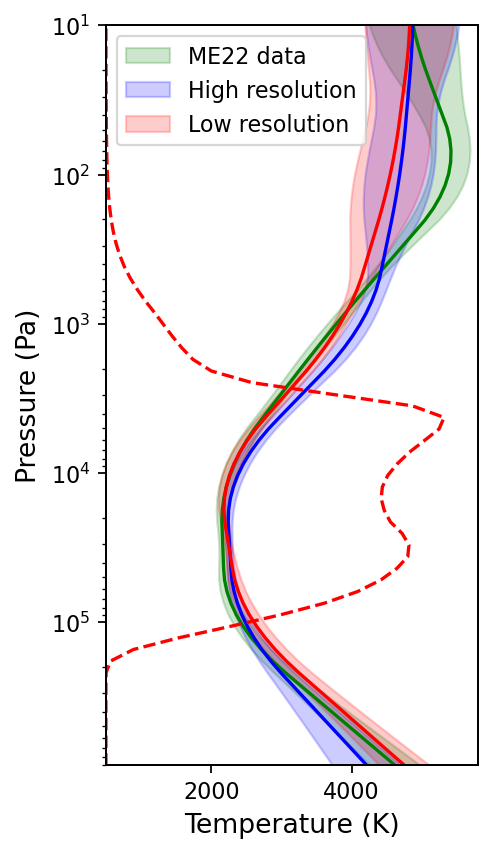}
    \caption{Temperature -- pressure profiles ($T-p$) obtained by the 
    1.5D retrievals on the spectra from different reductions. 
    Left panel: nightside. Middle panel: dayside. 
    Right panel: hotspot. 
    We show the extent of the radiative contribution function for 
    the {\it low}-resolution retrieval with dashed lines. 
    The retrievals on those different reductions are consistent 
    and provide a similar picture.}
    \label{fig:tp_reductions}
\end{figure}

\begin{figure}[H]
\centering
    \includegraphics[width = \textwidth]{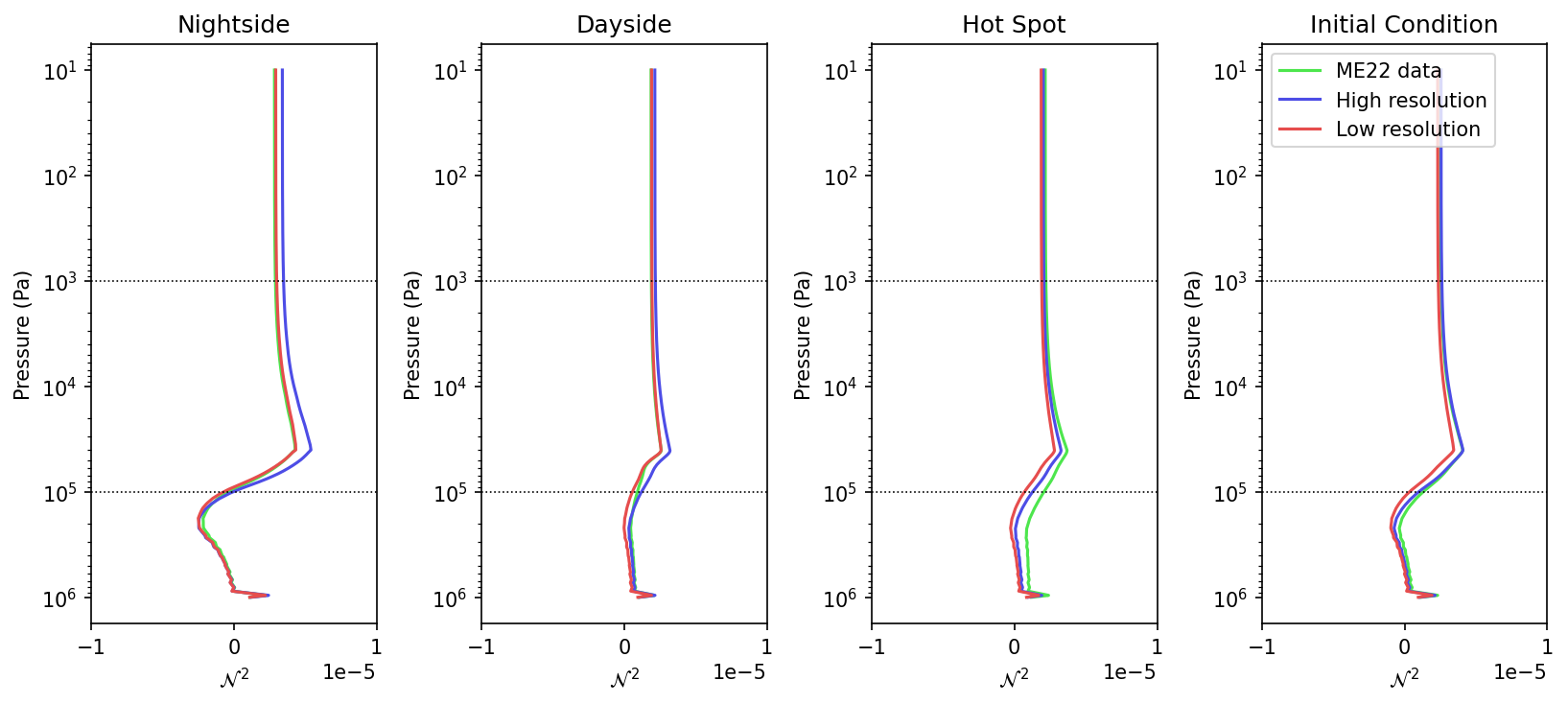}
    \caption{Profiles of Brunt-Väisälä frequency squared 
    $\mathcal{N}^2 = \left(g /\rho \right) \left({\rm d}\rho/{\rm d}z\right)$ 
    for the retrieved thermal profiles in Fig.\ref{fig:tp_reductions}. 
    Dotted lines at $p = 10^1$ and $p = 10^5$ Pa show the upper 
    and lower boundary of the GCM model. 
    In this region, flows are stably stratified ($\mathcal{N}^2 \geq 0$);
    hence, they satisfy the hydrostatic balance approximation of the model equations. 
    For $p > 10^6$, the nightside, hotspot and initial profiles exhibit 
    a jump in stratification. 
    }
    \label{fig:BV_profiles}
\end{figure}

\begin{figure}[H]
\centering
    \includegraphics[width = 0.99\textwidth]{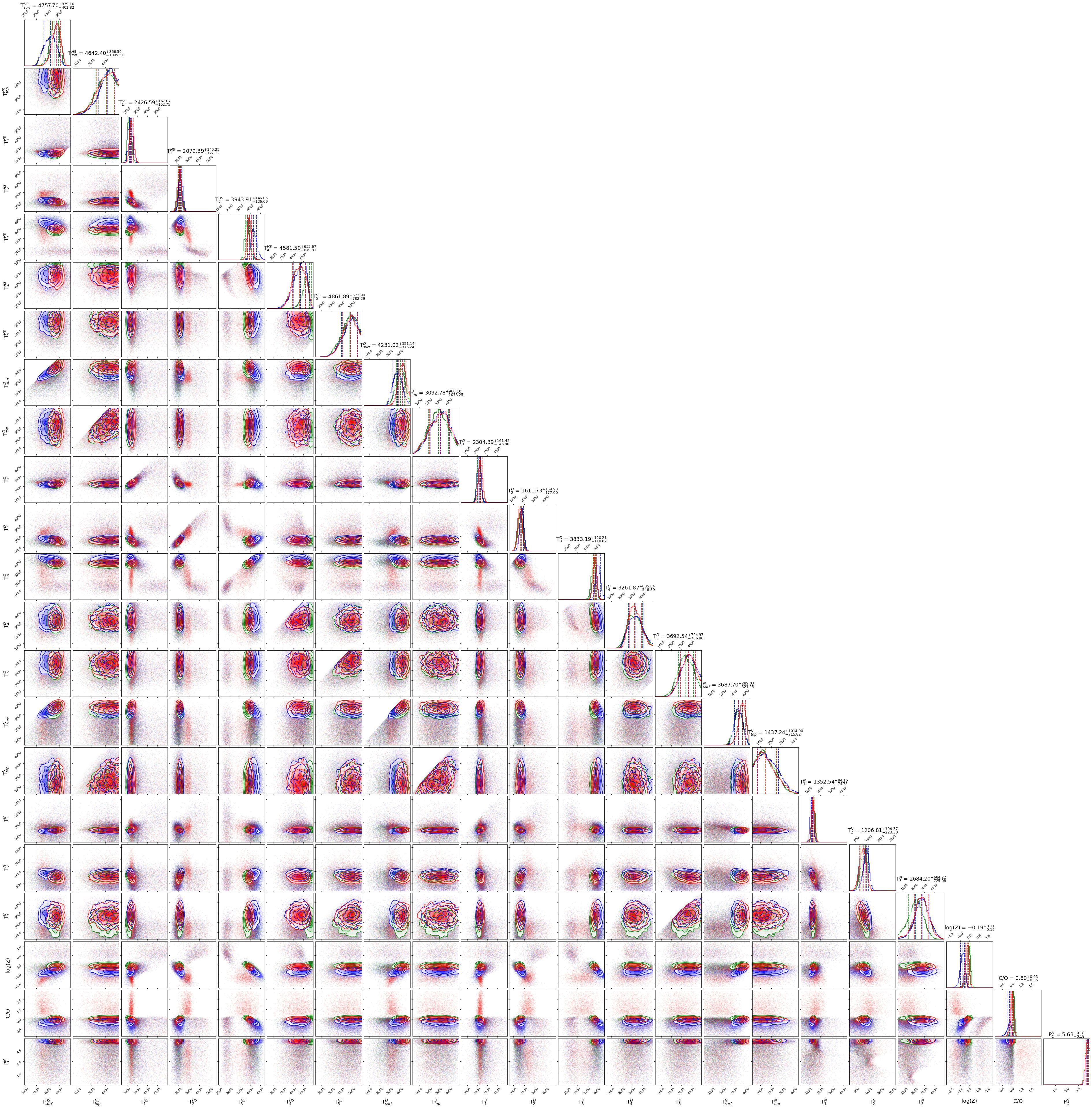}
    \caption{Posterior distributions for the atmosphere of WASP-121\,b 
    obtained by the 1.5D retrievals on different reduction methods. 
    Green: spectra at {\it low} resolution from ME22. 
    Red: spectra at {\it low} resolution obtained using the 4-short 
    instrument systematic model. 
    Blue: spectra at {\it medium} resolution obtained using the 4-short 
    instrument systematic model.}
    \label{fig:posteriors_full_compa}
\end{figure}

\begin{figure*}
    \includegraphics[width = 0.32\textwidth]{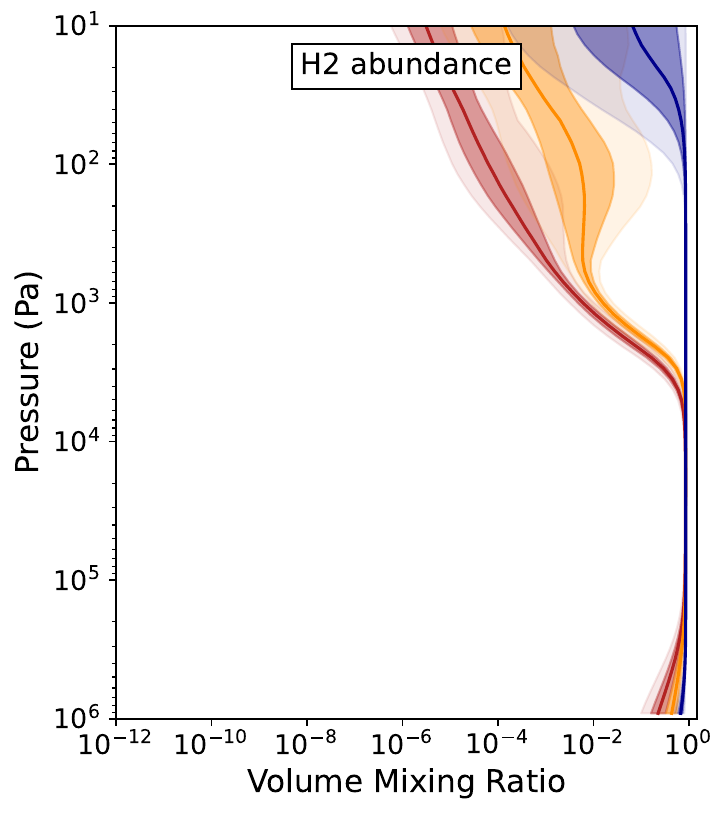}
    \includegraphics[width = 0.32\textwidth]{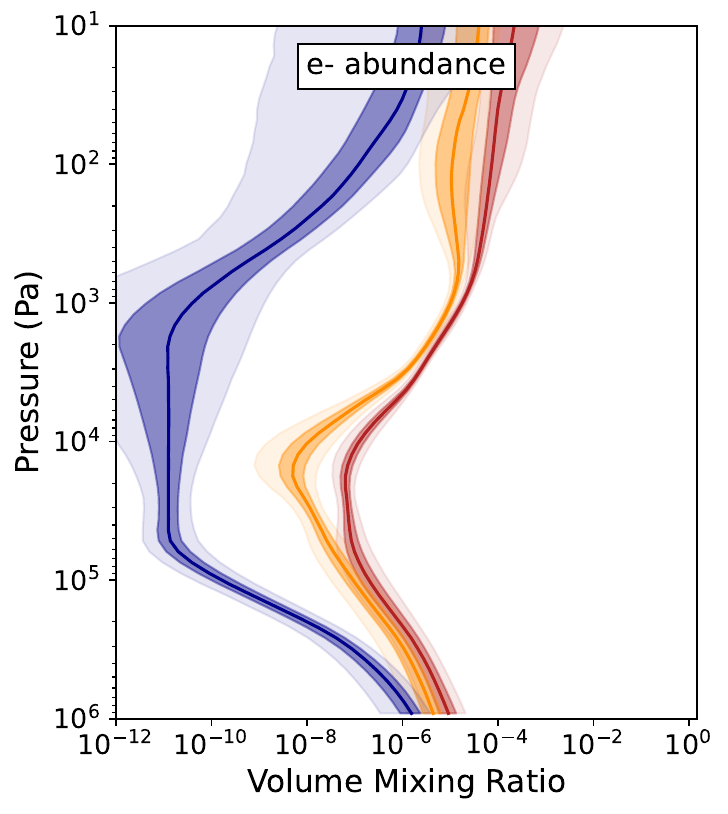}
    \includegraphics[width = 0.32\textwidth]{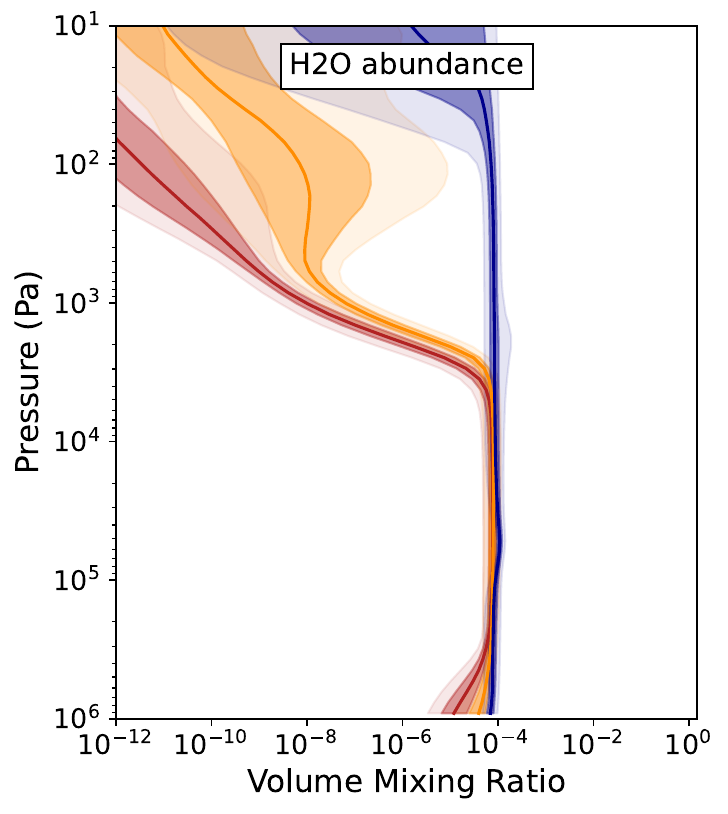}
    
    \includegraphics[width = 0.32\textwidth]{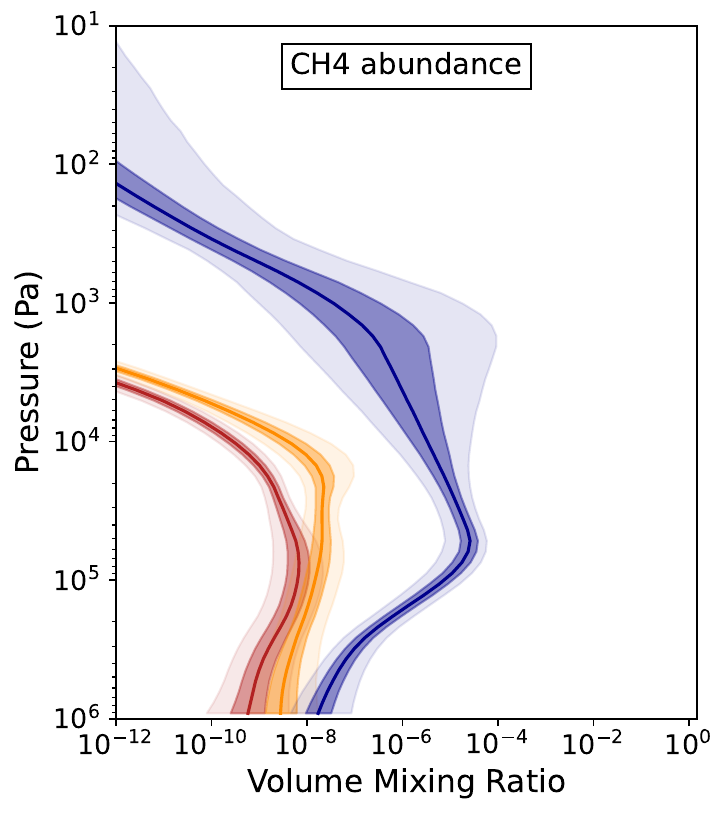}
    \includegraphics[width = 0.32\textwidth]{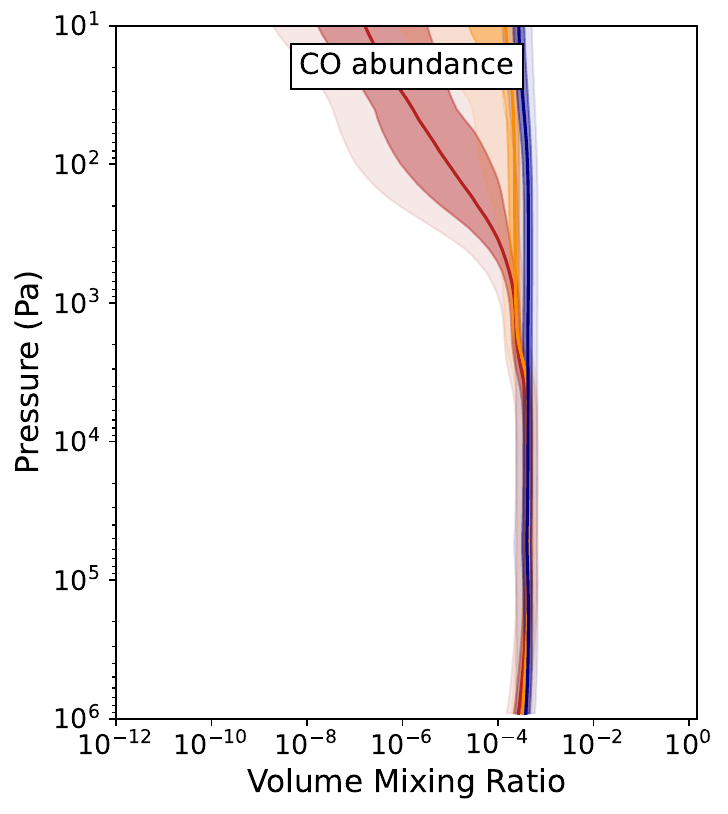}
    \includegraphics[width = 0.32\textwidth]{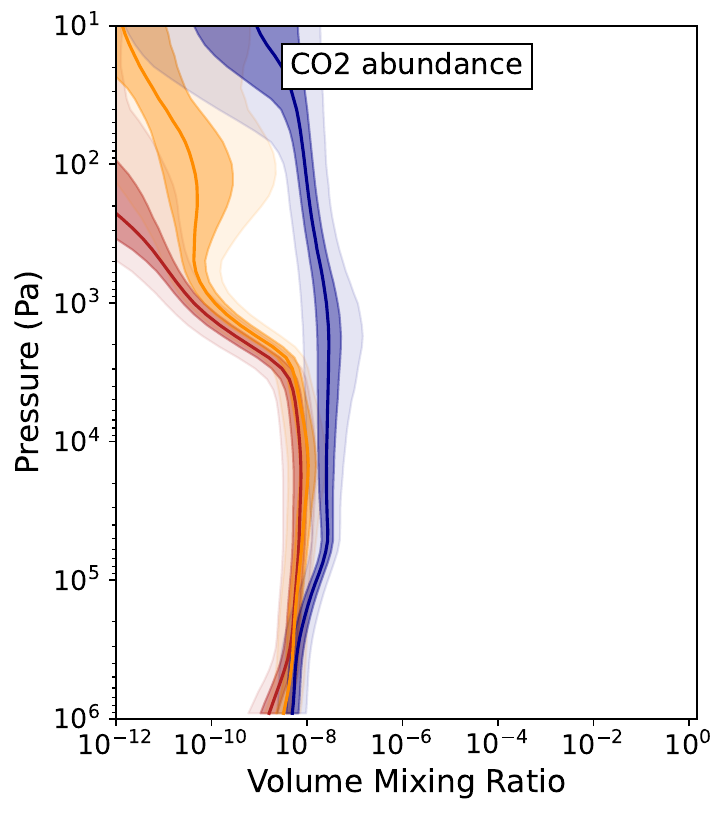}
    
    \includegraphics[width = 0.32\textwidth]{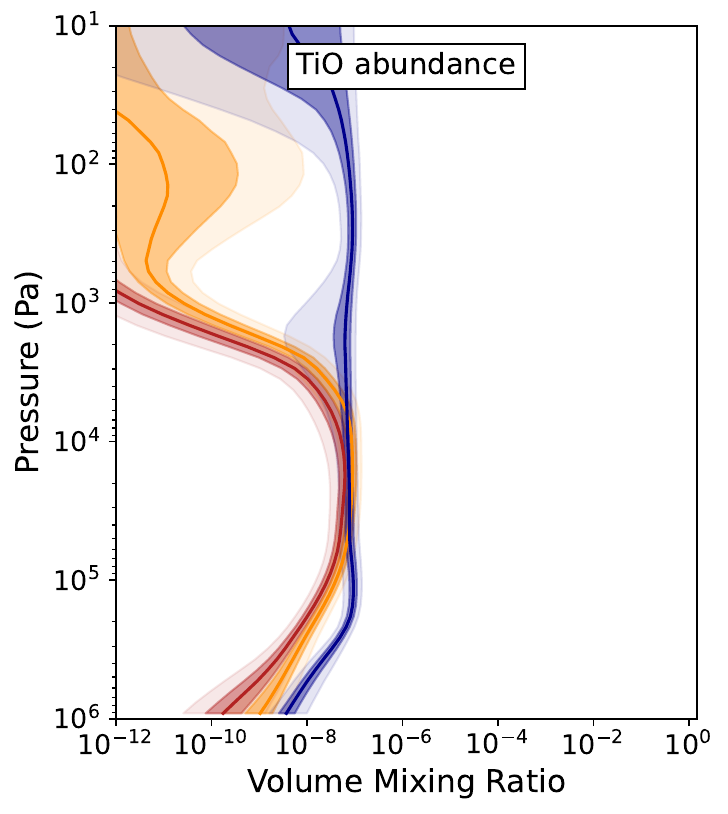}
    \includegraphics[width = 0.32\textwidth]{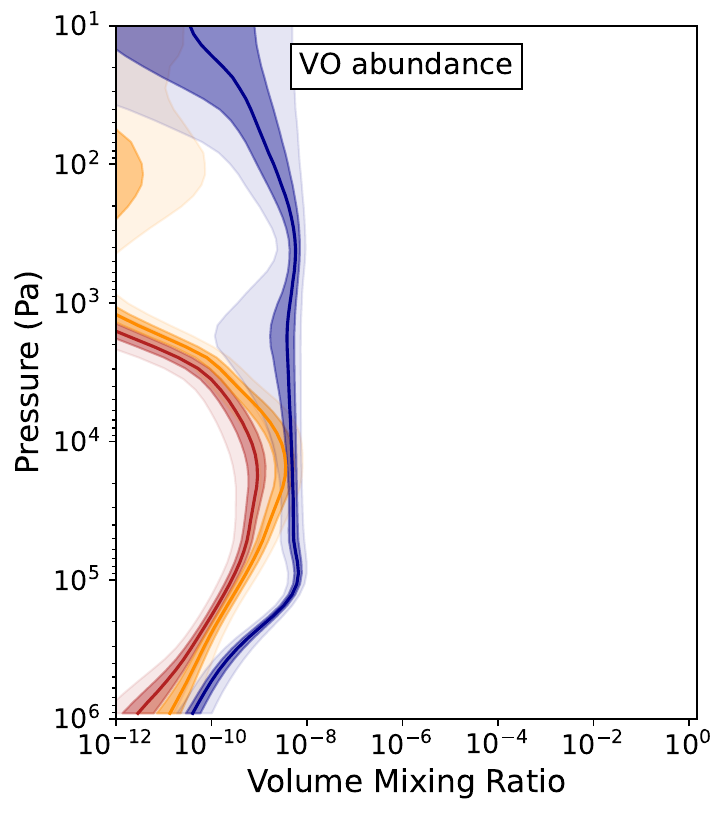}
    \includegraphics[width = 0.32\textwidth]{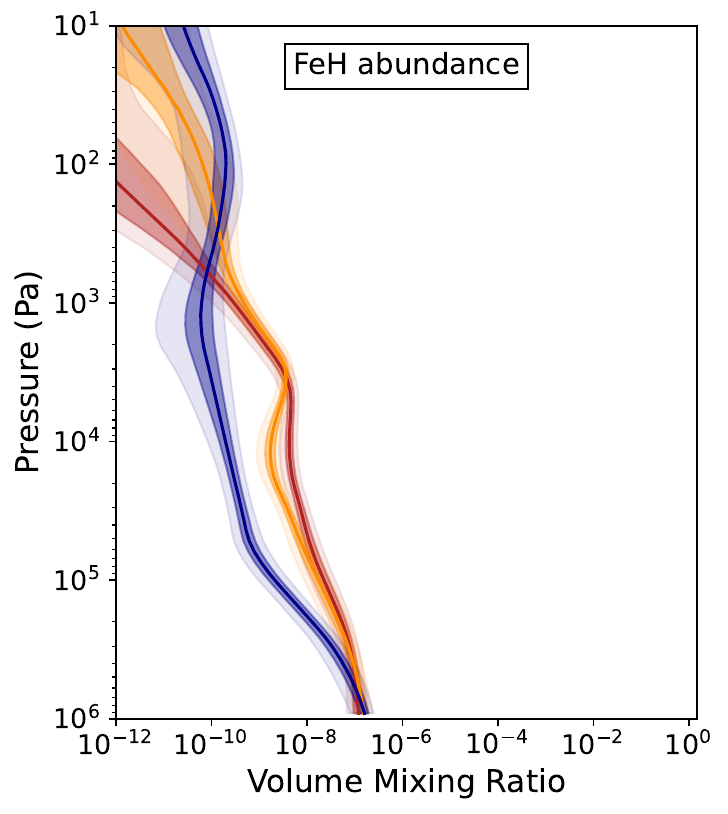}
    
    \caption{Chemistry of the main species recovered from the 
    phase-curve data ({\it low}-resolution reduction). 
    Red: hotspot; Orange: dayside; Blue: nightside. 
    Those chemical profiles show thermal dissociation of main 
    molecules such as H$_2$, H$_2$O, CO, CO$_2$, TiO and VO, 
    at the hotspot of WASP-121\,b.}
    \label{fig:chemical_profiles}
\end{figure*}

\begin{figure}[H]
\centering
    \includegraphics[width = 0.27\textwidth]{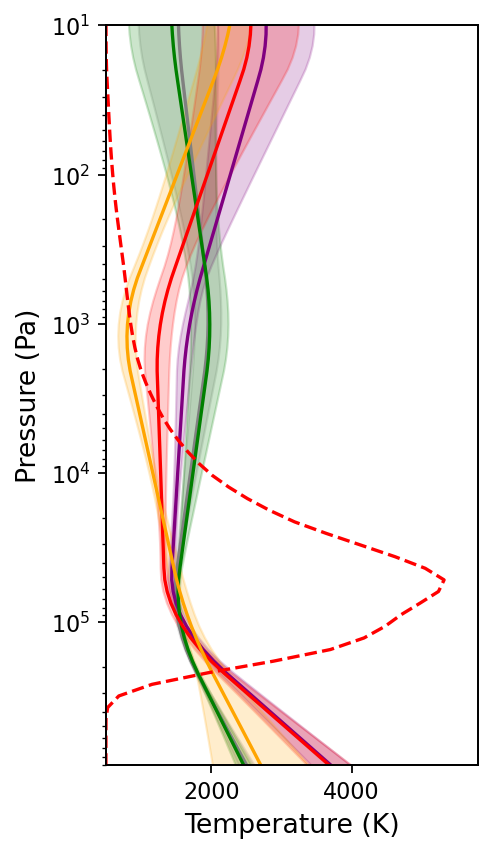}
    \includegraphics[width = 0.27\textwidth]{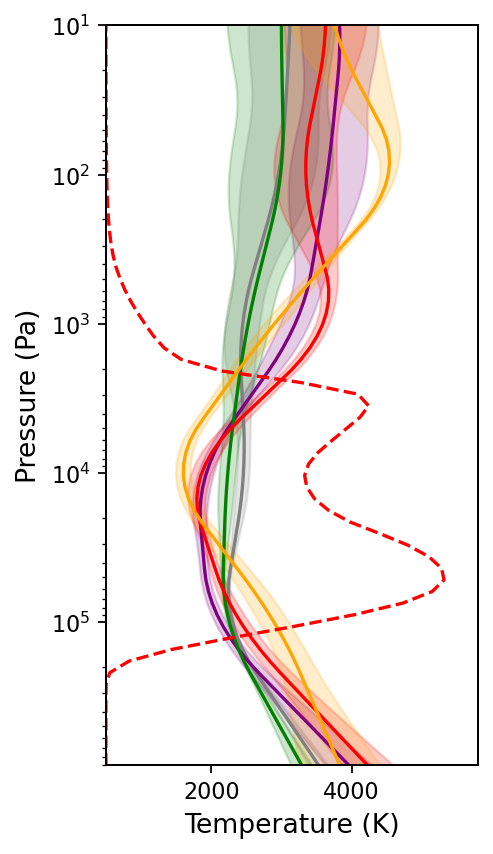}
    \includegraphics[width = 0.27\textwidth]{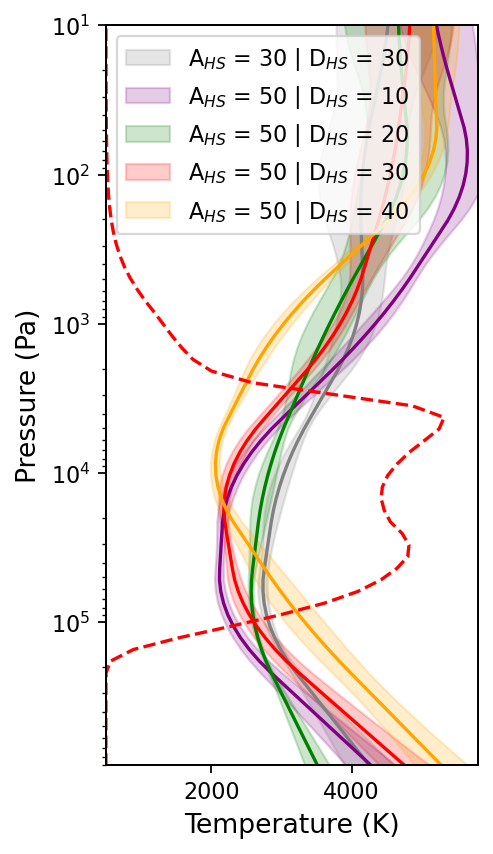}
    \caption{Temperature -- pressure profiles ($T-p$) obtained by 
    the 1.5D retrievals on the {\it low}-resolution spectra by varying 
    the hotspot size ($A_\mathrm{HS}$) and offset ($D_\mathrm{HS}$). 
    Left panel: nightside. Middle panel: dayside. 
    Right panel: hotspot. While the thermal structure are different, 
    the conclusions on the thermal structure of WASP-121\,b from those 
    runs would be the same, independently from the hotspot assumptions. 
    We also note that one model ($A_\mathrm{HS}$ = 50 and $D_\mathrm{HS}$ = 30) has 
    a significantly higher Bayesian evidence. 
    The models obtained: ln(E) = 1550.3 for $A_\mathrm{HS}$ = 30 and 
    $D_\mathrm{HS}$ = 30, ln(E) = 1510.6 for $A_\mathrm{HS}$ = 50 and 
    $D_\mathrm{HS}$ = 10, ln(E) = 1538.8 for $A_\mathrm{HS}$ = 50 and 
    $D_\mathrm{HS}$ = 20, ln(E) = 1554.5 for $A_\mathrm{HS}$ = 50 and 
    $D_\mathrm{HS}$ = 30, and ln(E) = 1537.5 for $A_\mathrm{HS}$ = 50 and 
    $D_\mathrm{HS}$ = 40.}
    \label{fig:tp_reductions_ADCompa}
\end{figure}

\begin{figure*}
\centering
    \includegraphics[width = 0.99\textwidth]{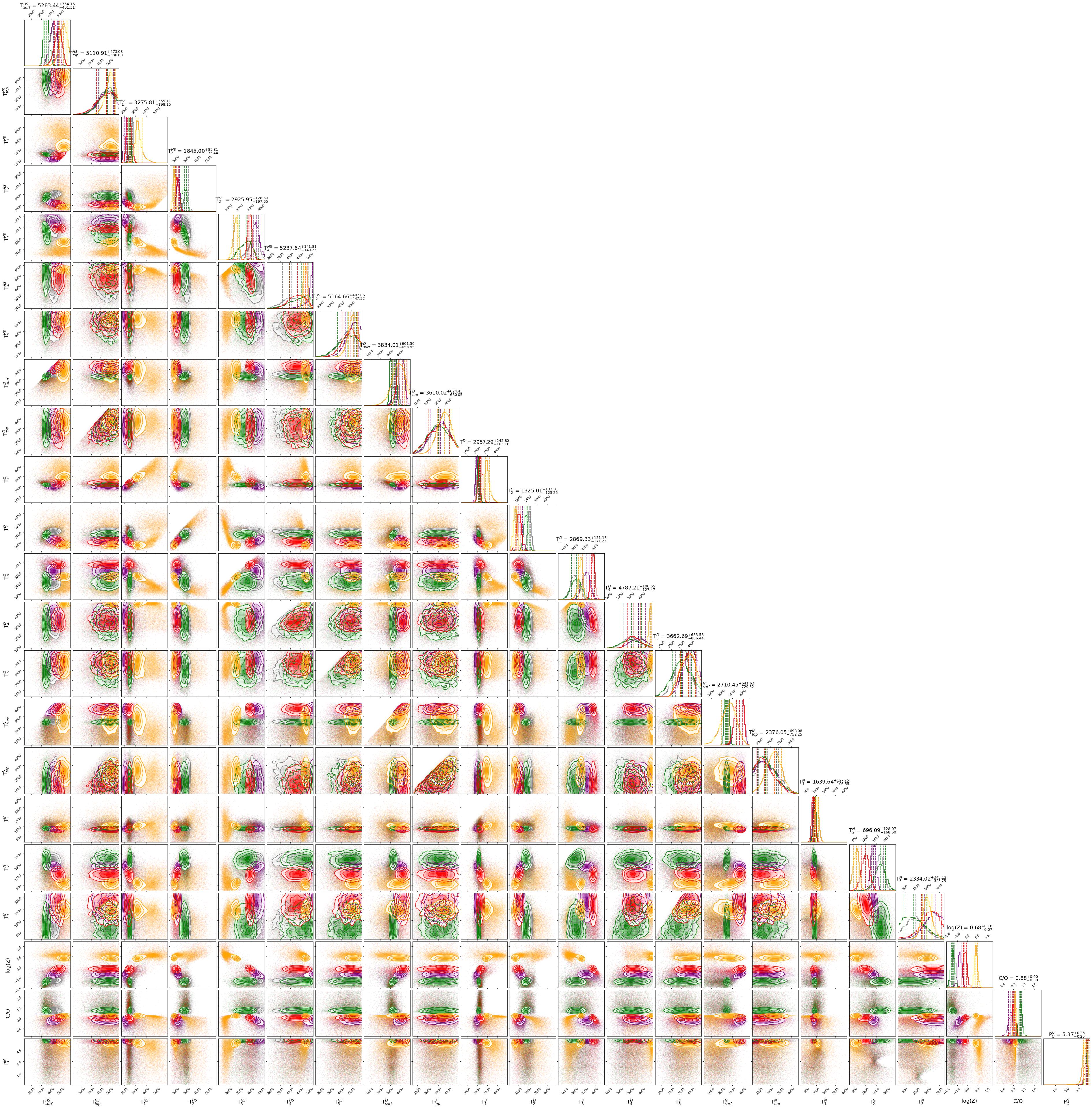}
    \caption{Posterior distributions for the atmosphere of 
    WASP-121\,b obtained on the {\it low}-resolution spectra by 
    varying the hotspot size ($A_\mathrm{HS}$) and offset ($D_\mathrm{HS}$). 
    The color code is the same as in 
    Figure~\ref{fig:tp_reductions_ADCompa}. 
    Independently from the hotspot assumptions, all those 
    retrievals of WASP-121\,b phase-curve data have a super-solar 
    C/O with 0.62 $<$ C/O $<$ 1.11, while the retrieved metallicity 
    is between C/O with -1.27 $<$ log(Z) $<$ 0.77. 
    For all cases, we do not find evidence of a fully opaque 
    nightside cloud deck. 
    One model ($A_\mathrm{HS}$ = 50 and $D_\mathrm{HS}$ = 30) has a 
    significantly higher Bayesian evidence. 
    The models obtained: ln(E) = 1550.3 for $A_\mathrm{HS}$ = 30 and 
    $D_\mathrm{HS}$ = 30, ln(E) = 1510.6 for $A_\mathrm{HS}$ = 50 and 
    $D_\mathrm{HS}$ = 10, ln(E) = 1538.8 for $A_\mathrm{HS}$ = 50 and 
    $D_\mathrm{HS}$ = 20, ln(E) = 1554.5 for $A_\mathrm{HS}$ = 50 and 
    $D_\mathrm{HS}$ = 30, and ln(E) = 1537.5 for $A_\mathrm{HS}$ = 50 and 
    $D_\mathrm{HS}$ = 40.}
    \label{fig:posteriors_full_compa_ADCompa}
\end{figure*}

\begin{figure*}
\centering
    \includegraphics[width = 0.47\textwidth]{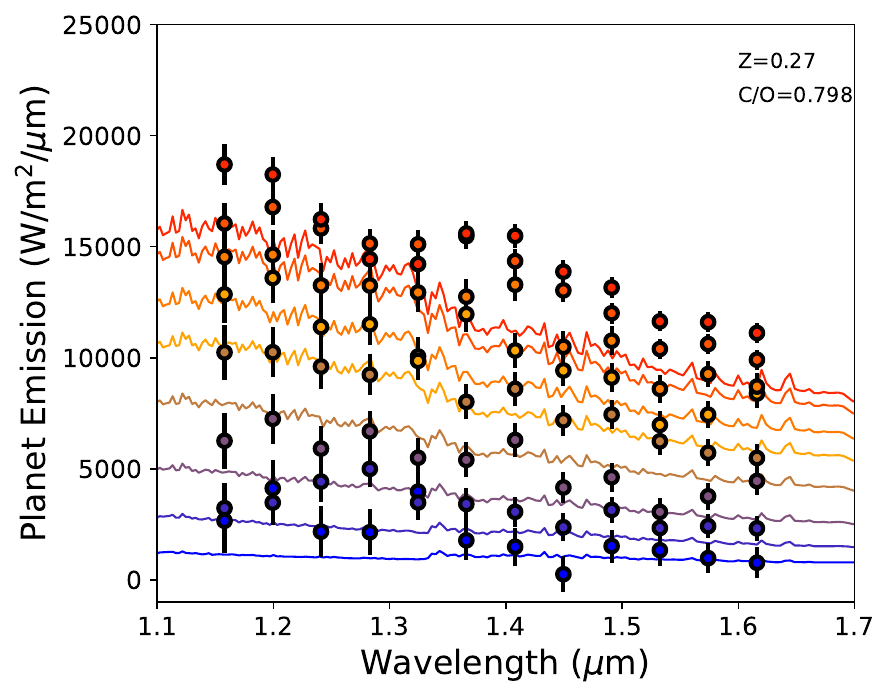}
    \includegraphics[width = 0.47\textwidth]{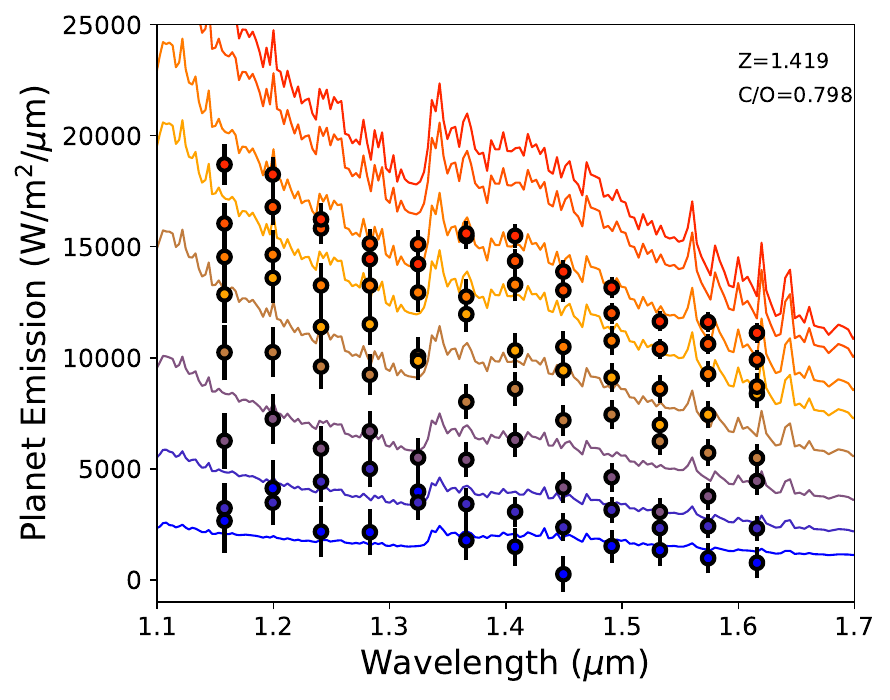}
    \includegraphics[width = 0.47\textwidth]{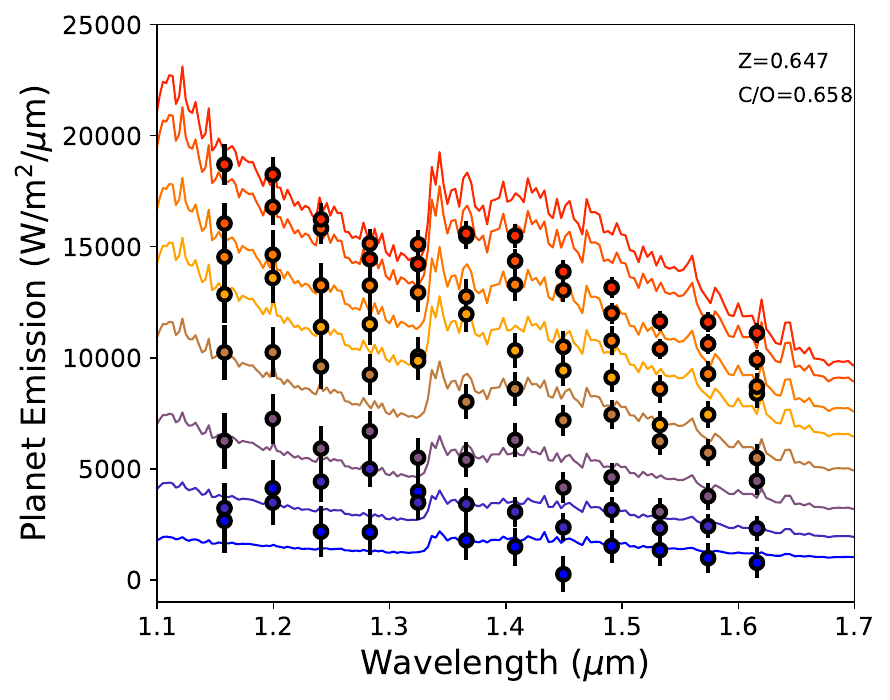}
    \includegraphics[width = 0.47\textwidth]{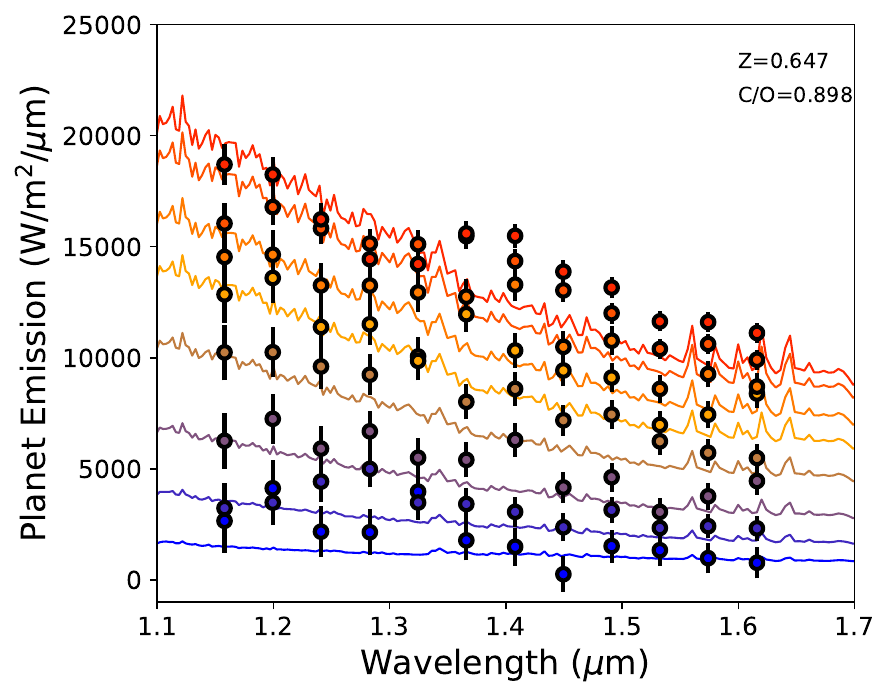}
    \caption{Sensitivity tests for four forward models, modified 
    at 3$\times\sigma$ from the best fit chemistry values inferred 
    in the retrieval (on the {\it low}-resolution data). 
    Top left: the metallicity is reduced to Z = 0.270. 
    Top right: the metallicity is increased to Z = 1.419. 
    Bottom left: the C/O ratio is reduced to C/O = 0.658. 
    Bottom right: the C/O ratio is increased to C/O = 0.898. 
    In all four cases, the simulated phase-curve spectra do not 
    match the observations, highlighting the sensitivity of those 
    datasets to chemistry parameters.}
    \label{fig:sensitivity}
\end{figure*}
\clearpage

\section{Complementary Figures to the Section 5}

This appendix contains the complementary figures to the main article 
Section 5, Figure~\ref{fig:spectra_ec_tr} to Figure~\ref{fig:posteriors_eclipses}.

\setcounter{figure}{0}
\renewcommand{\thefigure}{D\arabic{figure}}
\renewcommand{\theHfigure}{D\arabic{figure}}

\begin{figure}[H]
    \includegraphics[width = 0.49\textwidth]{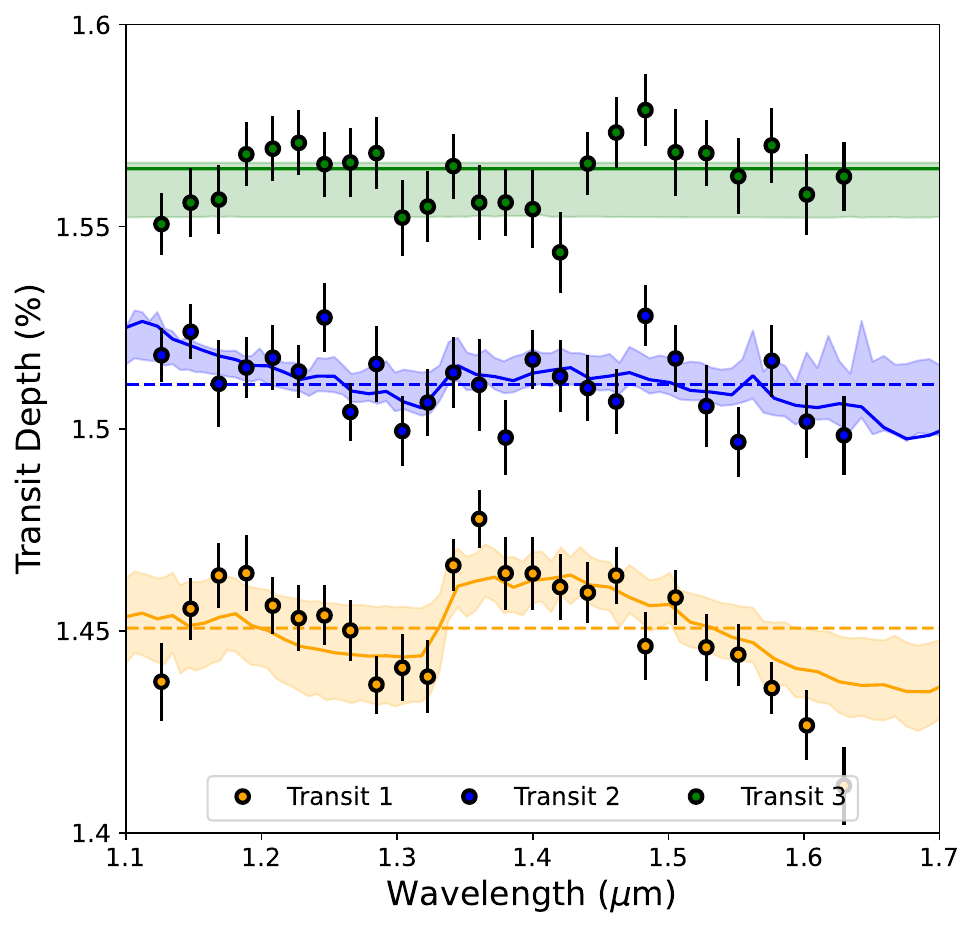}
    \includegraphics[width = 0.49\textwidth]{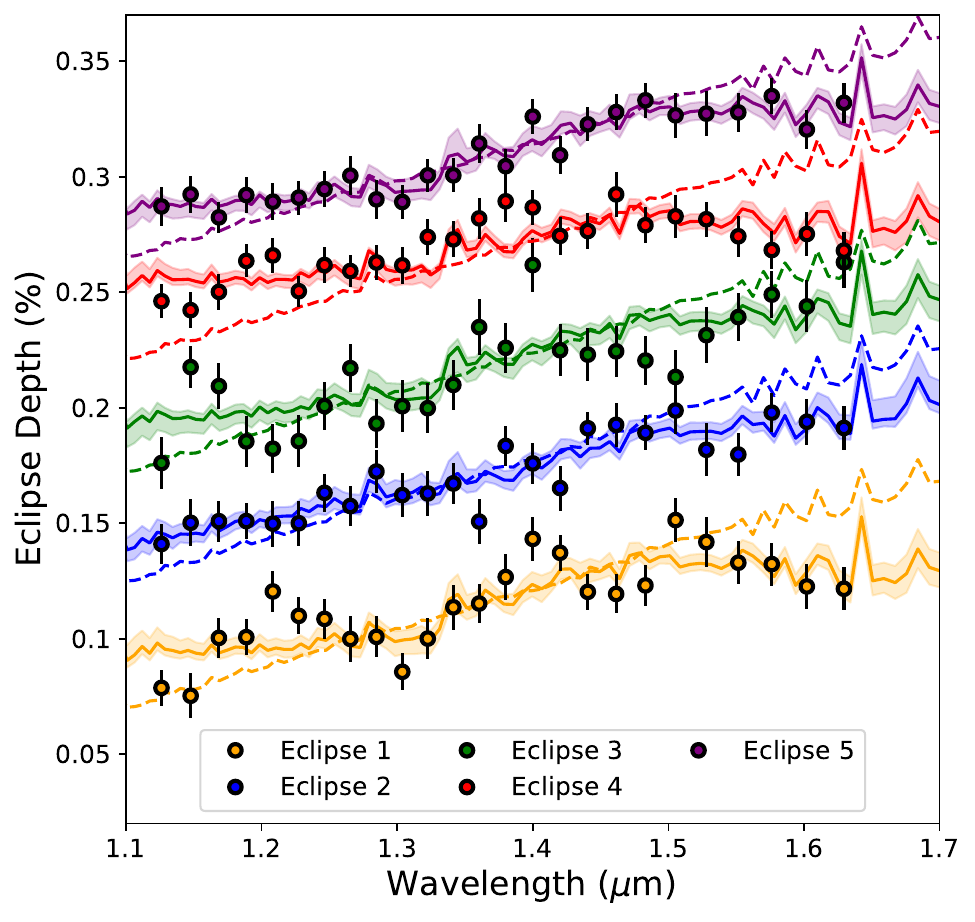}
    \caption{
    Transit (left) and eclipse (right) spectra of WASP-121\,b analyzed 
    in this work. Different observations are offset in the $y$-axis. 
    Best-fit models from the 1D retrievals are shown in solid lines. 
    Dashed lines show featureless models for visual comparison.}
    \label{fig:spectra_ec_tr}
\end{figure}

\begin{figure*}
    \centering
    \includegraphics[width = 0.7\textwidth]{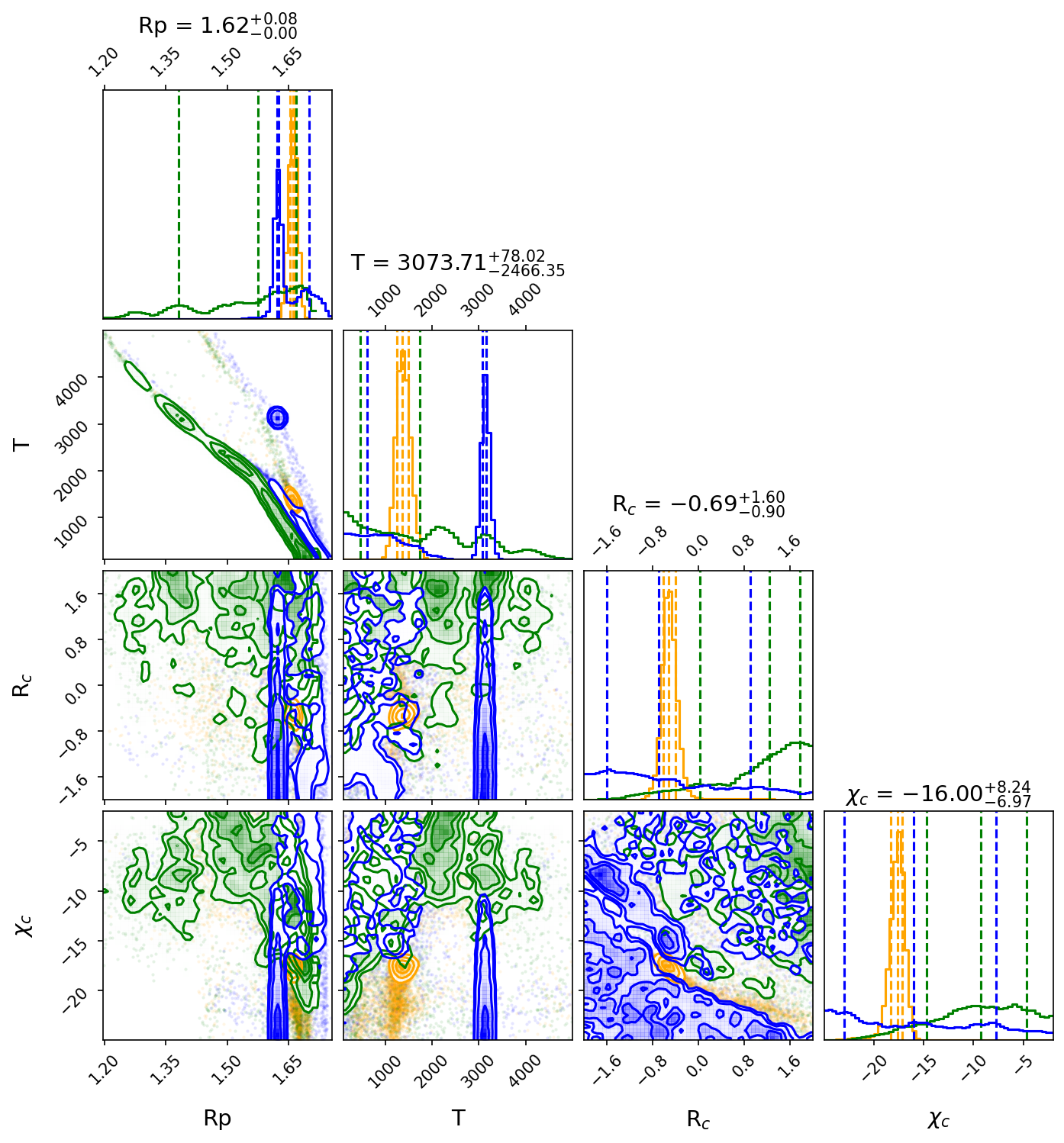}
    \caption{Posterior distributions obtained for the three transit 
    fits. 
    Yellow: Transit 1. 
    Blue: Transit 2. 
    Green: Transit 3. 
    Note that the chemistry parameters of the equilibrium model 
    \textsc{GGChem} in those fits are fixed to the median values obtained 
    by the 1.5D phase-curve retrieval (log Z = -0.19, C/O = 0.80). 
    Transit 1 shows an atmosphere with moderate hazes but absorption 
    at high altitude from water. 
    Transit 2 shows multimodal solutions involving either a fully 
    ionized atmosphere with un-physically high temperatures, or a cloudy/hazy atmosphere. 
    Transit 3 displays an atmosphere with opaque clouds. }
    \label{fig:posteriors_transits}
\end{figure*}

\begin{figure*}
    \centering
    \includegraphics[width = 0.9\textwidth]{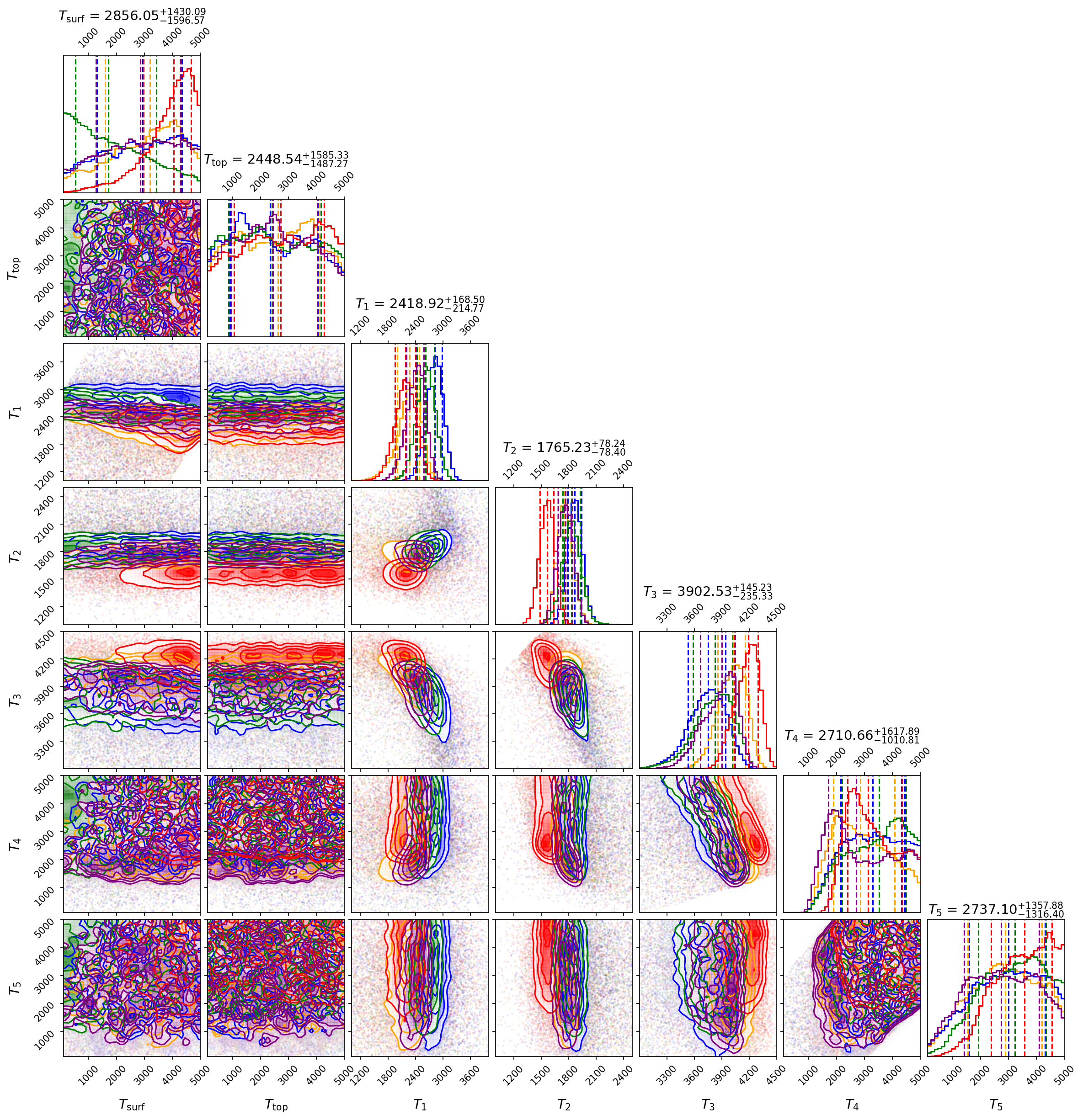}
    \caption{Posterior distributions obtained for the five eclipse 
    fits. Yellow: Eclipse 1. 
    Blue: Eclipse 2. 
    Green: Eclipse 3. 
    Red: Eclipse 4. 
    Purple: Eclipse 5. 
    Note that the chemistry parameters of the equilibrium model 
    \textsc{GGChem} in those fits are fixed to the median values obtained 
    by the 1.5D phase-curve retrieval (log Z = -0.19, C/O = 0.80). 
    The five eclipses are consistent with similarly inverted 
    thermal structures, however the posterior distributions are 
    not the same in the middle of the atmosphere (T$_1$ to T$_3$). 
    Large-scale atmospheric variability could creates those 
    observable departures in the thermal profiles. }
    \label{fig:posteriors_eclipses}
\end{figure*}
\clearpage

\section{Complementary Figures to the Section 6}

This appendix contains the complementary figures for Section 6 of 
the main article, Figure~\ref{fig:dynamics_press} to Fig.
\ref{fig:dynamics_chemistry_maps3}.

\setcounter{figure}{0}
\renewcommand{\thefigure}{E\arabic{figure}}
\renewcommand{\theHfigure}{E\arabic{figure}}

\begin{figure}[H]
    %\centering
    \begin{interactive}{animation}{gcm_model.mp4}
    \includegraphics[width = 0.82\textwidth]{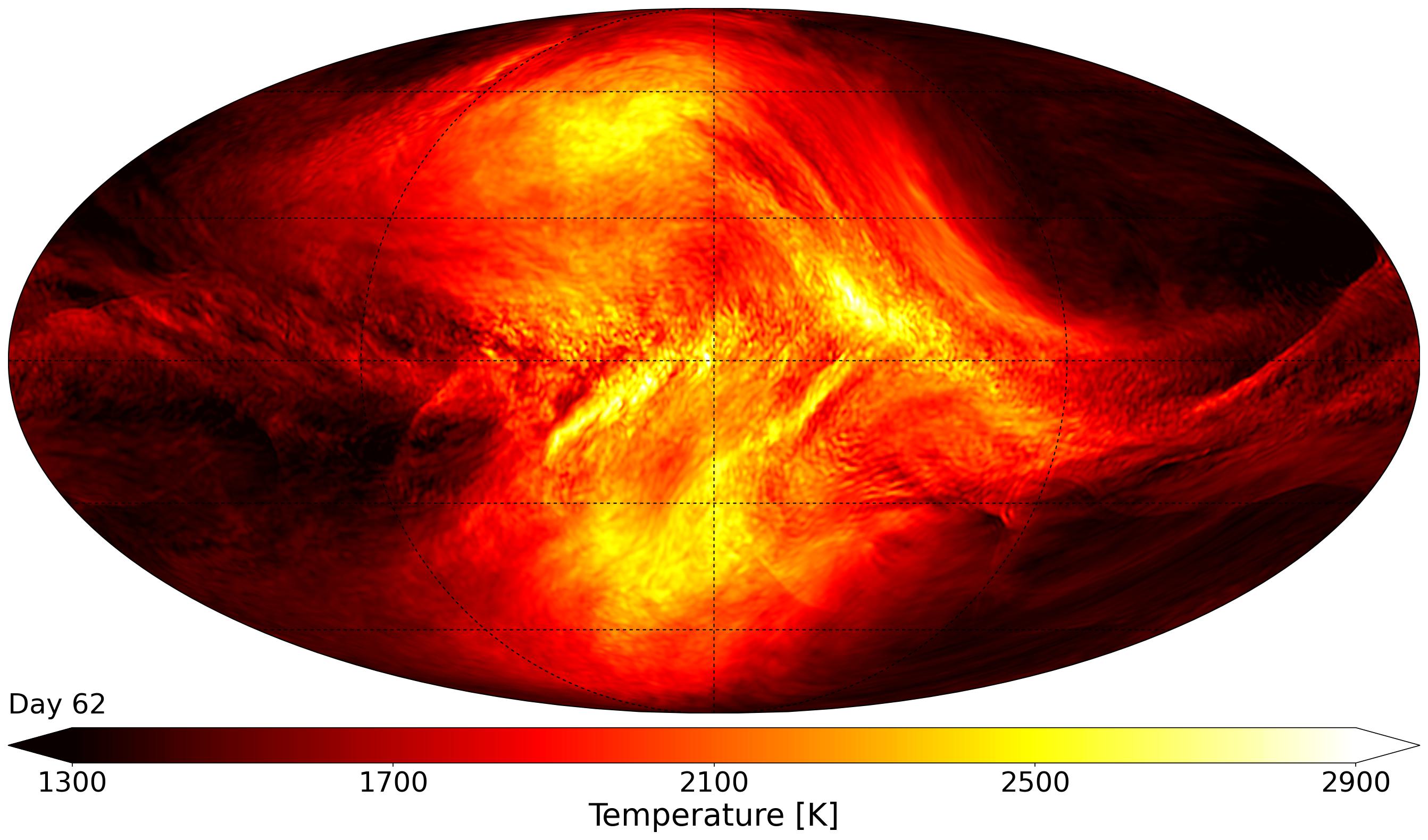}
    \includegraphics[width = 0.82\textwidth]{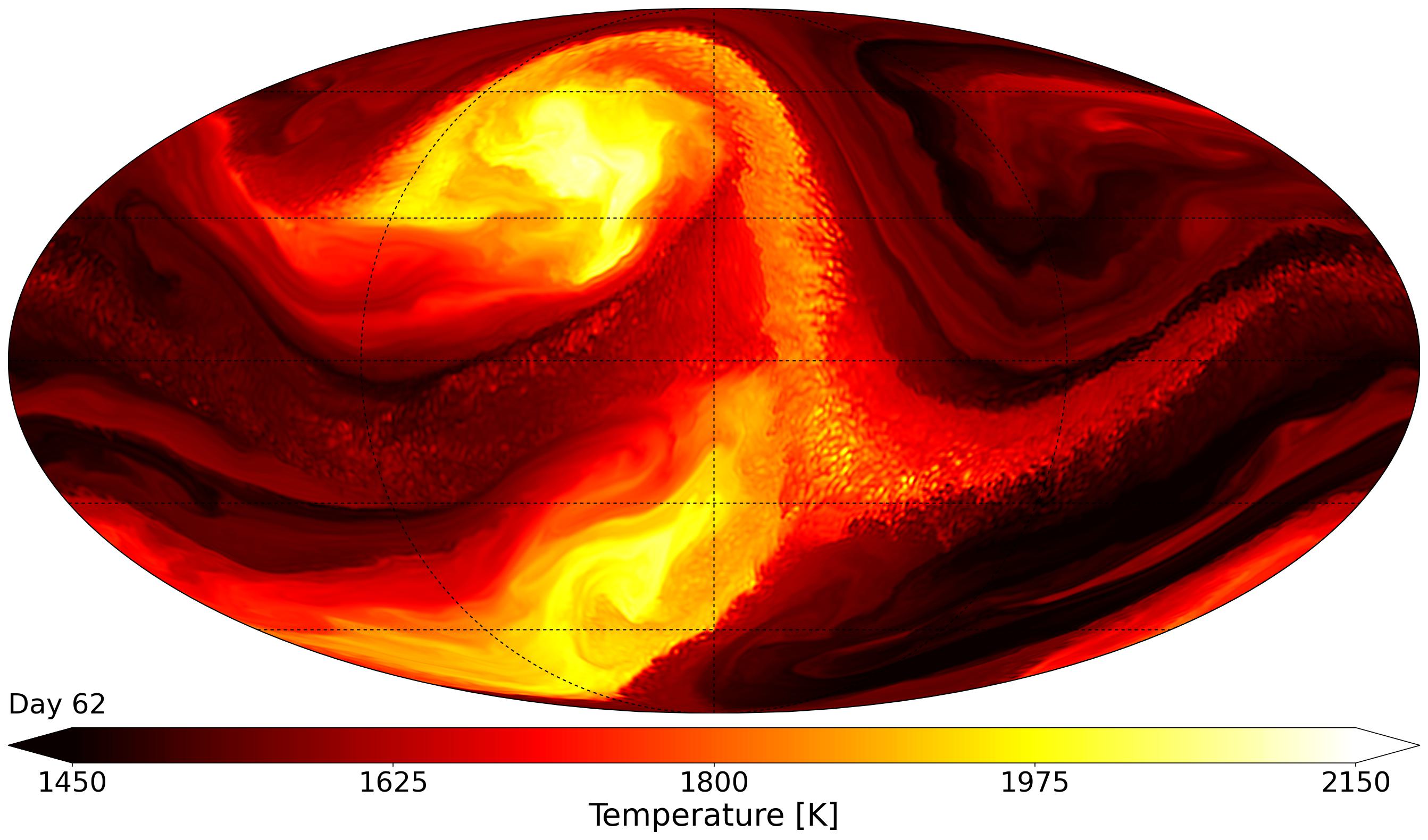}
    \end{interactive}
    \caption{Temperature maps of WASP-121\,b in Mollweide projection, 
    obtained at $p = 5\!\times\! 10^3 \, {\rm Pa}$ (top) and 
    $p = 10^5 \, {\rm Pa}$ (bottom) for $t \in [40, 185]$\,days. The figure is accompanied by two 1\,min 23\,s videos, available online at the journal, showing the evolution of the atmosphere during 145\,days in the simulations.
    The movie shows the highly time-variable atmosphere of 
    WASP-121\,b, expected from a high-resolution flow simulation.  
    Note that the temperature ranges are different. }
    \label{fig:dynamics_press}
\end{figure}

\clearpage

\centering
\begin{figure}[H]
    \centering
    \includegraphics[width = 0.90\textwidth]{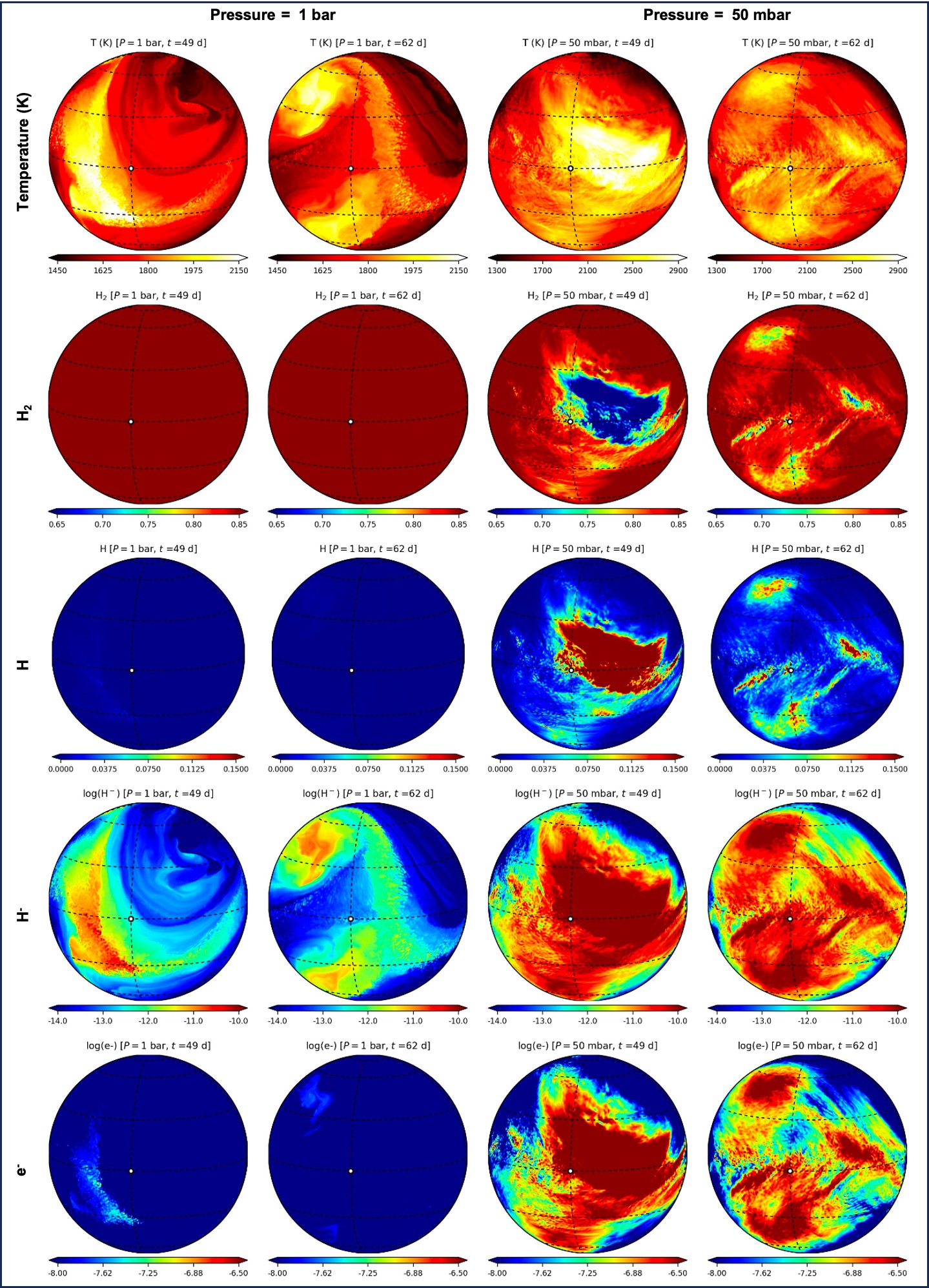}
    \caption{WASP-121\,b chemical maps centered around the 
    sub-stellar point for H$_2$, H, H$^-$ and e$^-$ at two 
    different times ($t = 49$ and $t = 62$\,days) and two 
    pressure levels ($p = 10^5$\,Pa and $p = 5\!\times\! 10^3$\,Pa). 
    Those maps are obtained by post-processing the temperature fields 
    (top row) from the \textsc{BoB} dynamics calculations with 
    the \textsc{TauREx} library.}
    \label{fig:dynamics_chemistry_maps1}
\end{figure}
\clearpage

\centering
\begin{figure}[H]
    \centering
    \includegraphics[width = 0.90\textwidth]{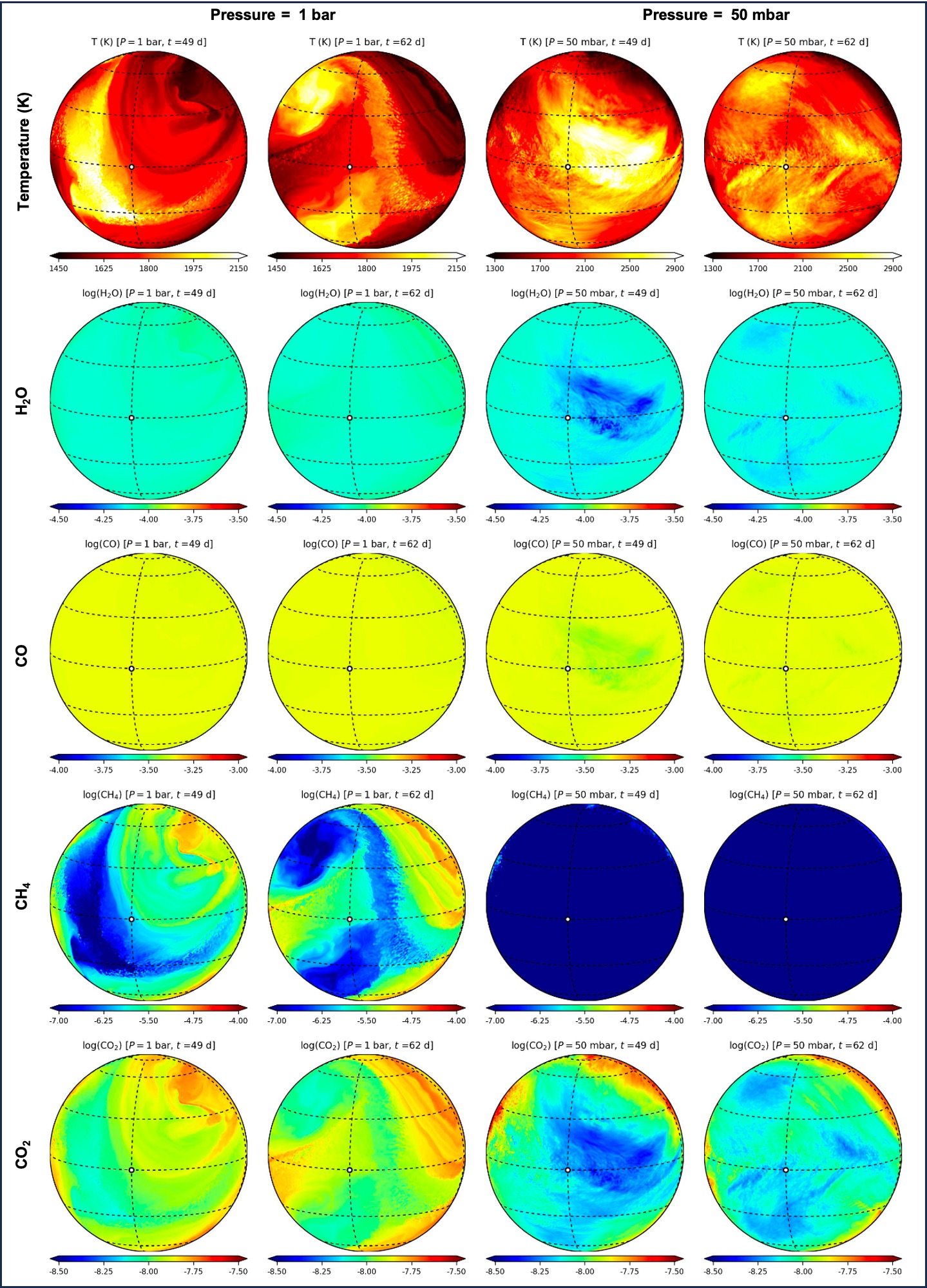}
    \caption{WASP-121\,b chemical maps centered around the 
    sub-stellar point for H$_2$O, CO, CH$_4$ and CO$_2$ at two 
    different times ($t = 49$ and $t = 62$ days) and two pressure 
    levels ($p = 10^5$ and $p = 5\times10^3$ Pa). 
    Those maps are obtained by post-processing the temperature 
    fields (top row) obtained from the \textsc{BoB} dynamics 
    calculations with the \textsc{TauREx} library.}
    \label{fig:dynamics_chemistry_maps2}
\end{figure}
\clearpage

\centering
\begin{figure}[H]
    \centering
    \includegraphics[width = 0.90\textwidth]{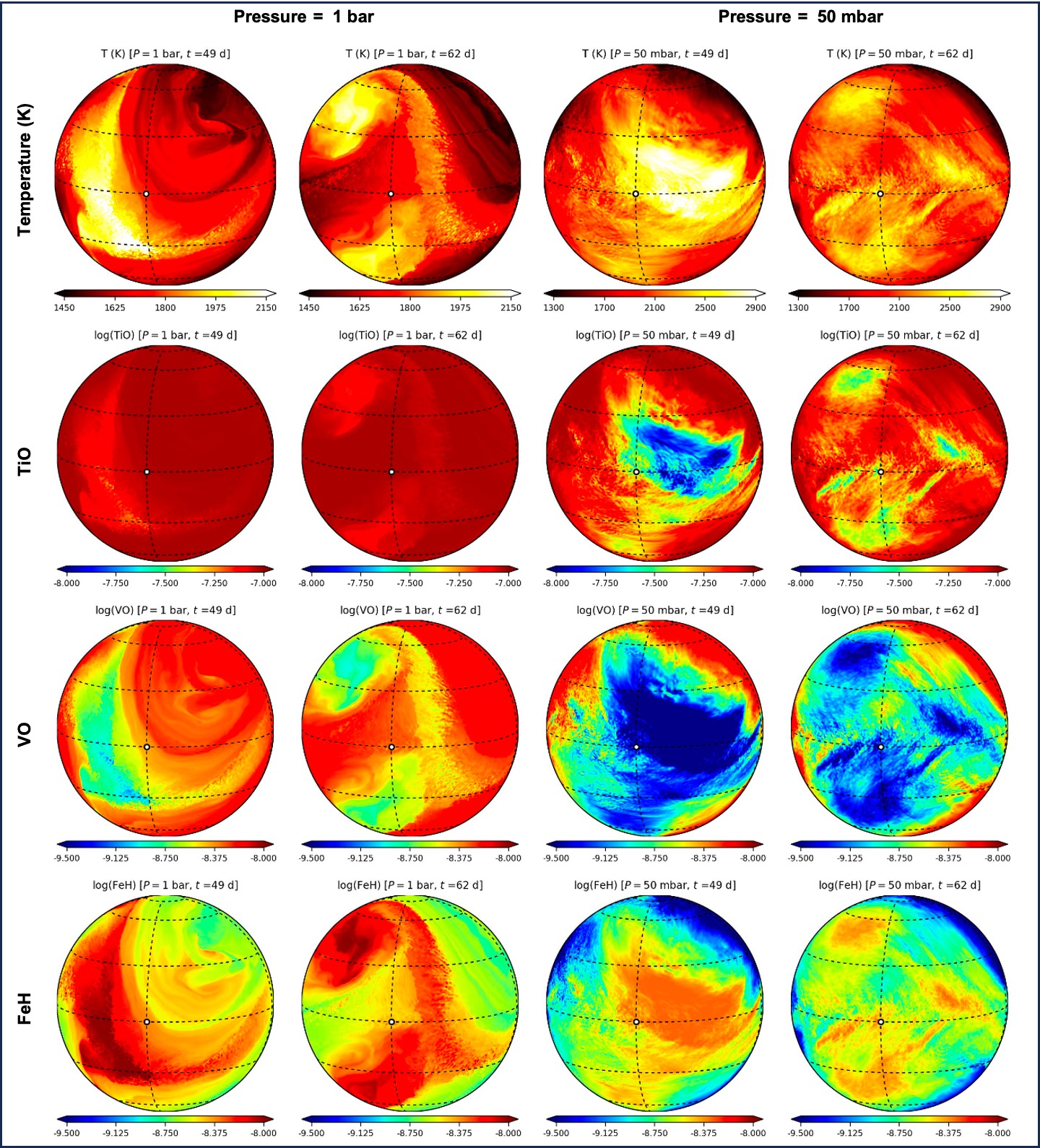}
    \caption{WASP-121\,b chemical maps centered around the 
    sub-stellar point for TiO, VO and FeH at two different times 
    ($t = 49$ and $t = 62$ days) and two pressure levels 
    ($p = 10^5$ and $p = 5\times10^3$ Pa). 
    Those maps are obtained by post-processing the temperature 
    fields (top row) obtained from the \textsc{BoB} dynamics 
    calculations with the \textsc{TauREx} library.}
    \label{fig:dynamics_chemistry_maps3}
\end{figure}
\clearpage

\end{document}